\documentclass{article}

    \PassOptionsToPackage{numbers, compress}{natbib}

\usepackage[preprint]{neurips_2024}

\usepackage{pifont}
\usepackage{xcolor} 
\newcommand{\greencheck}{{\color{green}\ding{51}}} 
\newcommand{\redcross}{{\color{red}\ding{55}}} 
\usepackage[utf8]{inputenc} 
\usepackage[T1]{fontenc}    
\usepackage{hyperref}       
\usepackage{url}            
\usepackage{booktabs}       
\usepackage{amsfonts}       
\usepackage{nicefrac}       
\usepackage{microtype}      
\usepackage{xcolor}         
\usepackage{graphicx} 
\usepackage{amsmath}
\usepackage{adjustbox}
\usepackage{multirow}
\usepackage{subcaption}
\usepackage{float}

\usepackage{makecell}

\title{WavChat: A Survey of Spoken Dialogue Models}

\author{
\hspace*{-0.6cm} 
Shengpeng Ji~$^{\spadesuit}$~\thanks{Equal contribution.}
~~Yifu Chen~$^{\spadesuit}$~\footnotemark[1]
~~Minghui Fang~$^{\spadesuit}$~\footnotemark[1]
~~Jialong Zuo~$^{\spadesuit}$~\footnotemark[1]
~~Jingyu Lu~$^{\spadesuit}$
~~Hanting Wang ~$^{\spadesuit}$
\\
\textbf{
\hspace*{-0.2cm} 
Ziyue Jiang~$^{\spadesuit}$
~~Long Zhou~$^{\diamondsuit}$ 
~~Shujie Liu~$^{\diamondsuit}$ 
~~Xize Cheng~$^{\spadesuit}$ 
~~Xiaoda Yang~$^{\spadesuit}$ 
~~Zehan Wang~$^{\spadesuit}$ 
}
\\
\textbf{
\hspace*{-0.4cm} 
Qian Yang~$^{\spadesuit}$ 
~~Jian Li~$^{\clubsuit}$
~~Yidi Jiang~$^{\heartsuit}$ 
~~Jingzhen He~$^{\heartsuit}$ 
~~Yunfei Chu~$^{\heartsuit}$ 
~~Jin Xu~$^{\heartsuit}$ 
~~Zhou Zhao~$^\spadesuit$\thanks{Corresponding author.}
} \\
$^\spadesuit$~Zhejiang University~\& $^\diamondsuit$~Microsoft~ \& $^{\heartsuit}$~Alibaba Group~\& $^{\clubsuit}$~Tencent YouTu Lab\\
\texttt{\{shengpengji,zhaozhou\}@zju.edu.cn}
}

\begin{document}

\maketitle

\begin{abstract}
Recent advancements in spoken dialogue models, exemplified by systems like GPT-4o, have captured significant attention in the speech domain. In the broader context of multimodal models, the speech modality offers a direct interface for human-computer interaction, enabling direct communication between AI and users. Compared to traditional three-tier cascaded spoken dialogue models that comprise speech recognition (ASR), large language models (LLMs), and text-to-speech (TTS), modern spoken dialogue models exhibit greater intelligence. These advanced spoken dialogue models not only comprehend audio, music, and other speech-related features, but also capture stylistic and timbral characteristics in speech. Moreover, they erate high-quality, multi-turn speech responses with low latency, enabling real-time interaction through simultaneous listening and speaking capability.
Despite the progress in spoken dialogue systems, there is a lack of comprehensive surveys that systematically organize and analyze these systems and the underlying technologies. To address this, \textbf{we have first compiled existing spoken dialogue systems in the chronological order and categorized them into the cascaded and end-to-end paradigms.} We then provide an in-depth overview of the core technologies in spoken dialogue models, covering aspects such as \textbf{speech representation, training paradigm, streaming, duplex, and interaction capabilities.} Each section discusses the limitations of these technologies and outlines considerations for future research. Additionally, we present a thorough review of \textbf{relevant datasets, evaluation metrics, and benchmarks} from the perspectives of training and evaluating spoken dialogue systems. We hope this survey will contribute to advancing both academic research and industrial applications in the field of spoken dialogue systems. The related material is available at \url{https://github.com/jishengpeng/WavChat}.

\end{abstract}

\section{Introduction}
Spoken dialogue models \cite{defossez2024moshi,zhang2023speechgpt,xie2024miniomni2opensourcegpt4ovision} represent one of the most direct methods of human-computer interaction, evolving from traditional voice assistants such as Alexa\footnote{\url{https://www.alexa.com/}}, Siri\footnote{\url{https://www.apple.com/siri/}}, and Google Assistant\footnote{\url{https://assistant.google.com/}} to the latest intelligent dialogue systems, such as GPT-4o\footnote{\url{https://openai.com/index/chatgpt-can-now-see-hear-and-speak/}}. The fundamental definition of a spoken dialogue model refers to a dialogue system capable of generating intelligent verbal responses based on the input speech. On the one hand, the \textbf{speech modality} serves as both the input and output interface for the human-computer interaction in the spoken dialogue models. On the other hand, the \textbf{dialogue system} \cite{dubey2024llama} requires the model to possess a certain level of textual intelligence, including the ability to comprehend the knowledge of human society and generating professional and intelligent responses. Recently, intelligent spoken dialogue systems, exemplified by GPT-4o and Moshi \cite{defossez2024moshi}, have garnered significant attention for their ability to extend speech intelligence capabilities beyond traditional text-based dialogue models~\cite{huang2024audiogpt}. These dialogue models can not only generate natural, human-like speech responses \cite{defossez2024moshi,speechteam2024funaudiollm} but also demonstrate an advanced understanding and generation of acoustic features beyond text, such as timbre, emotion, and style \cite{lin2024advancing,lin2024paralinguistics,xue2023chat}. Additionally, they exhibit strong performance in processing other speech-related representations, including music and audio events~\cite{chu2024qwen2,chu2023qwen,gong2023joint,tang2023salmonn}. Their realistic conversational interactivity \cite{fu2024vita,xie2024miniomni2opensourcegpt4ovision} and low-latency dialogue experiences~\cite{defossez2024moshi} further distinguish them among the traditional spoken dialogue models.

\begin{figure*}[t]
    \centering
    \includegraphics[width=0.95\linewidth]{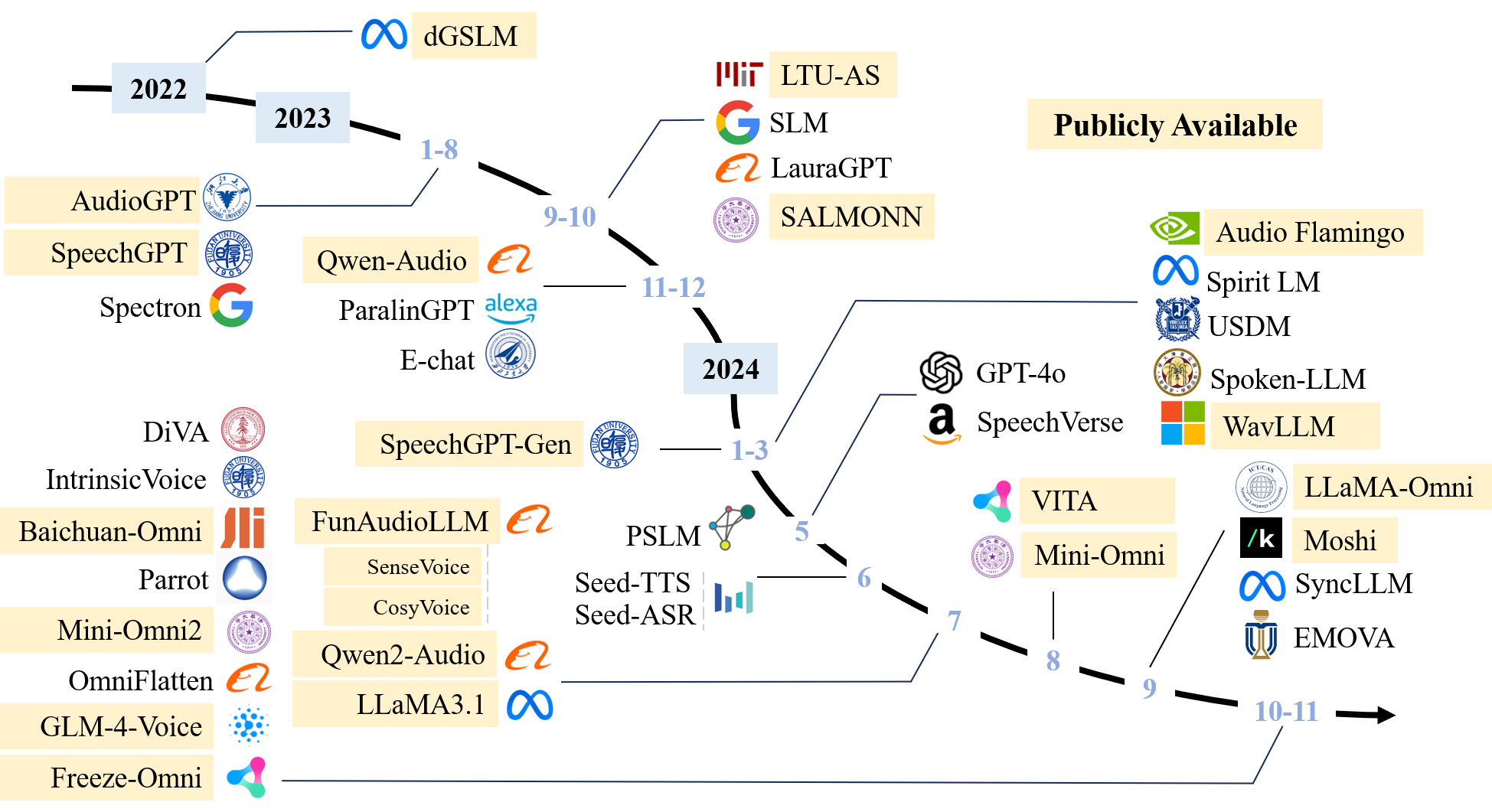}
    \caption{A timeline of existing spoken dialogue models in recent years. The timeline was
established mainly according to the release date (e.g., the submission date to arXiv) of the technical paper for each model. It is worth noting that certain works, such as Westlake-Omni, MooER-Omni, Hertz-dev, SpeechGPT2 and Fish-Agent do not have corresponding published papers. Therefore, we have not included them in the figure. We mark the publicly available model checkpoints in yellow color.}
    \label{fig:img1}
\end{figure*}

\begin{figure*}[t]
    \centering
    \includegraphics[width=1.0\linewidth]{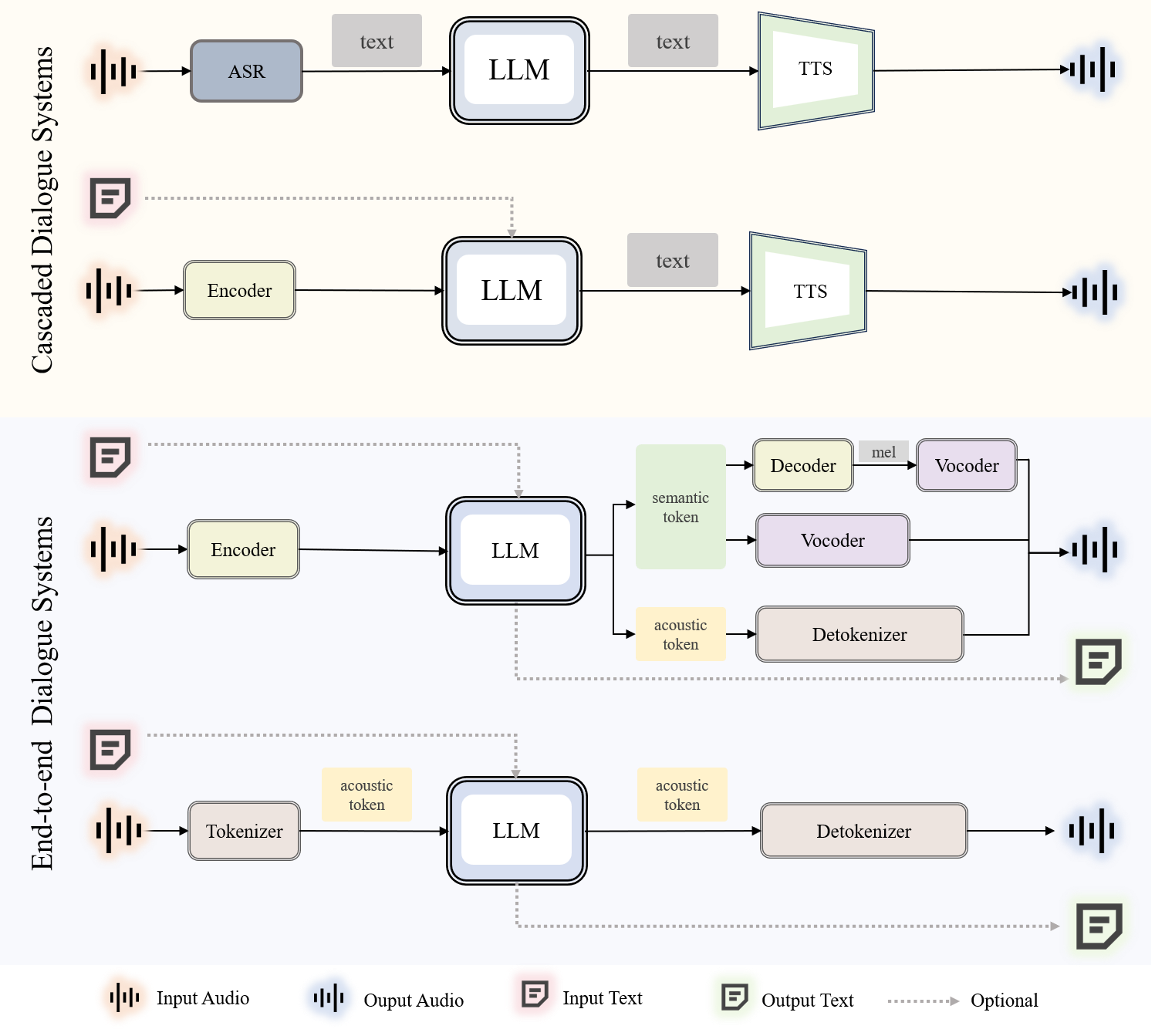}
    \caption{A general overview of current spoken dialogue systems. We categorize these systems into two paradigms, cascaded spoken dialogue models and end-to-end spoken dialogue models, based on whether the core language model can \textbf{directly} understand and generate speech representations. Additionally, we provide a visualization of the input and output methods used in different spoken dialogue systems.}
    \label{fig:img2}
\end{figure*}

The history of spoken dialogue models can be traced back to early systems like dGSLM~\cite{nguyen2023generative} and AudioGPT~\cite{huang2024audiogpt}, leading up to more recent advancements such as GPT-4o and Moshi~\cite{defossez2024moshi}. During this period, many notable spoken dialogue models have emerged. As shown in Figure \ref{fig:img1}, we have organized these models in chronological order. Broadly, they can be categorized into two types: cascaded spoken dialogue models~\cite{chu2024qwen2,chu2023qwen} and end-to-end~\cite{mentzer2023finite,xie2024mini,zhang2024omniflatten,zhang2024intrinsicvoice} spoken dialogue models. Given that most current spoken dialogue models rely on alignment with the text modality, the distinction between cascaded and end-to-end models is crucial. As illustrated in Figure \ref{fig:img2}, we classify all spoken dialogue models based on whether \textbf{the core language model can directly understand and generate speech representations}, dividing them into cascaded and end-to-end categories. Traditional cascaded spoken dialogue systems such as AudioGPT~\cite{huang2024audiogpt} are structured around text as the central intermediary, typically comprising three cascaded modules. First, the input audio is transcribed into text by an automatic speech recognition (ASR) module~\cite{radford2023robust}. The transcribed text is then fed into a large language model (LLM) such as ChatGPT to generate a textual response. Finally, this textual response is converted back into audio through a text-to-speech (TTS) module~\cite{kong2023vits2,ren2020fastspeech}. While this cascaded architecture leverages the strong in-context capabilities of large language models, it introduces several challenges, including high latency, limited interactivity, and the inability to process non-textual information. To address these issues, recent research has taken two primary directions. Some approaches~\cite{chu2023qwen,tang2023salmonn} focus on optimizing the understanding and generation components within the cascaded system to mitigate the aforementioned limitations. Some other approach~\cite{xie2024mini,xie2024miniomni2opensourcegpt4ovision,zhang2024speechgpt,zhang2024intrinsicvoice} seek to directly solve these problems by adopting end-to-end architectures for spoken dialogue systems. Although end-to-end spoken dialogue models exhibit various differences in terms of representations and model architectures, they share a common feature: they do not rely on text as the central intermediary. Instead, these models aim to directly comprehend and generate speech representations. We define such systems as end-to-end spoken dialogue models. 

When constructing spoken dialogue systems, we identify four core technologies closely related to spoken dialogue models, based on the different levels of intelligence involved. The first is the design of speech representations (i.e., tokenizers and detokenizers). The second concerns the paradigm for training, inference, and generation, specifically how to align the speech modality with the text modality while preserving or enhancing the intelligence of existing text-based dialogue models. This part also involves selecting different model architectures, generation strategies, and multi-stage training approaches. The third challenge involves the design of interactive, duplex, streaming for spoken dialogue systems. Lastly, the fourth challenge relates to data—specifically, how to construct training datasets for spoken dialogue systems and evaluate their performance.

\begin{figure}[H]
    \centering
    \includegraphics[width=\textwidth]{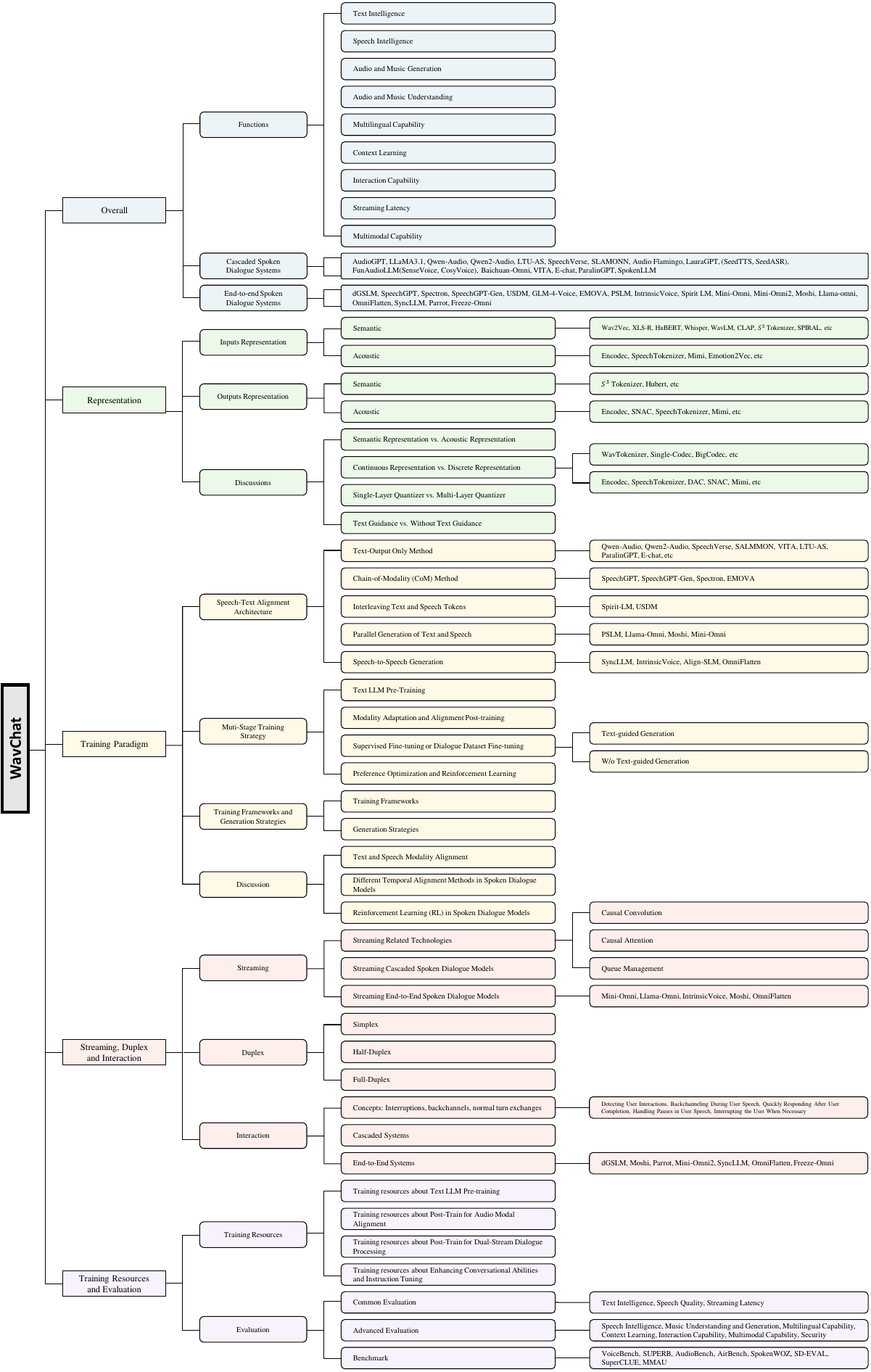}
    \caption{A general overview about the structure of WavChat }
    \label{fig:img4}
\end{figure}

Given these considerations, in the following sections of this paper, we address these four key technologies in the order outlined above. In Section 2, we provide an overview of spoken dialogue systems, including typical spoken dialogue scenarios (i.e., how to define a spoken dialogue model) and recent developments in the cascaded and end-to-end spoken dialogue models. Section 3 focuses on the speech representations used in spoken dialogue systems. In Section 4, we systematically discuss the training paradigms, with particular emphasis on how to align the speech modality with the text modality, as well as multi-stage training strategies, model architectures, and generation strategies. Section 5 highlights the unique characteristics of spoken dialogue systems, particularly their duplex, streaming nature, which distinguishes them from text-based dialogue systems. In Section 6, we examine the construction of training datasets and the evaluation methodologies specific to spoken dialogue models. At the end of each section, we include a summary and discussion to reflect on the key insights. Finally, in Section 7, we conclude the survey by summarizing the major findings and discussing open issues for future research. Given the complexity of the technical points, we provide an overview of the structure of this survey in Figure \ref{fig:img4}.

\section{Overall}
In this section, we will provide an overall overview of spoken dialogue models. we begin by defining what constitutes an intelligent spoken dialogue model by examining various dialogue scenarios. We then provide a comprehensive overview of spoken dialogue models, distinguishing between cascaded spoken dialogue models and end-to-end spoken dialogue models.

\subsection{Functions of Spoken Dialogue Systems}
\label{section21}
Based on the demos and inference interfaces of representative models such as GPT-4o, Moshi~\cite{defossez2024moshi}, Qwen2-Audio~\cite{chu2024qwen2}, and VITA~\cite{fu2024vita}, we categorize the usage scenarios of modern intelligent spoken dialogue models into the following nine representative categories: 1) Text Intelligence, 2) Speech Intelligence, 3) Audio and Music Generation, 4) Audio and Music Understanding, 5) Multilingual Capability, 6) Context Learning, 7) Interaction Capability, 8) Streaming Latency, and 9) Multimodal Capability. For the nine distinct use cases in spoken dialogue models, we provide corresponding examples for each scenario in Figure \ref{fig:main}. It is clear from these usage scenarios that a spoken dialogue model is not simply an extension of a text-based dialogue model to the speech modality (i.e., where the speech modality serves merely as an interface for converting speech into text). Rather, an intelligent spoken dialogue system must be capable of comprehending and generating acoustic information embedded in speech (such as timbre, style, and emotion) and of understanding and producing a wider range of audio representations, including information related to audio events and music. Additionally, unlike non-streaming text-based systems, spoken dialogue models need to support real-time, interactive streaming capabilities. These usage scenarios not only highlight the intelligence inherent in spoken dialogue systems but also present significant challenges for building end-to-end spoken dialogue models. Below, we provide a detailed examination of each of the nine usage scenarios.


\subsubsection{Text Intelligence}
As illustrated in Figure \ref{fig:main} (a), a spoken dialogue system must retain the fundamental capabilities of the original text-based dialogue models, such as ChatGPT. We define this usage scenario as textual intelligence. In this context, the spoken dialogue model can intelligently respond to user requests, generating appropriate responses such as travel itineraries, work plans, and scheduling. However, due to the limitations of voice-based interaction, the textual intelligence of current spoken dialogue systems is more focused on the daily scenarios. In certain contexts, such as complex mathematical theorem reasoning, the performance requirements for spoken dialogue models differ from those of text-based dialogue models~\cite{touvron2023llama}. These advanced aspects of textual intelligence warrant further exploration in unified multimodal dialogue models.

\subsubsection{Speech Intelligence}
\label{section212}
A distinguishing feature of spoken dialogue models, compared to text-based dialogue models~\cite{touvron2023llama}, is their ability to understand and generate acoustic information beyond mere textual content. In the speech modality, not only is the textual content present, but also additional acoustic information, such as timbre (speaker identity) and style (emotion, prosody, etc.). As illustrated in Figure \ref{fig:main} (b), an intelligent spoken dialogue system should be capable of \textbf{understanding} the timbre and style 

of conversational speech and, ideally, \textbf{generating} responses with specified timbre and style in a \textbf{zero-shot} manner.

This capability about speech intelligence involves several use cases. First, on the comprehension side, the spoken dialogue system should generate responses based on the speaker's vocal style. For example, in the E-chat~\cite{xue2023chat}, a classic example might be: if a user asks, "My phone won't turn on, what should I do?" in a cheerful tone, the system might respond, "It looks like you're excited about getting a new phone. What type of phone are you interested in?" Conversely, if the user asks the same question in a sad tone, the system might reply, "It's unfortunate your phone isn't working. If you're familiar with the repair policy, let's proceed with the next steps." This situation indicates that the spoken dialogue system may generate responses with different \textbf{content} based on varying acoustic information. Furthermore, the system should comprehend various acoustic cues, such as accents or emotional states, and adjust its responses of different \textbf{acoustic} information accordingly. For instance, if the speaker is an American, the system might reply with a native English accent, whereas if the speaker is a Shanghainese user, the system could respond using the corresponding dialect. Similarly, if the user speaks with a sad tone, the dialogue system should be able to generate a more encouraging and empathetic response.

On the generation side, speech intelligence is more prominently reflected in its controllability, such as voice cloning and style control. For example, the system could be instructed to mimic a specific voice or respond in a designated style (e.g., mimicking a grandmother's soft and gentle voice for a comforting interaction). Additionally, the system could use a voice prompt provided during the conversation to fully clone the timbre from the prompt and generate speech in that same voice. In summary, the ability to comprehend and generate acoustic information is one of the key characteristics of an intelligent spoken dialogue model.
\subsubsection{Audio and Music Generation}
In the spoken dialogue models, beyond basic spoken dialogue capabilities, an intelligent spoken dialogue system may be required to generate music and audio. For example, a user might instruct the system to generate a one-minute piano piece or a ten-second recording of a dog barking. Additionally, users might provide lyrics and a musical melody, asking the spoken dialogue model to create a pop song. The system should thus inherit the generative capabilities of large-scale music~\cite{agostinelli2023musiclm,copet2024simple,lam2024efficient,ma2024foundation} and audio~\cite{huang2023make,liu2023audioldm,liu2024audioldm} models on the output side.

\subsubsection{Audio and Music Understanding}
Complementing its music and audio generation capabilities, a spoken dialogue model should also be able to understand music and audio on the input side~\cite{chu2024qwen2,tang2023salmonn}. For instance, when given an audio clip, the intelligent system should identify both its content and acoustic characteristics, such as recognizing whether the sound is a bird chirping or a cat meowing, or whether the music is calm or energetic. Moreover, the system could extend its understanding by creating literary works—like poetry or songs—based on the given music or audio.

\begin{figure*}[htbp]
    \centering
    \captionsetup[subfigure]{font=scriptsize}

    \vspace{0.2cm}
    \begin{subfigure}[b]{0.30\textwidth}
        \centering
        \includegraphics[width=\linewidth]{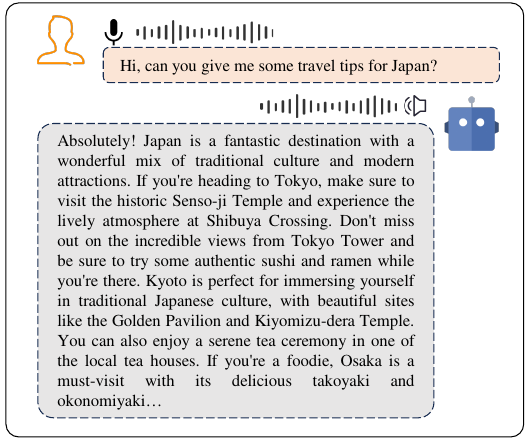}
        \caption{Text Intelligence}
    \end{subfigure}
    \begin{subfigure}[b]{0.30\textwidth}
        \centering
        \includegraphics[width=\linewidth]{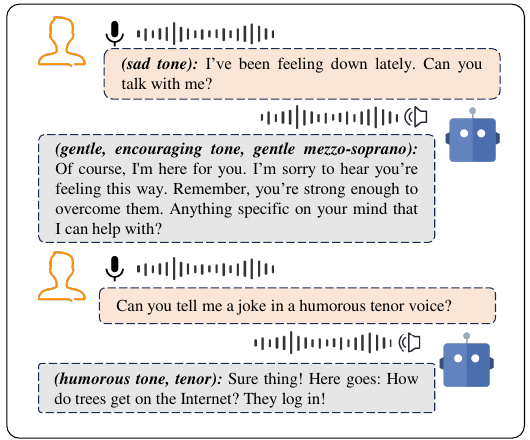}
        \caption{Speech Intelligence}
    \end{subfigure}
    \begin{subfigure}[b]{0.30\textwidth}
        \centering
        \includegraphics[width=\linewidth]{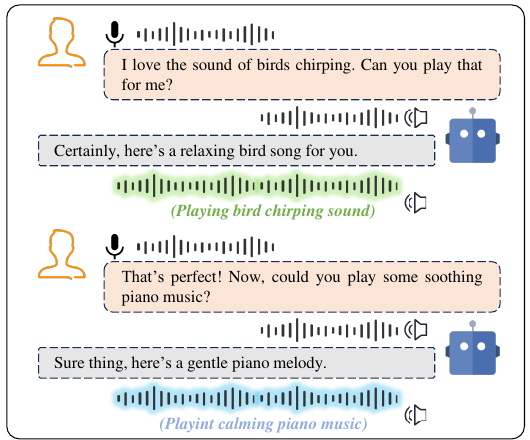}
        \caption{Audio and Music Generation}
    \end{subfigure}
    
    \vspace{0.2cm}
    \begin{subfigure}[b]{0.30\textwidth}
        \centering
        \includegraphics[width=\linewidth]{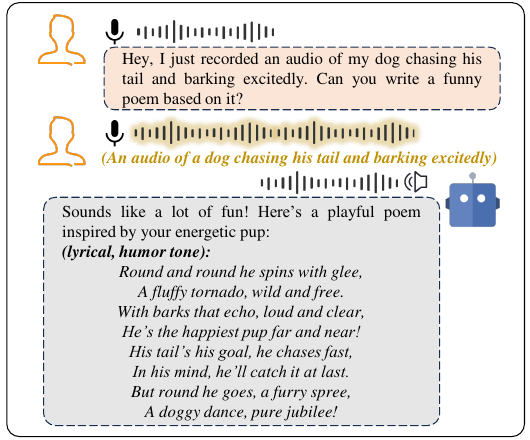}
        \caption{Audio and Music Understanding}
    \end{subfigure}
    \begin{subfigure}[b]{0.30\textwidth}
        \centering
        \includegraphics[width=\linewidth]{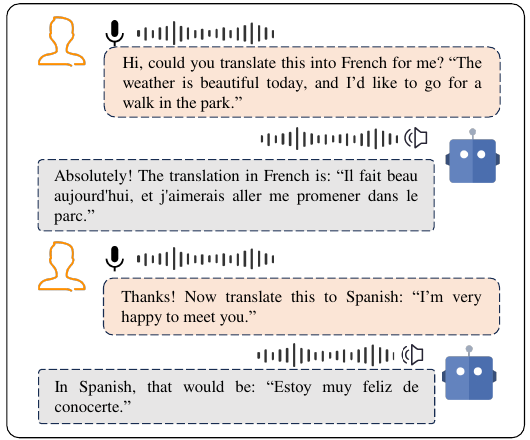}
        \caption{Multilingual Capability}
    \end{subfigure}
    \begin{subfigure}[b]{0.30\textwidth}
        \centering
        \includegraphics[width=\linewidth]{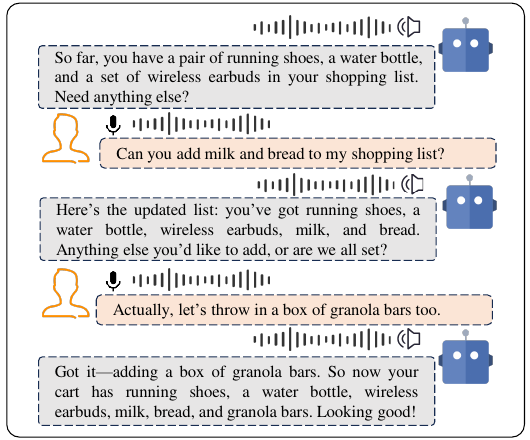}
        \caption{Context Learning}
    \end{subfigure}

    \vspace{0.2cm}
    \begin{subfigure}[b]{0.30\textwidth}
        \centering
        \includegraphics[width=\linewidth]{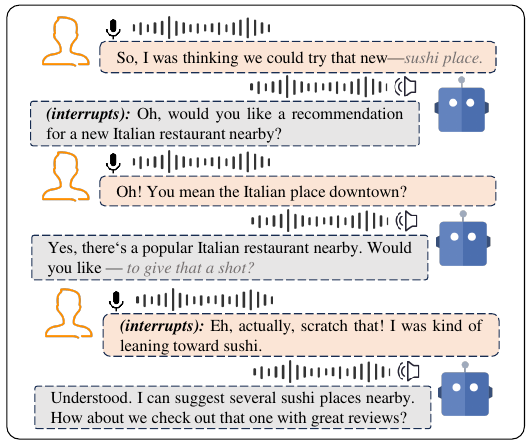}
        \caption{Interaction Capability}
    \end{subfigure}
    \begin{subfigure}[b]{0.30\textwidth}
        \centering
        \includegraphics[width=\linewidth]{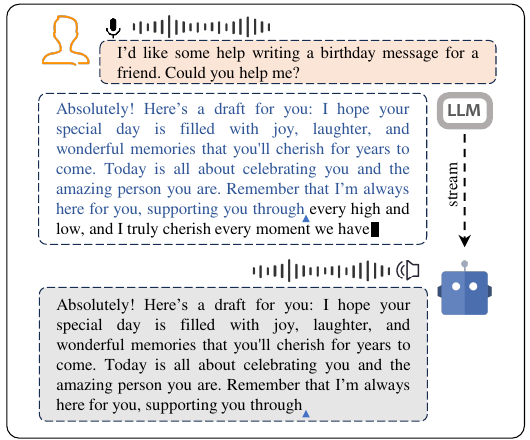}
        \caption{Streaming Latency}
    \end{subfigure}
    \begin{subfigure}[b]{0.30\textwidth}
        \centering
        \includegraphics[width=\linewidth]{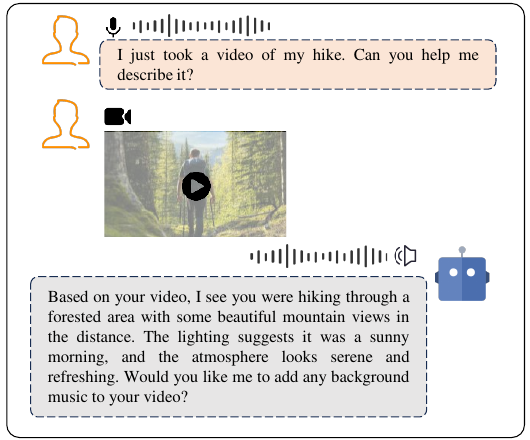}
        \caption{Multimodal Capability}
    \end{subfigure}

    \caption{An overall demonstration of the functions of the spoken dialogue systems. We describe the ideal capabilities of such systems from nine different perspectives: Text Intelligence, Speech Intelligence, Audio and Music Generation, Audio and Music Understanding, Multilingual Capability, Context Learning, Interaction Capability, Streaming Latency, and Multimodal Capability. Each function is illustrated with corresponding dialogue examples.}
    \label{fig:main}
\end{figure*}

\subsubsection{Multilingual Capability}
Similar to text-based dialogue models, spoken dialogue systems are expected to possess multilingual capabilities. Specifically, these models should be able to perform multilingual content translation, such as translating a spoken segment in Japanese into French speech clips, effectively inheriting the capabilities of simultaneous interpretation. In addition to multilingual content translation, the system should also handle multilingual acoustic information. This means that the intelligent spoken dialogue model should be able to generate responses in various languages and accents, replying in the corresponding accent of the target language based on the different input speech.

\subsubsection{Context Learning}
In the spoken dialogue models, the ability to handle long-form and multi-turn conversations is a key benchmark for evaluating performance~\cite{defossez2024moshi}. This requires that spoken dialogue models not only support long-duration audio inputs but also generate extended audio outputs. Moreover, they must be capable of engaging in multi-turn conversations based on historical context. An important aspect of multi-turn dialogue is the ability to revise previous responses based on new user instructions. As shown in Figure \ref{fig:main} (f), an intelligent spoken dialogue model should be able to continuously modify its previous replies according to the user’s evolving requests.

\subsubsection{Interaction Capability}
A distinguishing feature of spoken dialogue systems compared to the text-based dialogue models is their duplex and interactive nature~\cite{defossez2024moshi}. In text-based dialogue, interactions typically follow a half-duplex structure, where the response can only be provided after the question has been completed, and the user is unable to interrupt the reply in real-time. However, in the spoken dialogue systems, full-duplex interaction is common. This means that a conversation does not need to be fully completed before a response can be generated. Both the system and the user can interrupt and interact in real time. For example, if the user is unsatisfied with the system's response, they can immediately interrupt, causing the system to halt its current generation and respond to the new input. Additionally, to emulate more natural conversational settings, the system can also interrupt the user when appropriate, such as when clarifying the user’s intent. Beyond the ability to interrupt, interactive dialogue often includes the use of conversational fillers, such as "okay," "haha," or "oh," which signal acknowledgment or agreement. Including these within spoken dialogue models enhances the realism and natural flow of conversations. The underlying requirement for interaction capabilities is that the system should be able to listen and speak simultaneously, responding dynamically to the flow of the interaction.

\subsubsection{Streaming Latency}
Streaming comprehension and generation are also fundamental functionalities of spoken dialogue models~\cite{xie2024miniomni2opensourcegpt4ovision,zhang2024intrinsicvoice,fang2024llama}. In the real-world scenarios, a model cannot wait until an entire minute-long audio segment has been processed before generating a response. Instead, the model must operate on a chunk-based mechanism, dynamically processing and generating audio in real time, one chunk at a time. Additionally, the streaming requirement means that the entire system must operate in a causal manner—understanding and generating audio based solely on past information, without relying on future information. Streaming function is often closely tied to the need for low latency. In practical conversational experiences, the latency of the first token generated by the spoken dialogue model (i.e., the wait time for the user) and the average latency of the generation process are critical factors that influence the overall responsiveness and usability of the spoken dialogue system.

\subsubsection{Multimodal Capability}
Multimodal dialogue capability~\cite{chen2024emova,fu2024vita} represents an advanced feature of spoken dialogue models. In existing systems, this typically refers to the ability to process inputs from multiple modalities, such as video, images, and text, while generating intelligent speech responses. A spoken dialogue model equipped with this capability achieves the ability to “hear, see, and speak” simultaneously. Multimodal inputs significantly enhance the potential of these systems; for instance, users can employ various gestures to improve the quality of the model’s generated responses, and the system can develop a deeper understanding of the physical world. Beyond multimodal inputs, the future of dialogue systems lies in large multimodal models that unify the comprehension and generation capabilities across all modalities, with spoken dialogue serving as the foundational modality.

\subsection{Cascaded Spoken Dialogue Systems}
The earliest prototype of cascaded spoken dialogue systems can be traced back to AudioGPT~\cite{huang2024audiogpt}. To achieve speech-to-speech dialogue functionality, the system first employed an Automatic Speech Recognition (ASR) model to convert speech into text, followed by ChatGPT for text-based dialogue, and finally, a Text-to-Speech (TTS) model to convert the generated text back into speech. In this primitive version, speech was used solely as an input-output interface, retaining only the most basic textual intelligence. For example, in the Huggingface’s open-source Speech-To-Speech framework\footnote{\url{https://github.com/huggingface/speech-to-speech}}, an additional Voice Activity Detection (VAD) module\footnote{\url{https://github.com/snakers/silero-vad}} was further layered onto the traditional cascaded modules to distinguish between speech and silent segments, as well as between different speakers.

After the basic textual intelligence had been established in the cascaded spoken dialogue models, researchers began incorporating paralinguistic features, such as emotion and style, to enhance the speech intelligence in the cascaded spoken dialogue models. For instance, ParalinGPT~\cite{lin2024paralinguistics} and E-chat~\cite{xue2023chat} integrate conversational context, speech embeddings, and paralinguistic attributes into an autoregressive model via a sliding window, allowing the model to generate more accurate text responses by combining historical text and emotional representations. Similarly, Spoken-LLM~\cite{lin2024advancing} introduces an Emotion2Vec~\cite{ma2023emotion2vec} module to provide style vectors to the Llama2-Chat model. Through LoRA~\cite{hu2021lora} fine-tuning, Llama2-Chat is trained not only to generate content-based text responses but also to produce text responses with specific stylistic attributes (e.g., <cheerful, fast, normal>), which can guide downstream TTS systems in generating expressive speech.

In addition to understanding acoustic information within cascaded spoken dialogue models, there have been efforts to directly input speech representations while retaining text as the output modality~\cite{das2024speechverse,chu2023qwen,kong2024audio}. This forces cascaded spoken dialogue systems to process input speech directly. A common approach involves integrating frozen speech encoders (such as Whisper~\cite{radford2023robust}) with trainable encoder adapters, allowing the speech input to be interpreted as a specialized form of text by the large language model. By extending the vocabulary of the text-based dialogue model, the large language model can process speech as if it were a unique form of text, enabling the generation of appropriate text responses in the cascaded spoken dialogue models.

Notably, these cascaded spoken dialogue models have further advanced beyond the comprehension of human speech alone and can now understand a variety of audio modalities, including music and audio~\cite{gong2023joint,tang2023salmonn}. For example, SALMONN~\cite{tang2023salmonn} models both speech and audio information by freezing the Whisper~\cite{radford2023robust} and BEATs~\cite{chen2022beats} encoder and bridging them to a large language model via a Window-Level Q-Former~\cite{li2023blip}. As a result, these cascaded spoken dialogue systems are capable of further performing a wide range of tasks on the comprehension side. For instance, models like Qwen-audio~\cite{chu2024qwen2,chu2023qwen} can handle multiple tasks such as Automatic Speech Recognition (ASR), Speech-to-Text Translation (S2TT), Automatic Audio Captioning (AAC), Acoustic Scene Classification (ASC), Speech Emotion Recognition (SER), Audio Question Answering (AQA), Vocal Sound Classification (VSC), and Music Note Analysis (MNA). Consequently, these cascaded models are often regarded as part of multitask speech-text large language models.

It is worth noting that the aforementioned cascaded spoken dialogue models generate text only and then directly feed it into a pre-trained TTS module. However, more recent cascaded spoken dialogue models, such as Llama3.1, have begun integrating trainable TTS modules as part of the decoder within the large language model (LLM). While these models have made progress in incorporating low-latency streaming functionalities, they are still fundamentally based on generating text content first, which is then converted into speech. They do not directly generate speech-related representations within the LLM itself. Therefore, we classify these models as cascaded spoken dialogue systems.

In addition, some recent efforts have focused on enhancing models like Qwen2-Audio~\cite{chu2024qwen2} by incorporating multimodal comprehension capabilities, thereby enabling a degree of multimodal dialogue functionality. For instance, models such as VITA~\cite{fu2024vita} and Baichuan-Omni\cite{li2024baichuan} integrate various encoders or tokenizers for images, audio, and video into the LLM, allowing the model to understand multimodal inputs and generate corresponding text responses.

The above developments concern the comprehension side of cascaded spoken dialogue systems. On the generation side, two main types of speech synthesis work are relevant to cascaded spoken dialogue systems. Firstly, there has been a recent surge of advanced speech synthesis systems that can produce highly expressive and natural audio based on textual input, such as VALL-E (X)~\cite{wang2023neural,zhang2023speak}, MegaTTS1/2~\cite{jiang2024mega,jiang2023mega}, CosyVoice~\cite{du2024cosyvoice}, ChatTTS\footnote{\url{https://github.com/2noise/ChatTTS}}, FishSpeech\footnote{\url{https://github.com/fishaudio/fish-speech}}, ParlerTTS~\cite{lyth2024natural}, MaskGCT~\cite{wang2024maskgct} and F5-TTS~\cite{chen2024f5}. In addition, there has been significant progress in the field of text-style controllable TTS, with systems like TextrolSpeech~\cite{ji2024textrolspeech}, PromptTTS~\cite{guo2023prompttts}, PromptTTS2~\cite{leng2023prompttts}, InstructTTS~\cite{yang2024instructtts}, and ControlSpeech~\cite{ji2024controlspeech}. These TTS systems can generate highly natural audio based both on the content and style of the text output produced by the cascaded spoken dialogue models.

\subsection{End-to-End Spoken Dialogue Systems}
Ideally, end-to-end spoken dialogue models should enable \textbf{only} speech input and output during both training and inference, thereby achieving multiple intelligent dialogue functions. However, considering that speech modal is a low-density (contains a lot of acoustic information) modality compared to text modal, and that the volume of available text data far exceeds that of available speech data, many end-to-end spoken dialogue models choose to align the speech modality with the text modality to leverage pre-trained language models (LLMs). Consequently, as showed in the Figure \ref{fig:img2}, as long as the large language models can directly understand and generate speech representations, we classify such systems as end-to-end spoken dialogue models. In contrast, if the large language models can only generate text, we categorize the system as cascaded spoken dialogue systems.

The earliest end-to-end spoken dialogue system can be traced back to dGSLM~\cite{nguyen2023generative}, which was trained on thousands of hours of dual-track data~\cite{cieri2004fisher} using self-attention and cross-attention mechanisms to simulate duplex interactions. Although dGSLM lacks integration with LLMs and even basic textual intelligence, it is notable as the first fully end-to-end spoken dialogue system that does not rely on text while maintaining excellent conversational interactivity.

Following the release of dGSLM~\cite{nguyen2023generative}, the progress in the domain of end-to-end spoken dialogue systems stagnated for a few months. However, with the advent of ChatGPT, this field experienced rapid development. A representative approach is SpeechGPT~\cite{zhang2023speechgpt}, which employs autoregressive language modeling by using a sequence of speech tokens, text tokens, text tokens, and speech tokens. This method enables the direct generation of speech tokens using textual intelligence, inspiring subsequent end-to-end spoken dialogue systems such as Spectron~\cite{marge2022spoken}, SpeechGPT-Gen~\cite{zhang2024speechgpt}, GLM-4-Voice\footnote{\url{https://github.com/THUDM/GLM-4-Voice}}, and EMOVA~\cite{chen2024emova}. These systems continue to use an autoregressive framework, generating the text tokens followed by the speech tokens. Although this approach allows LLMs to generate speech tokens directly, it introduces latency issues since speech token generation cannot begin until the generation of text tokens is complete. This leads to problems in multi-turn dialogue and overall system delay.

Beyond the design of SpeechGPT~\cite{zhang2023speechgpt}, another intuitive approach is to directly use the hidden states before the LLM’s softmax layer to predict both text tokens and speech tokens through different projection layers. This allows the network to share weights up to the projection layer, thereby aligning the speech and text modalities. The PSLM~\cite{mitsui2024pslm} model is a typical example of this design. Another method, proposed by Meta, is the interleaving approach, as seen in Spirit-LM~\cite{nguyen2024spirit}, where speech and text sequences are concatenated into a single token stream and trained using a word-level interleaving method with a small, automatically curated speech-text parallel corpus. However, this approach requires precise alignment between speech and text.

Recently, several new end-to-end spoken dialogue systems have emerged. For instance, Moshi~\cite{defossez2024moshi}, which is based on a global-local transformer, can simultaneously generate text and speech acoustic tokens from a multi-layer quantizer. Starting from a text-based language model backbone, Moshi generates speech tokens from the residual quantizer of a neural audio codec while modeling both the user’s speech and the system’s responses in parallel streams. This design eliminates the need for explicit speaker turns and allows for the modeling of arbitrary conversational dynamics. Moreover, Moshi extends previous hierarchical semantic-to-acoustic token generation by first predicting time-aligned text tokens as a prefix to audio tokens. Similarly, Mini-Omni~\cite{xie2024mini} uses a MusicGen-based~\cite{copet2024simple} method to simultaneously generate text and speech codec tokens. It introduces two strategies: autoregressive generation without strict temporal alignment by padding text tokens and batch-parallel inference strategies to boost performance. Mini-Omni2~\cite{xie2024miniomni2opensourcegpt4ovision} further enhances this by incorporating multimodal understanding and duplex functionality. At the same time, Llama-Omni~\cite{fang2024llama}, Freeze-Omni~\cite{wang2024freezeomnismartlowlatency} and IntrinsicVoice~\cite{zhang2024intrinsicvoice} design an LLM for real-time voice interaction. Their commonality lies in the fact that, at the generation stage, the hidden states of the LLM are further fed into the corresponding decoder model. LLaMA-Omni~\cite{fang2024llama} integrates a pretrained speech encoder, a speech adapter, an LLM, and a streaming speech decoder. It eliminates the need for speech transcription, and can simultaneously generate text and speech responses directly from speech instructions with low latency.
Freeze-Omni ~\cite{wang2024freezeomnismartlowlatency} designed 3-stage training strategies both for the modeling of speech input and output, enabling it to obtain speech-to-speech
dialogue ability noly by using text-speech paired data. The core idea of Freeze-Omni lies in transferring the functionalities of spoken dialogue models to the encoder (ASR) and decoder (TTS), rather than assigning these tasks to the large language model. IntrinsicVoice~\cite{zhang2024intrinsicvoice} facilitates the transfer of textual capabilities from pre-trained LLMs to the speech modality by reducing the modality gap between text and speech. By using a GroupFormer to generate HuBERT tokens from the LLM’s hidden states, IntrinsicVoice effectively reduces speech sequences to lengths comparable to text sequences, generating high-quality audio while significantly speeding up inference and mitigating long-text modeling issues. Additionally, some end-to-end spoken dialogue models align speech and text through multi-stage training, eliminating the need to generate text during inference. For example, Omni-Flatten~\cite{zhang2024omniflatten} employs modality alignment, half-duplex dialogue learning, and full-duplex dialogue learning, along with a flattening-style standardization of text and speech tokens, to achieve duplex, text-free speech dialogue during inference. Similar approaches include SyncLLM~\cite{veluri2024beyond}. 

In this section, we have provided a general overview of current end-to-end spoken dialogue systems. However, these systems differ significantly in their speech representations, training paradigm, model architectures and generation strategy. In Section \ref{section3} and \ref{section4}, we will present a detailed classificationfollowed by our discussions at the end of each section.


\section{Representations of Spoken Dialogue Models}
\label{section3}
Representations play a critical role in spoken dialogue systems as they determine how the spoken dialogue system comprehends, processes, and generates speech signals. Additionally, they serve as a bridge between speech and other modalities, thereby directly influencing the system’s performance, functionality, and range of applications. Compared to text and visual representations, speech representations possess a unique complexity. Text representations primarily rely on a well-defined symbolic system, conveying meaning through structured elements like vocabulary and syntax. Visual representations, on the other hand, focus on capturing spatial relationships and visual features in images. In contrast, speech signals contain both dynamic acoustic features (such as timbre, prosody and emotion) and rich semantic content, requiring representations that not only capture temporal variations but also preserve an understanding of the underlying meaning.


The unique nature of speech has led to the development of two types of representation models. The representations obtained by these two modeling approaches are often classified as semantic tokens and acoustic tokens. \textbf{One category (semantic) is prediction-based modeling}, these models are trained for representation learning by predicting future frames in an autoregressive manner \cite{chung2020vector, shain2020acquiring} or by using surrounding frames to predict masked frames \cite{chi2021audio, hsu2021hubert, liu2020mockingjay}. This approach tends to prioritize capturing linguistic information within speech, making it particularly useful for recognition and understanding tasks. \textbf{The other category (acoustic) focuses on speech compression and reconstruction} \cite{ji2024wavtokenizer, defossez2022high, kumar2024high, zeghidour2021soundstream}. These models quantify speech features (which are downsampled from raw wavforms by one encoder) into a series of discrete tokens, then use one decoder to upsample these discrete tokens into the speech, calculating the reconstruction loss against the original signal. By this approach, we can get discrete acoustic tokens with impressive compression rates and high-fidelity acoustic information, making it more suitable for tasks such as speech synthesis and emotion analysis.

In the spoken dialogue systems, as illustrated in Figure \ref{fig:img2}, different spoken dialogue models employ various approaches for representation selection. In the following part, we will enumerate the commonly used speech representations in spoken dialogue models from both the input and output perspectives. At the end of this section, we will thoroughly discuss the advantages and limitations of these representations, as well as the future trends in the development of representations used in spoken dialogue models.

\subsection{Speech Representations at the Inputs}
\textbf{Semantic.} To enhance language models' ability to understand speech representations and align multimodal data at input, using pretrained models such as Wav2Vec \cite{schneider2019wav2vec}, HuBERT \cite{hsu2021hubert}, Whisper \cite{radford2023robust}, and WavLM \cite{chen2022wavlm} to extract high-level semantic features from speech has become a core strategy for many spoken dialogue systems.

\quad$\bullet$ \emph{Wav2Vec.} Wav2Vec \cite{schneider2019wav2vec} is a foundational work in the field of speech representation learning, pioneering the extraction of self-supervised speech representations from unlabeled speech data. This approach has driven technological advancements in tasks such as speech recognition, speaker identification, and other speech processing applications. Wav2Vec employs a multi-layer, one-dimensional convolutional neural network directly on raw speech waveforms to progressively extract temporal speech features. Training is accomplished through contrastive learning: the model selects a "correct" target (from the current speech frame) alongside several "incorrect" targets (negative samples). By learning to distinguish positive samples from negatives, the model effectively learns to represent speech features in latent space. As an improved version of Wav2Vec, Wav2Vec 2.0 \cite{baevski2020wav2vec} introduces the Transformer architecture and masked modeling. Wav2Vec 2.0 quantizes the latent speech representations extracted by the CNN and then uses a Transformer to model semantic information, similar to BERT \cite{devlin2018bert}. It also employs a contrastive learning objective, requiring the model to distinguish the correct quantized representations from multiple candidate representations. ParalinGPT \cite{lin2024paralinguistics} aims to incorporate emotional expression in conversational interactions, choosing Wav2Vec 2.0 for its proven capability to encode rich prosodic information, beneficial for speech emotion recognition \cite{li2023exploration}. Specifically, ParalinGPT uses Wav2Vec 2.0’s intermediate layer (the 12th layer) for frame-by-frame feature extraction, as this layer has shown optimal results in linear probing tasks for emotion analysis. Additionally, ParalinGPT applies mean pooling and a linear feature projector to extract utterance embeddings.

\quad$\bullet$ \emph{XLS-R.} XLS-R \cite{babu2021xls} is a multilingual self-supervised speech representation model based on the Wav2Vec 2.0 architecture. It extends and optimizes Wav2Vec 2.0 to support a broader range of languages, particularly low-resource languages. During cross-lingual training, XLS-R employs multilingual data augmentation and denoising techniques, enhancing the model's adaptability when processing speech in various languages. 
USDM \cite{kim2024unified} uses XLS-R to obtain continuous intermediate representations at 50Hz, followed by a quantizer \cite{barrault2023seamless} with $K$=10000 to generate speech tokens.

\quad$\bullet$ \emph{HuBERT.} HuBERT \cite{hsu2021hubert} is a commonly used unsupervised learning model that performs K-Means clustering on the MFCC \cite{zheng2001comparison} features of speech to assign pseudo-labels to each frame. It uses a convolutional encoder to generate a sequence of features at a 20ms frame rate from 16kHz sampled speech. Finally, it randomly masks a portion of features from consecutive frames as input to the Transformer \cite{vaswani2017attention}. HuBERT generates masked content based on surrounding context, enabling it to capture temporal and semantic information within speech and gain a deeper understanding of contextual details. Spoken dialogue systems, such as E-Chat \cite{xue2023chat}, SpeechGPT \cite{zhang2023speechgpt}, PSLM \cite{mitsui2024pslm}, IntrinsicVoice \cite{zhang2024intrinsicvoice}, widely use HuBERT as their speech encoder. E-Chat extracts the weighted sum of the 24 layers from the HuBERT to serve as speech embeddings, and incorporates an additional set of weighted parameters to extract emotion embeddings, thereby enabling emotion-aware capabilities. SpeechGPT applies K-Means clustering to quantize the continuous features extracted from HuBERT, converting them into discrete unit sequences. These discrete units are then integrated into the vocabulary of the large language model, enabling direct alignment between the text and speech modalities. To more effectively integrate the language model with speech streams, PSLM adds an additional embedding layer after extracting features with HuBERT. IntrinsicVoice uses HuBERT as the speech tokenizer, grouping speech tokens to reduce sequence length. An embedding layer then converts these tokens into dense embeddings, which are subsequently mapped into the language model's embedding space using a trainable speech adapter. Spirit-LM \cite{nguyen2024spirit} extracts semantic features using HuBERT, employing a K-Means model with 500 units as the basic unit. It trains a feedforward quantizer with data augmentation techniques \cite{gat2022augmentation} to produce discrete speech tokens. In the Align-SLM \cite{lin2024alignslmtextlessspokenlanguage}, HuBERT is used and the cluster number K is set to 500. Notably, when continuous representations are clustered into discrete units, they primarily capture content information, which can be leveraged for modeling and understanding. This process first extracts 25Hz frame-level continuous representations from the 11-th layer of the HuBERT model, assigns each frame to its closest cluster index, and then de-duplicates consecutive identical indices to shorten the sequence.

\quad$\bullet$ \emph{Whisper.} Whisper \cite{radford2023robust}, based on the classic encoder-decoder architecture, has gained widespread attention in the field of speech recognition. The encoder transforms input speech into high-level feature representations, while the decoder generates the corresponding text output from these representations. Pretrained on large-scale data across various speech environments with text as the target, Whisper demonstrates strong capabilities in extracting semantic information from speech. Qwen-Audio \cite{chu2023qwen}, Qwen-Audio 2 \cite{chu2024qwen2} use Whisper’s encoder to convert speech into continuous representations, which are then combined with text representations and fed into the large language model. Mini-Omni \cite{xie2024mini}, Mini-Omni 2 \cite{xie2024miniomni2opensourcegpt4ovision}, and LLama-Omni \cite{fang2024llama} follow a similar approach, connecting a speech adapter after the Whisper encoder. Their shared objective is to map speech representations into the text embedding space of the large language model, enhancing the model's ability to understand speech by forcibly aligning them through vocabulary expansion. 

\quad$\bullet$ \emph{WavLM.} WavLM \cite{chen2022wavlm} is a pretrained model designed for comprehensive speech processing tasks, playing a critical role in advancing speech technology. Specifically, WavLM employs a masked speech denoising and prediction framework, where some inputs consist of simulated noise or overlapping speech with masked sections. The goal is to predict pseudo-labels of the original speech in the masked areas. This approach enables the model to learn ASR-related information through masked speech prediction, while also gaining knowledge relevant to non-ASR tasks through speech denoising modeling. The masking and prediction pipeline for speech frames in WavLM is similar to that of HuBERT. However, WavLM introduces an additional gated relative position bias to enhance the model's sensitivity to temporal information in speech. SpeechVerse \cite{das2024speechverse} leverages the pretrained WavLM Large as its backbone speech encoder, encoding all intermediate layer features from WavLM to capture various forms of semantics and achieve better generalization performance. To address the significant length disparity between speech features and text tokens, SpeechVerse applies a learnable convolutional module for downsampling the speech features. 


\quad$\bullet$ \emph{$S^3$ Tokenizer.} CosyVoice \cite{du2024cosyvoice} proposes using a supervised automatic speech recognition module to generate a supervised semantic speech($S^3$) tokenizer. Unlike a standard ASR model, the $S^3$ tokenizer splits the encoder into two parts and introduces a vector quantization layer in between. The first encoder converts the mel spectrogram into context-aware representations, while the second encoder transforms discrete speech units into continuous hidden states. Finally, a Transformer-based ASR decoder predicts the posterior probabilities of text labels. Through supervision in multilingual ASR tasks, the $S^3$ tokenizer can convert speech into semantically consistent tokens that facilitate both speech understanding and generation. OmniFlatten \cite{zhang2024omniflatten} uses the $S^3$ tokenizer to extract discrete speech tokens, which are then directly fed into a text-speech pre-trained Transformer.

\quad$\bullet$ \emph{SPIRAL.} SPIRAL \cite{huang2022spiral} aims to learn representations from speech data that are robust to noise and perturbations. It uses a teacher-student network, where various perturbations—such as noise addition, gain adjustment, and time-frequency warping—are applied to the speech input of the student model. The teacher model then guides the student model to produce consistent representations despite these perturbations. EMOVA \cite{chen2024emova} utilizes the SPIRAL’s architecture as a speech encoder to process speech, and employs the finite scalar quantization \cite{mentzer2023finite} to discretize these features. This process aligns speech with the text vocabulary, allowing for a more natural integration into the LLM.

\quad$\bullet$ \emph{Others.} Some spoken dialogue systems do not use pre-trained representation models; instead, they process input features by stacking fundamental modules. VITA \cite{fu2024vita} initially decomposes the speech signal using mel filter banks, mimicking the nonlinear perception of sound in humans. It then processes the input features with a 4-layer CNN downsampling module followed by a 24-layer Transformer. To align with the subsequent language model, VITA employs a simple 2-layer MLP as an adapter. Freeze-Omni \cite{xiong2024freeze} utilizes a chunk-wise streaming speech encoder to transform input speech features into high-dimensional representations. An adapter module then maps these high-dimensional representations into the embedding space of the main LLM, ensuring a quick, low-latency response to the input speech. The speech encoder module consists of several downsampling convolutional layers and Transformer blocks, while the adapter includes only a few downsampling convolutional layers. Downsampling layers are used to reduce the frame rate of speech features, increase the LLM's processing speed during the prefill phase, and minimize latency.

\textbf{Acoustic.} Considering that semantic features are insufficient to capture the emotion, timbre, and style of speech, some representation models, such as Emotion2Vec \cite{ma2023emotion2vec}, attempt to extract acoustic information through self-supervised training. Others focus on reconstruction objectives to ensure high-fidelity speech, including models like Encodec \cite{defossez2022high}, SpeechTokenizer~\cite{zhang2023speechtokenizer}, Mimi \cite{defossez2024moshi}. 

\quad$\bullet$ \emph{Encodec.} EnCodec \cite{defossez2022high} is a straightforward, streaming, convolution-based encoder-decoder architecture. Raw speech is downsampled through a series of convolutional layers, mapping it to latent feature representations. Residual vector quantization \cite{zeghidour2021soundstream} then discretizes the encoder’s continuous latent features. The quantization objective is to map continuous features to a predefined set of discrete tokens (known as a "codebook") for subsequent compression and transmission. The decoder restores the discrete features to a waveform close to the original speech through a series of de-convolution layers. LauraGPT \cite{du2023lauragpt} employs an enhanced version of EnCodec as its speech encoder with specific modifications: (1) adding a reconstruction loss in the magnitude spectral domain to improve mid-to-high frequency signal quality; (2) stacking five strided convolutional blocks with strides of (8, 5, 4, 2, 2) to address the challenges of long sequence lengths, resulting in a token rate of 25Hz per token group; and (3) using 32 quantizers with structured dropout in the Residual Vector Quantization (RVQ) module, each with a vocabulary size of 1024. This revision increases speech quality by incorporating more quantizers while preserving most information in the shallow quantizers. LauraGPT ultimately selects the output from the first quantizer layer as the speech token, balancing performance with sequence length efficiency. The remaining quantizers are used only during the training of the encoder-decoder model.

\quad$\bullet$ \emph{SpeechTokenizer.} 
SpeechTokenizer \cite{zhang2023speechtokenizer} unifies semantic and acoustic tokens, hierarchically decomposing different aspects of speech information across various RVQ layers. It is built on the framework of RVQ-GANs, following the same pattern as SoundStream \cite{zeghidour2021soundstream} and EnCodec \cite{defossez2022high}. Notably, SpeechTokenizer has substituted the two-layer LSTM, originally following the convolution blocks in the EnCodec encoder, with a two-layer BiLSTM to augment the semantic modeling ability. SpeechTokenizer uses HuBERT as a semantic teacher, given HuBERT’s proven capacity to encode substantial content information \cite{mohamed2022self}. During training, it introduces two types of distillation: continuous representation distillation and pseudo-label prediction. For continuous representation distillation, SpeechTokenizer employs the 9th layer HuBERT representation or the average representation across all HuBERT layers as semantic teachers. The training objective is to maximize the cosine similarity at the dimension level across all timesteps between the outputs of RVQ first layer and semantic teacher representations. For pseudo-label prediction, SpeechTokenizer adopts HuBERT units as the target label. In dialogue systems, SpeechGPT-Gen uses SpeechTokenizer RVQ-1 to process raw speech, primarily enhancing the large language model's ability to model the semantics of speech.

\quad$\bullet$ \emph{Mimi.} Taking inspiration from previous work on SpeechTokenizer, Mimi \cite{defossez2024moshi} uses distillation to transfer non-causal, high-level semantic information into the tokens produced by a causal model, allowing for streaming encoding and decoding of semantic-acoustic tokens. To improve the ability of Mimi to encode speech into compact representations while reconstructing high-quality speech, Transformer modules are added in the encoder and decoder. Mimi uses WavLM to distill RVQ-1, enriching it with semantic information. Notably, performing distillation significantly enhances the speech discrimination capability of the first quantizer; however, it can also negatively impact speech quality. Mimi hypothesizes that this is due to distilling semantic information into the first level of a single RVQ: As higher-order quantizers operate on the residual of the first one, the latter needs to trade speech quality for phonetic discriminability. Mimi addresses this issue by introducing a split-RVQ approach. Instead of using a single 8-level RVQ, it extracts semantic information into a simple VQ and applies a parallel 7-level RVQ, combining their outputs at the end. This removes the constraint that acoustic information must be preserved in the residuals of the semantic quantizer. After careful design, Mimi serves as the speech encoder in Moshi \cite{defossez2024moshi}, this approach enhances the model's ability to capture both semantic and acoustic details.

\quad$\bullet$ \emph{Emotion2Vec.} Emotion2Vec \cite{ma2023emotion2vec} is a versatile speech emotion representation model designed to extract emotional features from speech. During the pre-training phase, Emotion2Vec conducts online distillation with a teacher network and a student network. When a specific downstream task is performed, Emotion2Vec is frozen and a lightweight downstream model is trained. Emotion2Vec introduces an utterance-level loss to control global emotion and employs a frame-level loss to build a frame-wise pretext task, enabling it to learn contextual emotions. Spoken-LLM~\cite{lin2024advancing} uses features extracted by Emotion2Vec as input for the large language model, aiming to enable the model to understand and respond to emotions.

\subsection{Speech Representations at the Outputs}


\textbf{Semantic.} At the output stage, Most spoken dialogue systems choose to autoregressively model semantic tokens, such as $S^3$ tokens \cite{du2024cosyvoice} and HuBERT \cite{hsu2021hubert} units. It is worth noting that these semantic tokens lack acoustic conditioning and therefore require a vocoder \cite{kong2020hifi, polyak2021speech} or decoder, which futher takes semantic discrete units as input to synthesize speech consistent with the speakers encountered during training.

\quad$\bullet$ \emph{$S^3$ Tokenizer.} OmniFlatten \cite{zhang2024omniflatten} uses the LLM to autoregressively predict $S^3$ tokens at the speech output stage. When converting discrete tokens back into speech, it adopts the same optimal transport conditional flow matching model (OT-CFM) as used in CosyVoice \cite{du2024cosyvoice}. OT-CFM transforms the speech token sequence into Mel spectrogram, which is then used to generate the final speech with the HiFi-GAN vocoder \cite{kong2020hifi}.

\quad$\bullet$ \emph{Hubert.} Speech tokens extracted by the pre-trained HuBERT \cite{hsu2021hubert} are widely used as generation targets for large language models in the spoken dialogue systems. SpeechGPT \cite{zhang2023speechgpt} and Spirit-LM \cite{nguyen2024spirit} use LLaMA \cite{touvron2023llama} to autoregressively predict a sequence of units and are trained with a HuBERT unit-based HiFi-GAN \cite{kong2020hifi} to decode the speech signal from discrete representations. PSLM \cite{mitsui2024pslm} introduces an additional speech projection layer after the Transformer layers to process the hidden states, obtaining semantic tokens via the softmax layler. The speech decoder in LLama-Omni \cite{fang2024llama} operates in a non-autoregressive manner, taking the output hidden states of the large language model as input to generate a discrete HuBERT unit sequence corresponding to the speech response. The discrete units can be converted into waveform with an additional unit-based vocoder \cite{polyak2021speech}. IntrinsicVoice \cite{zhang2024intrinsicvoice} introduces Group-Former to enhance the large language model’s capability in sequence modeling. When the large language model predicts the $<speech>$ token, the global embedding is passed through a projection layer and delivered, along with a set of learnable queries, to the group model, which then predicts units. IntrinsicVoice uses HiFi-GAN \cite{kong2020hifi}, a non-autoregressive neural vocoder that efficiently generates high-fidelity waveforms, for speech detokenization to reduce overall latency. Align-SLM \cite{lin2024alignslmtextlessspokenlanguage} also uses a HiFiGAN-based \cite{kong2020hifi} model to convert discrete units back into waveforms, utilizing model checkpoints from the textlesslib \cite{kharitonov2022textless} library.

\quad$\bullet$ \emph{Others.} USDM \cite{kim2024unified} does not generate speech directly from input speech; instead, it first transcribes the speech, generates the response text, and then produces corresponding speech token in an end-to-end pipeline. By inserting text-related tasks between speech input and output, the model benefits from both pre-trained LLMs and chain-of-thought \cite{wei2022chain} reasoning in the intermediate modality. Since each stage in the pipeline processes all input and output tokens generated by the previous stage. USDM is more robust to transcription errors and better able to produce contextually relevant spoken responses compared to a cascaded approach with separate modules. USDM uses the Voicebox \cite{le2024voicebox} architecture to train a unit-to-speech model for reconstructing speech from units. EMOVA \cite{chen2024emova} generates a response in the form of speech units when given an image or speech input, which is then converted into an output waveform using the U2S detokenizer. The U2S detokenizer follows the VAE architecture: it uses a speech unit encoder to convert the predicted speech units into continuous embeddings, combines these with style embeddings predicted by the large language model to determine duration, and finally reconstructs the speech waveform through the decoder.

\textbf{Acoustic.} Many spoken dialogue systems choose to directly generate tokens from acoustic representation models, such as EnCodec \cite{defossez2022high}, SpeechTokenizer \cite{zhang2023speechtokenizer}, and Mimi \cite{defossez2024moshi}. These acoustic tokens are then upsampled into the raw waveform through the frozen codec decoder directly.

\quad$\bullet$ \emph{Encodec.} LauraGPT \cite{du2023lauragpt} uses Qwen-1.8B \cite{bai2023qwen} to predict speech tokens. When synthesizing speech, it conditions the predictor not only on the speech tokens predicted by the LLM but also on text and speech inputs. Such text and speech conditionings allow the model to generate high-quality speech signals by leveraging the diverse information in prompt and noisy speeches, which is lacked in the discrete tokens (output from the first quantizer of the Encodec). The predicted speech tokens and conditioning inputs are delivered together to the codec vocoder. An encoder-only Transformer models these inputs into dense embeddings, which are then reconstructed into speech by the codec decoder.

\quad$\bullet$ \emph{SNAC.} SNAC \cite{siuzdak2024snac} encodes speech into hierarchical tokens, similar to EnCodec \cite{defossez2022high} and DAC \cite{kumar2024high}, by introducing quantization at different time resolutions to form a multi-scale discrete representation of speech. In this approach, shallow RVQ layers have a lower sampling frequency, covering a broader time span, while deeper RVQ layers sample at higher frequencies. SNAC introduces modest enhancements over RVQ-GAN by incorporating residual noise blocks, deep convolutions, and local window attention. 
The Mini-Omni \cite{xie2024mini, xie2024miniomni2opensourcegpt4ovision} series continues the parallel generation method introduced by MusicGen\cite{copet2024simple}, utilizing SNAC \cite{siuzdak2024snac} as the speech encoder, which comprises seven complementary token layers. In a single step, it generates eight tokens, including text, while maintaining a one-step delay between layers. Furthermore, Mini-Omni and Mini-Omni 2 incorporates a batch approach that involves two samples: one requiring both text and speech responses and the other necessitating a text-only response. By discarding the text token from the first sample and embedding the output from the second sample into the first, it effectively transfer the model’s text-based capabilities to speech tasks, significantly enhancing reasoning abilities with minimal resource overhead. 

\quad$\bullet$ \emph{SpeechTokenizer.} On the output side, SpeechGPT-Gen synthesizes speech tokens using flow matching\cite{lipman2022flow}. Flow matching effectively models the transformation from a simple prior distribution to complex data distributions, yielding promising results in speech generation. SpeechGPT-Gen \cite{zhang2024speechgpt} applies flow matching for perceptual modeling, generating speech tokens that align with those of SpeechTokenizer \cite{zhang2023speechtokenizer}. Specifically, given speech $S$, semantic representation $V_1$, perceptual representation $V_{2:8}$ and the complete information representation $V_{1:8} = V_1 + V_{2:8}$ extracted by SpeechTokenizer, perceptual modeling refers to predicting the complete representation $V_{1:8}$ given the prompt speech a and the semantic representation $V_1$. SpeechGPT-Gen synthesizes response speech by concatenating the output of SpeechGPT \cite{zhang2023speechgpt} with the prompt speech and using a flow matching model.

\quad$\bullet$ \emph{Mimi.} Mimi \cite{defossez2024moshi} has eight codebooks at a frame rate of 12.5Hz, which requires 100 autoregressive steps to generate one second speech. This results in high computational costs and incompatibility with streaming inference. To address these issues, Moshi \cite{defossez2024moshi} proposes the RQ-Transformer, comprising a temporal Transformer and a deep Transformer. The RQ-Transformer breaks down a flattened sequence of length $K \cdot S$ into $S$ timesteps for a large temporal Transformer which produces a context embedding used to condition a smaller depth Transformer over $K$ steps. This allows scaling to longer sequences by increasing $S$ or to a higher depth by increasing $K$ than modeling the flattened sequence with a single model.

\quad$\bullet$ \emph{TiCodec.} Ti-Codec~\cite{ren2024fewer} is a decoupled codec model which can separate the time-varying and time-invariant information in speech and quantize them separately. Inspired by VALL-E \cite{wang2023neural}, Freeze-Omni \cite{xiong2024freeze} uses a token-based speech decoder which contains NAR prefill and AR generate stage to achieve speech output capabilities. The speech decoder mainly consists of the NAR decoder, the AR decoder, and the frozen decoder of a codec model \cite{ren2024fewer}. Both the NAR decoder and AR decoder are built upon transformer blocks. The NAR decoder is used to model the semantic features from the output of LLM, and then the AR decoder generates speech tokens based on the output of the NAR decoder. Finally, the decoder of the codec model converts the speech tokens into a speech stream.




\begin{table}[]
\centering
\caption{The comparison of semantic and acoustic representations.}
\scalebox{0.65}{
\label{comparison_of_rep}
\begin{tabular}{c|c|c|c|c|c}
\toprule
                        & \begin{tabular}[c]{@{}c@{}}Advantages of the \\ comprehension side\end{tabular} & \begin{tabular}[c]{@{}c@{}}Performance of \\ unify music and audio\end{tabular} & \begin{tabular}[c]{@{}c@{}}Compression rate \\ of speech\end{tabular}  & \begin{tabular}[c]{@{}c@{}}Emotional and \\ acoustic information\end{tabular} & \begin{tabular}[c]{@{}c@{}}Pipeline for \\ post-processing\end{tabular} \\
\midrule
Semantic & Strong                                                                          & Weak                                                                              & High & Less                                                            & Cascade                                                                 \\
\midrule
Acoustic & Weak                                                                            & Strong                                                                                & Low                                                                  &  More                                                              & End-to-end    \\
\bottomrule
\end{tabular}}
\end{table}

\subsection{Discussions about Representation used in Spoken Dialogue Systems}

\subsubsection{Semantic Representation vs. Acoustic Representation}
Current dialogue systems typically choose different approaches for the understanding (input) and generation (output) sides based on task requirements. For example, Spirit-LM \cite{nguyen2024spirit} uses semantic representations (HuBERT \cite{hsu2021hubert}) consistently on both ends, while Mini-Omni \cite{xie2024mini} uses semantic representations (Whisper \cite{radford2023robust}) on the input side and acoustic representations (SNAC \cite{siuzdak2024snac}) on the output side. Each combination offers unique advantages and trade-offs, and a consensus on a unified speech representation approach has yet to be reached in practical applications.

We revisited the differences between semantic and acoustic representations, as shown in Table~\ref{comparison_of_rep}. Benefiting from specific task objectives, models such as Wav2Vec \cite{schneider2019wav2vec}, HuBERT \cite{hsu2021hubert}, WavLM \cite{chen2022wavlm}, and Whisper \cite{radford2023robust} focus on extracting semantic information embedded within the spoken content. This inherent advantage allows speech to be directly mapped into the embedding space of large language models (LLMs), facilitating alignment with other modalities and fully leveraging the LLM’s strengths. In contrast, acoustic representations extracted by models like EnCodec \cite{defossez2022high} and DAC \cite{kumar2024high} are less conducive to LLM understanding, which is why SpeechTokenizer \cite{zhang2023speechtokenizer} and Mimi \cite{defossez2024moshi} opt for semantic distillation. In addition, semantic representations offer higher compression rates. By configuring various downsampling parameters in convolutional layers, models like HuBERT and Whisper easily achieve frame rates of 25Hz to 50Hz. Spirit-LM \cite{nguyen2024spirit}, for instance, uses 25Hz HuBERT units, meaning that only 25 tokens are needed to represent one second of speech. In contrast, acoustic features are designed with compression and reconstruction in mind, where the constraints of signal transmission make extreme compression and high-quality reconstruction challenging to achieve simultaneously. Although Mimi \cite{defossez2024moshi} has achieved a frame rate of 12.5Hz, its use of 8 codebooks means that autoregressively predicting one second of speech requires 100 steps. Finally, in certain scenarios, semantic representations hold distinct advantages. 


However, we must acknowledge that purely semantic representations fall short in naturalness and expressiveness, especially in tasks involving emotional expression or complex speech dynamics, where acoustic representations provide more nuanced information. For instance, HuBERT \cite{hsu2021hubert} cannot extract prosodic and stylistic features as effectively as EnCodec \cite{defossez2022high} or Emotion2Vec \cite{ma2023emotion2vec}. Notably, using acoustic representations allows for flexible handling of various data types—speech, audio, music, and sound—making dialogue systems more unified and versatile. Moreover, when acoustic representations are used as the output of a language model, they can seamlessly connect to the codec decoder for speech synthesis. In contrast, dialogue systems using semantic features often require separately trained vocoders \cite{nguyen2024spirit, kim2024unified} or rely on additional text-to-speech toolkits \cite{fang2024llama}. This gap is crucial for dialogue systems, as the resulting latency directly impacts the user experience.

Given the unique advantages of semantic and acoustic features across different tasks, future research may shift toward integrating these features. A valuable perspective is that models like SpeechTokenizer \cite{zhang2023speechtokenizer} and Mimi \cite{defossez2024moshi} have already attempted to distill semantic representations from HuBERT \cite{hsu2021hubert} or WavLM \cite{chen2022wavlm} into RVQ-1, ensuring a balanced representation of both semantic and acoustic information in the system. With technological advancements, we look forward to more unified and refined modeling approaches. A promising direction would be to design new training objectives for speech tokenizers, exploring both data-driven and objective-driven methods, thus avoiding the need for additional pre-trained models. As spoken dialogue Systems are still evolving, exploring more robust hybrid representations is indeed valuable.


\subsubsection{Continuous Representation vs. Discrete Representation}
There is still no consensus on whether to use continuous or discrete representations in the spoken dialogue systems. Considerations on the input side mainly depend on the type of representation model chosen by the system. Some systems \cite{xie2024mini, xie2024miniomni2opensourcegpt4ovision, fang2024llama} use models like HuBERT \cite{hsu2021hubert} or Whisper \cite{radford2023robust} to extract continuous speech representations, which requires adding a speech adapter and an additional training phase focused on modality alignment. Another systems \cite{zhang2023speechgpt, chen2024emova, defossez2024moshi} use models like EnCodec \cite{defossez2022high} or Mimi \cite{defossez2024moshi} to extract discrete speech representations, adding speech tokens directly to the LLM’s vocabulary, thereby shifting the training burden onto the LLM itself. Despite the different approaches, the key is to enable large language models to effectively understand speech features. For autoregressive models, using discrete inputs may appear more manageable; however, whether this truly outperforms continuous inputs in terms of performance remains to be explored. 

Language models trained with next-token prediction objectives tend to favor discrete modalities. Using discrete features on the output side naturally supports simple codec decoders \cite{xie2024mini, xie2024miniomni2opensourcegpt4ovision, defossez2024moshi, xiong2024freeze} for reconstructing high-fidelity speech, enhancing speech quality and acoustic control while enabling an end-to-end system. In contrast, continuous features may require additional text-to-speech toolkits \cite{fu2024vita} or vocoders \cite{fang2024llama}, resulting in a cascaded pipeline and making it difficult to preserve detailed acoustic information. Another notable advantage of using discrete representations as output is the ability to quickly feed them into the input of the next dialogue round, as demonstrated in OmniFlatten \cite{zhang2024omniflatten}. In the field of computer vision, a range of work \cite{zhou2024transfusion, xie2024show} has emerged that combines discrete and continuous representations, aiming to fully integrate these modes without information loss, and has already achieved success in certain areas. These approaches may provide valuable insights for the next generation of spoken dialogue systems.

\subsubsection{Single-Layer Quantizer vs. Multi-Layer Quantizer}
As previously mentioned regarding compression rates, the number of quantizers must be carefully considered when using the speech codec. Currently, dialogue systems commonly use multi-layer quantizers, such as those in EnCodec \cite{defossez2022high}, SpeechTokenizer \cite{zhang2023speechtokenizer}, SNAC \cite{siuzdak2024snac} and Mimi \cite{defossez2024moshi}. This inevitably introduces generation latency, as residual vector quantization requires each quantizer’s input to depend on the output of the previous quantizer. Mini-Omni \cite{xie2024mini} and Mini-Omni 2 \cite{xie2024miniomni2opensourcegpt4ovision} adopt an approach similar to MusicGen \cite{copet2024simple}, introducing delayed steps to enable parallel generation across multiple quantizers. Moshi \cite{defossez2024moshi} proposes splitting the RVQ, allowing the eight VQs to generate independently in parallel. These strategies help mitigate latency issues to some extent but still fall short of the efficiency achieved with semantic representations. 

Recently, research on single-layer quantizers has shown promising breakthroughs. Models like WavTokenizer \cite{ji2024wavtokenizer}, Single-Codec \cite{li2024single}, and BigCodec \cite{xin2024bigcodec} advocate using a single VQ to discretize speech, achieving competitive results in both reconstruction and generation tasks. Notably, WavTokenizer \cite{ji2024wavtokenizer} has already achieved an impressive compression rate of 40Hz. Integrating a single-layer quantizer with dialogue systems is promising, as it allows for rapid extraction of speech features on the input side and significantly reduces the burden of autoregressive modeling.

\subsubsection{With Text Guidance vs. Without Text Guidance}
In practice, researchers have found direct speech-to-speech generation challenging \cite{xie2024mini, xie2024miniomni2opensourcegpt4ovision, fang2024llama} due to complex mapping relationships, so intermediate texts are often generated to achieve higher generation quality. Current end-to-end dialogue systems commonly adopt one of two strategies: one \cite{fang2024llama, zhang2024intrinsicvoice} generates the hidden states corresponding to the text response first, which are then post-processed to obtain speech tokens; the other \cite{xie2024mini, xie2024miniomni2opensourcegpt4ovision, defossez2024moshi} generates text and speech tokens in parallel. These approaches leverage the text modeling capabilities of large language models, essentially guiding the synthesis of semantically consistent speech by first generating text. However, this comes at the expense of response speed.

Although directly performing speech-to-speech generation presents challenges such as increased model complexity and inference difficulty, we believe it remains a promising direction for future research. One approach is to retrain large spoken language models to adapt to specific speech representations. However, this faces challenges related to data resources, as large-scale and high-quality conversational datasets remain scarce. Additionally, this method cannot completely eliminate text prompts and requires multi-stage training, starting with text-speech pairs to allow the model to progressively acquire conversational capabilities. Another approach could begin with speech codecs, as demonstrated by SpeechTokenizer and Mimi’s extensive work in semantic distillation. We envision a novel speech codec that aligns text and speech during the encoding phase, thereby reducing the generation burden on large language models. By aligning speech representations with the text representation space earlier in the process, the autoregressive modeling would no longer require text guidance, giving rise to an entirely new paradigm for conversational systems.

\section{Training Paradigm of Spoken Dialogue Model}
\label{section4}

Existing text-based large language models have demonstrated strong contextual understanding and reasoning abilities in the field of natural language processing, such as GPT-4 \cite{achiam2023gpt}, Llama 3.1 \cite{dubey2024llama}, and Qwen-2 \cite{yang2024qwen2}. Due to their training on large-scale corpora, these models achieve exceptional accuracy when handling complex contexts. To further expand the capabilities of large language models, some research \cite{chen2024emova,chu2024qwen2,fu2024vita,xie2024miniomni2opensourcegpt4ovision} has explored enabling them to understand other modalities, thereby building multimodal interaction abilities. The spoken dialogue model, also known as the speech-text dialogue model, allows users to interact with LLMs naturally and straightforwardly through speech. However, the transition from text intelligence to speech intelligence involves two inherent hurdles: one core issue is the insufficient amount of speech data compared to the massive datasets used for pre-training text-based large language models. For instance, Llama 3.1 \cite{dubey2024llama}  uses 800 billion training tokens, and Qwen-2 \cite{yang2024qwen2} is trained on over 7 trillion tokens, whereas pure speech pre-training data often amounts to hundreds of thousands or millions of hours. For example, Moshi's \cite{defossez2024moshi} pre-training speech data comprises 7 million hours, and the amount of labeled speech data is even smaller, making it difficult to support LLMs in achieving powerful speech intelligence comparable to text. Another challenge is that speech information density is not as compact as text. Text commonly uses byte-pair encoding (BPE) \cite{gage1994new,sennrich2015neural} encoding to compress it into a tight token space, whereas the speech modality includes not only semantic information but also acoustical information, which is less dense. This undoubtedly increases the difficulty for LLMs to learn. Understanding and generating the inherent knowledge of the speech modality more effectively is a significant challenge. 

Consequently, existing spoken dialogue models aim to build upon text-based LLMs by incorporating the speech modality into these large language models. \cite{zhang2023speechgpt,chen2024emova,xie2024mini,defossez2024moshi} support speech-in and speech-out capabilities for LLMs, forming the foundation of basic speech dialogue capabilities. Some of the latest advanced approaches \cite{defossez2024moshi,zhang2024omniflatten,veluri2024beyond} attempt to transition from traditional turn-based spoken dialogue systems to full-duplex systems, aiming to simulate the natural spontaneity of human conversation. While these advancements are promising, achieving low latency and natural interaction in full-duplex systems remains a significant challenge. Moreover, enhancing LLMs to effectively handle the speech modality—mastering both speech comprehension and generation—while maintaining robust natural language text processing capabilities, is hindered by the limited size of labeled speech datasets. These datasets are far smaller compared to the vast amounts of pure text data available, which risks diminishing the models' original text processing capabilities. Thus, building a truly end-to-end conversational model that meets real-world requirements necessitates careful consideration of model architecture, training paradigms, and training data. Overall, we believe that several key aspects are crucial in the training paradigm of spoken dialogue models: aligning speech-text modalities to ensure consistent understanding, designing multi-stage training strategies for gradual adaptation, and optimizing training structures and inference paradigms for efficient performance.
\subsection{Architecture Paradigm about Modal Alignment of Speech and Text}
\begin{figure*}[ht]
    \centering
    \includegraphics[width=0.95\linewidth]{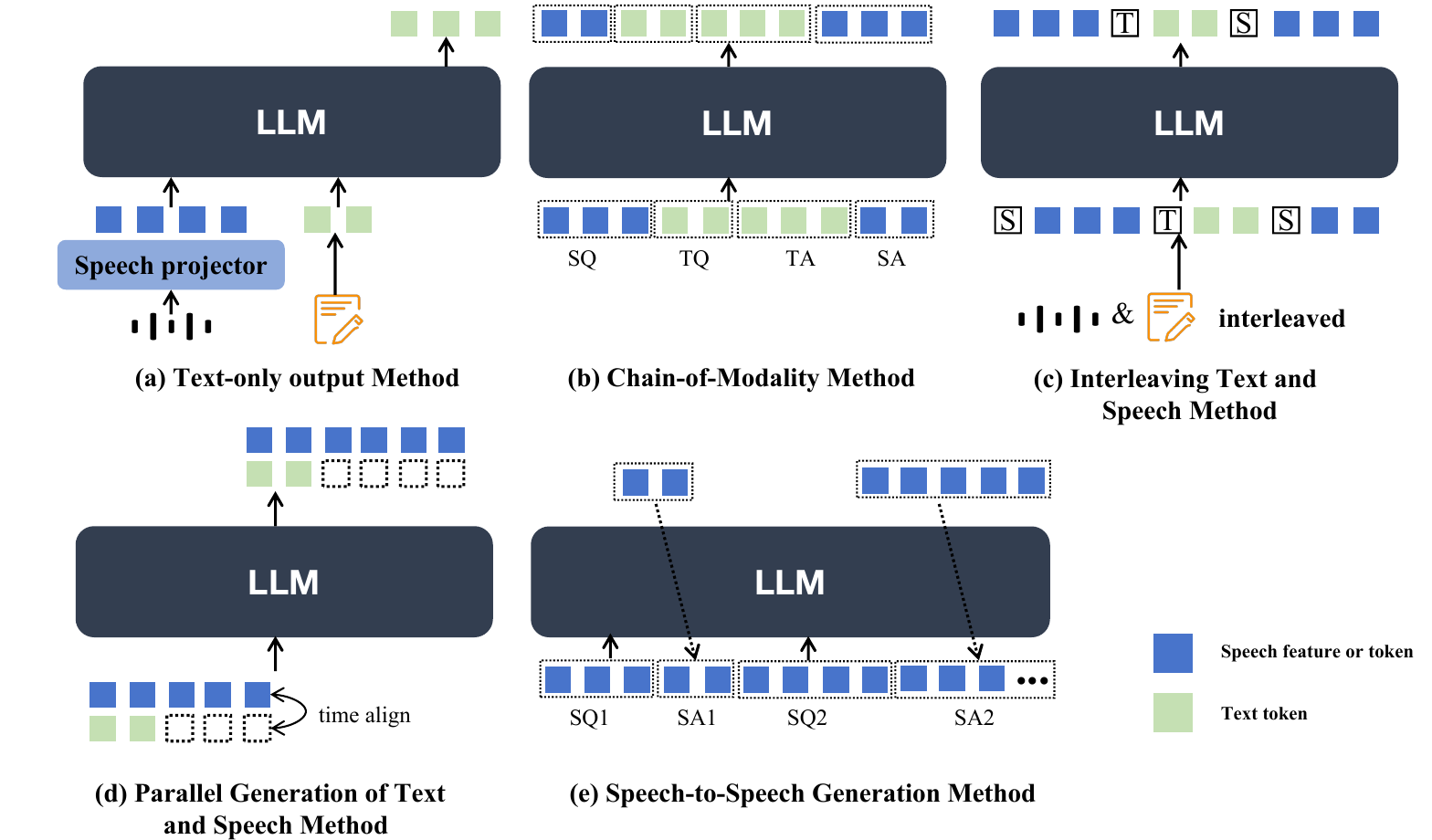}
    \caption{Categorization Diagram of Spoken Dialogue Model Architectural Paradigms.}
    \label{fig:archi_img1}
\end{figure*}

To enable large language models (LLMs) to handle both speech input and output, a significant amount of prior work \cite{rubenstein2023audiopalm,dubey2024llama,fang2024llama,xie2024mini,defossez2024moshi} has focused on adapting text-based foundation models into robust spoken dialogue models. Based on different architectural paradigms, these approaches can be broadly categorized into five types, as shown in Figure ~\ref{fig:archi_img1}.

\textbf{Text-Output Only Method.} These systems \cite{chu2024qwen2,chu2023qwen,gong2023joint,xue2023chat,tang2023salmonn,hu2024wavllm,das2024speechverse,fu2024vita} maintain the text-based LLM’s foundational structure unchanged, \textbf{using an audio encoder and adaptor to map speech input into the LLM's pre-trained text latent space directly.} This method of direct embedding alignment, combined with a multi-task training strategy, equips the LLM with the ability to 'listen,' thus enabling it to understand and process speech modality inputs effectively and perform exceptionally well in various audio understanding tasks. Nevertheless, the output remains text-based, which necessitates the use of an external text-to-speech (TTS) system \cite{casanova2024xtts,du2024cosyvoice} to generate speech output. LTU-AS \cite{gong2023joint} uses Whisper \cite{radford2023robust} and the Time and Layer-Wise Transformer (TLTR) as its audio encoder, allowing it to recognize both speech and audio events. Qwen-Audio 1 \cite{chu2023qwen} scales up audio-language pre-training to cover over 30 tasks and various audio types, facilitating universal audio understanding abilities. It employs a unified encoder for all audio inputs, bridging the gap between audio and textual modalities, and uses the large language model Qwen-7B \cite{bai2023qwen} as its foundational component. Qwen-Audio 2 \cite{chu2024qwen2} simplifies the pre-training process by utilizing natural language prompts for different data and tasks, with DPO \cite{rafailov2024direct} optimizing the model’s performance in terms of factuality and adherence to desired behavior. SALMMON \cite{tang2023salmonn} employs dual auditory encoders: a speech encoder from the Whisper model and a non-speech BEATs \cite{chen2022beats} audio encoder. The auditory features from these two encoders are complementary, making them suitable for general audio inputs that contain both speech and non-speech information. These inputs are then connected to a well-trained LLM using Q-former style attention to generate responses.  VITA \cite{fu2024vita} implements a duplex solution through two independent modules: one generates text responses to user queries, while the other continuously monitors environmental input to selectively provide updated interaction content, although it still requires an external TTS system. All the aforementioned methods frequently overlook paralinguistic information, including emotion, prosody, and non-verbal elements, rendering them insufficient for scenarios that involve emotional speech dialogue. ParalinGPT \cite{lin2024paralinguistics} utilizes an ASR model to obtain text and a speech encoder to extract emotion embeddings, thereby more accurately simulating both the linguistic content and paralinguistic attributes of spoken responses. E-chat \cite{xue2023chat} employs a Hubert speech encoder \cite{hsu2021hubert} to extract speech and emotion features, using a connection module to map these features to the textual space within the LLM decoder. Although these approaches have explored emotional responses within spoken dialogue systems, they require additional systems to synthesize speech from text and suffer from high latency, making real-time dialogue challenging to achieve.

\textbf{Chain-of-Modality (CoM) Method.} This method tokenizes speech into discrete tokens and extends the LLM’s vocabulary to handle both speech input and output. To address alignment issues between speech and text modalities, Recent works \cite{zhang2023speechgpt,zhang2024speechgpt,nachmani2023spoken,chen2024emova} utilize a prompting approach called Chain-of-Modality (CoM), which first generates response text autoregressively before producing the corresponding speech. This technique allows the text LLM's output to guide speech generation, thereby enhancing the quality of the response content. However, it is not suitable for live interactions, as the model must complete the entire text response before beginning speech generation, leading to increased response latency. SpeechGPT \cite{zhang2023speechgpt} and SpeechGPT-gen \cite{zhang2024speechgpt} employ the SpeechTokenizer \cite{zhang2023speechtokenizer} model as a speech token extractor, breaking down speech generation into the prediction of semantic tokens followed by acoustic tokens. Spectron \cite{nachmani2023spoken} performs speech continuation by predicting spectrograms frame-by-frame, optimizing the LLM with a combination of cross-entropy loss for text and reconstruction loss for speech frames. EMOVA \cite{chen2024emova}, on the other hand, utilizes the FSPIRAL \cite{huang2022spiral} architecture for its speech encoder to capture phonetic and tonal information, which is then discretized using finite scalar quantization (FSQ) \cite{mentzer2023finite}. Its speech response procedure is divided into three primary steps: 1) transcribing user instructions into text, 2) generating textual responses based on these instructions, and 3) producing style labels and response speech units from the textual responses. This process enables EMOVA to facilitate emotional speech dialogue.

\textbf{Interleaving Text and Speech Tokens.} Some earlier models \cite{rubenstein2023audiopalm,maiti2024voxtlm} employed supervised training methods, using specific input and output sequences, and trained on mixed speech-text tasks, including text-to-speech (TTS), automatic speech recognition (ASR), and speech-to-speech translation. Spirit-LM \cite{nguyen2024spirit} leverages the temporal alignment between speech and its transcription, continuing training on a pre-trained text-based LLM using alternating text and speech tokens. This significantly improves the model’s performance in both speech understanding and generation. However, it employs discrete Hubert units \cite{hsu2021hubert} as speech representations, which results in some loss of paralinguistic information. USDM \cite{kim2024unified} continues pretraining Mistral-7B \cite{chaplot2023albert} with interleaved speech-text data to capture multimodal semantics. For dialogue finetuning, it constructs templates using both speech and transcripts of user input as instruction data.

\textbf{Parallel Generation of Text and Speech.} PSLM \cite{mitsui2024pslm} proposes generating speech and text tokens in parallel to reduce latency; however, this approach may compromise response quality. Additionally, this method still relies on speech recognition for input \cite{radford2023robust}, which introduces further delay. Llama-Omni \cite{fang2024llama} introduces a novel streaming speech decoder that can simultaneously generate text responses and discrete speech unit sequences, significantly reducing latency and meeting real-time interaction needs. Moshi \cite{defossez2024moshi} and Mini-Omni \cite{xie2024mini} adopt similar approaches, introducing dual streams that generate both speech tokens and corresponding text tokens simultaneously on the assistant side, facilitating the transfer of the pre-trained LLM’s textual capabilities to the speech modality, enabling the model to directly engage in reasoning through speech. The key difference lies in how speech-text alignment is handled: Moshi \cite{defossez2024moshi} uses explicit alignment information to supervise the model’s learning, while Mini-Omni \cite{xie2024mini} allows the LLM to learn implicit alignment information. On the input side, Mini-Omni feeds continuous speech embeddings from the Whisper encoder \cite{radford2023robust} into the LLM, enhancing the model's ability to understand spoken instructions without requiring text input. However, inconsistencies between speech input and output introduce additional computational overhead, increasing latency in multi-turn dialogue scenarios. In contrast, Moshi allows users to input speech without relying on text, and generates both text and speech tokens in parallel on the assistant side. Moshi further extends its architecture to model several speech streams in parallel, allowing for conceptually and practically simple handling of full-duplex dialogues with arbitrary dynamics.

\textbf{Speech-to-Speech Generation.} This approach aims to remove the dependency on intermediate text, thereby reducing latency and making the system closer to real-time interaction. SyncLLM \cite{veluri2024beyond} achieves real-time full-duplex interaction through time chunking methods, integrating time information into LLMs to enable synchronous operation with the real-world clock. IntrinsicVoice \cite{zhang2024intrinsicvoice} utilizes a specific model to generate multiple speech tokens in a single step, effectively reducing speech token sequences to lengths comparable to text sequences while producing high-quality audio. Align-SLM \cite{lin2024alignslmtextlessspokenlanguage} utilizes a pre-trained self-supervised Hubert model \cite{hsu2021hubert} with K-means clustering \cite{hassid2024textually} to convert continuous speech representations into discrete units. It employs LoRA adapter \cite{hu2021lora} fine-tuning on a pre-trained Twist \cite{hassid2024textually} to produce multiple speech continuations from a given prompt and uses semantic metrics to generate preference data for Direct Preference Optimization (DPO) \cite{rafailov2024direct}. Experimental results indicate that integrating the preference optimization method significantly improves the semantic comprehension of the Spoken LLM.

\subsection{Multi-stage Training strategy}
\begin{figure*}[ht]
    \centering
    \includegraphics[width=0.95\linewidth]{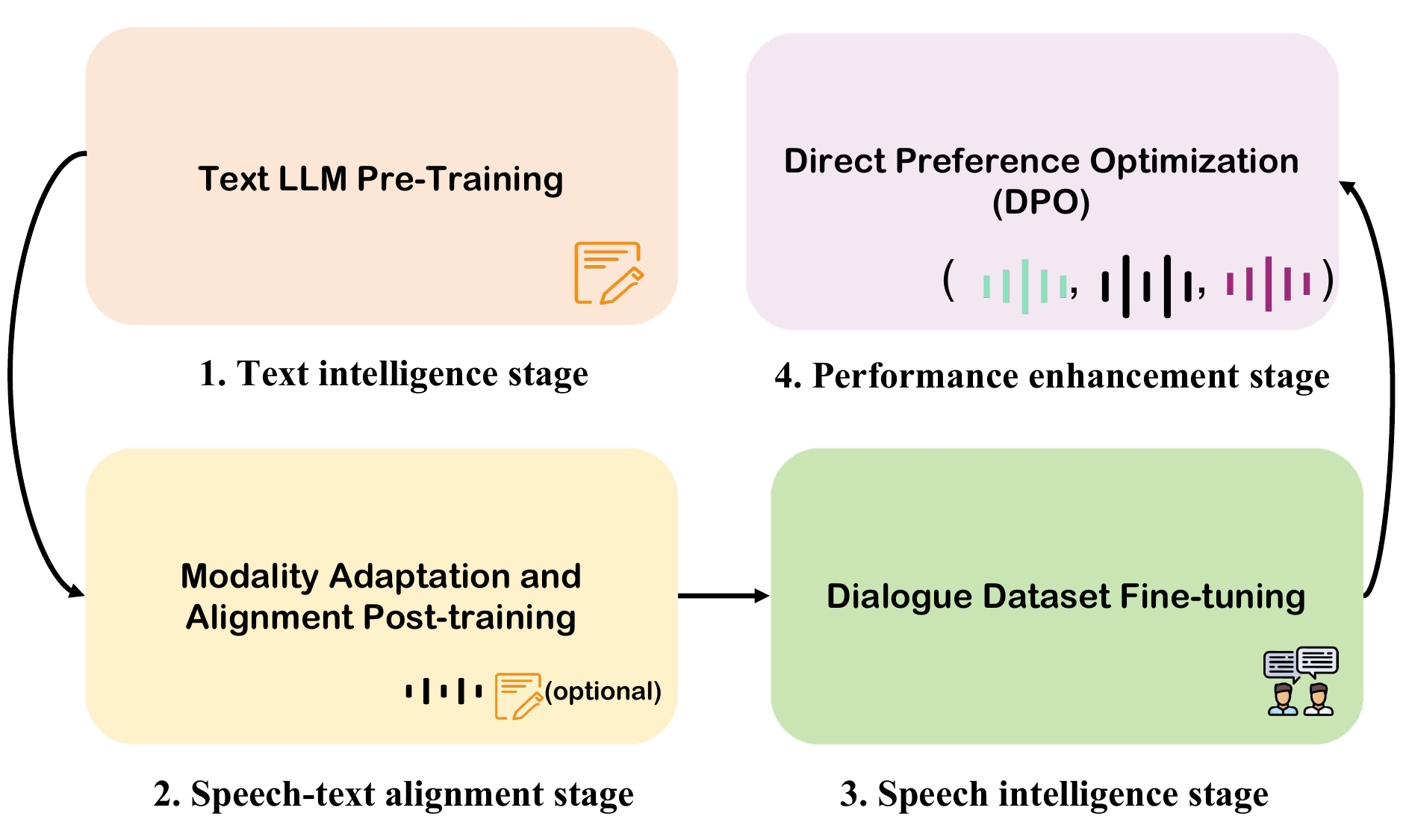}
    \caption{Diagram of Multi-stage Training Steps.}
    \label{fig:archi_img2}
\end{figure*}

This section primarily discusses the training process of the Spoken Dialogue Model, building upon previous work on spoken dialogue systems. Generally, this process consists of four stages: text LLM pre-training, modality adaptation and alignment post-training, followed by supervised fine-tuning, and optionally, preference optimization. The primary goal in training most spoken dialogue systems is to preserve the model's original capabilities while integrating the speech modality for voice interaction into the LLM. The diagram of multi-stage training can be referred to in Figure ~\ref{fig:archi_img2}.

\subsubsection{Text LLM Pre-Training}
The goal is to develop a text-intelligent LLM model capable of handling complex contexts and possessing knowledge reasoning abilities, thus preparing it for integration with speech-intelligent LLMs. Most spoken dialogue systems utilize pre-trained large language models as foundational models rather than pre-training with separate text data themselves. A series of approaches \cite{zhang2023speechgpt,zhang2024speechgpt,nguyen2024spirit,chen2024emova,fang2024llama,veluri2024beyond} use the LLaMA model and its variants as their foundational language model. On the other hand, \cite{du2023lauragpt,xie2024mini,xie2024miniomni2opensourcegpt4ovision,zhang2024omniflatten} employ the Qwen \cite{bai2023qwen,yang2024qwen2} family of large language models as their backbone. Meanwhile, Moshi \cite{defossez2024moshi} employs an RQ-Transformer for hierarchical autoregressive modeling of speech, utilizing a unique structure that involves pre-training a text-only language model with datasets from the internet (e.g., Wikipedia \footnote{\url{https://dumps.wikimedia.org/}} and StackExchange \footnote{\url{https://archive.org/details/stackexchange/}}). The collected data was filtered using a comprehensive preprocessing pipeline to ensure quality and relevance, which included deduplication to remove redundant entries, language identification to retain text in the desired language, and quality filtering to exclude low-quality or irrelevant content based on criteria such as coherence and completeness. VITA \cite{fu2024vita} utilizes Mixtral 8x7B1 \cite{jiang2024mixtral}, a representative LLM with a sparse mixture of experts (SMoE) architecture, and performs pure-text instruction tuning for its extended Chinese vocabulary.

\subsubsection{Modality Adaptation and Alignment Post-training}
\begin{figure*}[ht]
    \centering
    \includegraphics[width=0.95\linewidth]{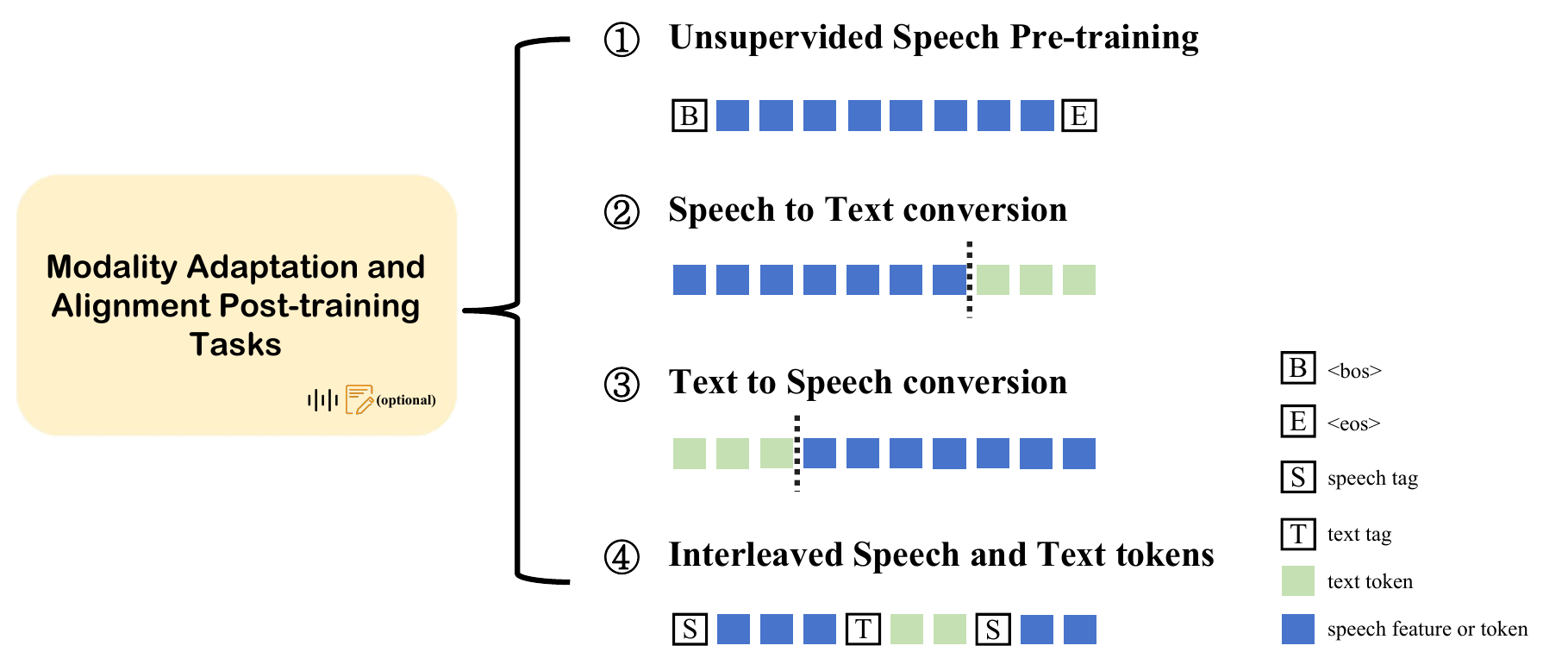}
    \caption{Alignment Post-training Methods.}
    \label{fig:archi_img3}
\end{figure*}
This phase explores strategies to adapt text-based large language models (LLMs) for speech modality input, focusing on aligning text and audio modalities effectively. The primary goal is to enhance the models' ability to understand and generate speech by bridging the gap between these two modalities. Common approaches include multimodal training techniques, leveraging unlabeled speech corpora, and employing multi-task learning frameworks. These methods typically involve fine-tuning existing LLMs with speech-related tasks and integrating speech-specific modules, such as speech adaptors and decoders, to facilitate seamless interaction between text and speech modalities. Different training tasks for modality adaptation and alignment are shown in Figure ~\ref{fig:archi_img3}. Spirit-LM \cite{nguyen2024spirit} continuously pretrains on text LLM checkpoints using interleaved text and speech tokens to improve the model's performance in speech understanding and generation. LLaMA-Omni \cite{fang2024llama} adopts a two-stage training strategy: the first stage jointly trains a speech adaptor and LLM with speech input and text responses, while the second stage uses the same dataset to train a streaming speech decoder independently. Consequently, this LLM primarily possesses the capability for speech input understanding, with speech generation handled by a separate decoder module. SpeechGPT \cite{zhang2023speechgpt}, Moshi \cite{defossez2024moshi}, and VITA \cite{fu2024vita} utilize unlabeled speech corpora to train models in a next-token prediction task. In the first phase, VITA focuses on training the audio encoder and connector, while in the second phase, it optimizes both the connector and the LLM model through multimodal training. Although capable of processing speech input, it outputs only text. Spectron \cite{nachmani2023spoken} addresses the alignment issue between text and speech representations by jointly supervising multiple objectives. IntrinsicVoice \cite{zhang2024intrinsicvoice} employs a two-stage training approach, constructing multiple cross-modal tasks from a single dataset to enable the model to better learn the semantic consistency between speech and text. Mini-Omni \cite{xie2024mini}, EMOVA \cite{chen2024emova}, and OmniFlatten \cite{zhang2024omniflatten} adopt similar methodologies, commencing with supervised multi-task fine-tuning of the text LLM backbone to achieve speech-text modality alignment and develop a multimodal LLM~\cite{jin2024efficientmllm, li2024surveybenchmarksmultimodallarge} using Automatic Speech Recognition (ASR) and Text-to-Speech (TTS) tasks. Notably, Mini-Omni divides the training of various modules into three phases: the first phase utilizes data from speech recognition and synthesis to enhance the model’s abilities in these aspects, training only the ASR and TTS adapters. The second phase focuses exclusively on enhancing the model’s text capabilities when given speech inputs, updating only the LLM parameters while freezing other modules. Through these two training phases, the original language LLM’s capabilities are maximally preserved, while adapting to speech modality input and output, thereby addressing the primary modality alignment tasks.
\subsubsection{Supervised Fine-tuning or Dialogue Dataset Fine-tuning}
During this stage, most models use instruction-following datasets or dialogue data for supervised fine-tuning of the LLM, enhancing natural conversational abilities. \cite{zhang2023speechgpt,zhang2024speechgpt} propose a two-stage instruction-tuning process that includes cross-modal instruction fine-tuning and chain-of-modality instruction fine-tuning. Ultimately, the model follows the A-T-T-A method to achieve end-to-end speech input and output. EMOVA \cite{chen2024emova} employs a similar chain-of-modality concept to construct instruction-tuning datasets, empowering it to respond accurately to speech instructions. Moshi \cite{defossez2024moshi}, Mini-Omni \cite{xie2024mini}, OmniFlatten \cite{zhang2024omniflatten}, and SyncLLM \cite{veluri2024beyond} utilize spoken dialogue datasets for fine-tuning, endowing the models with conversational interaction capabilities. Remarkably, Moshi constructs a more natural and realistic dialogue dataset that incorporates elements such as noise and overlap, enabling the model to learn authentic multi-stream interactions. OmniFlatten fine-tunes the speech-text LLM using interleaved and serialized dialogues across three stages to progressively train the model in acquiring half-duplex and full-duplex communication capabilities. Similarly, SyncLLM employs a three-stage training procedure that predominantly uses synthetic spoken dialogue data along with a relatively small amount of real-world spoken dialogue data to develop a full-duplex voice agent.
\subsubsection{Preference Optimization and Reinforcement Learning}
The research on leveraging preference optimization to align a spoken dialogue model with human preferences is virtually absent. Recently, \cite{anastassiou2024seed,zhang2024speechalign,chen2024enhancing} adopted preference optimization for Text-to-Speech (TTS) models to align speech synthesis quality with human preferences but not for spoken dialogue models. Align-SLM \cite{lin2024alignslmtextlessspokenlanguage} pioneers the integration of Direct Preference Optimization (DPO) \cite{rafailov2024direct} in textless Spoken Language Models (SLMs) to enhance semantic understanding. It transforms continuous speech into discrete units using a pre-trained Hubert model and K-means clustering. LoRA fine-tuning on a Spoken LLM generates multiple speech continuations from prompts. Semantic metrics create preference data offline, making DPO training efficient and stable, eliminating the need for an external reward model. Coupled with curriculum learning \cite{bengio2009curriculum}, Align-SLM progressively refines preference data selection, optimizing semantic feedback, and improving SLM performance.

\subsection{Training Frameworks and Generation Strategies}
Recent advanced methods in spoken dialogue models employ a variety of innovative techniques to achieve more natural speech output and lower latency. In this part, we explore various approaches that exemplify these advancements:

    \quad$\bullet$ \emph{LLama-Omni.} LLama-Omni \cite{fang2024llama} adds a streaming speech decoder that operates after the LLM. This decoder runs in a non-autoregressive manner, taking the output hidden states from the LLM as input and generating the discrete unit sequence corresponding to the speech response. To model the variable-length mapping between input and output, LLama-Omni employs an upsample factor, denoted as $\lambda$, along with Connectionist Temporal Classification (CTC) loss \cite{graves2006connectionist}. This ensures that the model can generate speech responses simultaneously with text responses. Additionally, a predefined chunk size is set to further enable vocoder streaming synthesis of speech waveforms, facilitating real-time interaction and reducing latency.
    
    \quad$\bullet$ \emph{Mini-Omni.} Mini-Omni \cite{xie2024mini} selects SNAC \cite{siuzdak2024snac}, a music-grade encoder, to discretize one second of audio into hundreds of tokens, which significantly increases the burden on the LLM for modeling speech tokens. Delay Pattern language model decoding strategies are often applied in modeling multiple parallel streams of acoustic tokens in speech tasks like MusicGen \cite{copet2024simple}, VoiceCraft \cite{peng2024voicecraft}, and Parler-TTS \cite{lyth2024natural}. Compared with traditional sequential step decoding, this strategy can effectively reduce the time steps required for LLM decoding and generating speech tokens. Inspired by this, Mini-Omni innovatively applies text-instructed delayed parallel generation to address the issue of long SNAC codebook sequences, simultaneously producing audio and text tokens. This effectively leverages and preserves the original capabilities of the language model. Moreover, Mini-Omni proposes a Batch Parallel Decoding method. Specifically, it generates two samples in parallel for a single input: the first predicts text tokens, and the second predicts both text and speech tokens simultaneously. The text output from the first sample is embedded into the corresponding positions of the second sample, while the second sample's text output is discarded. This further enhances the model’s reasoning capabilities during dialogue, maximizing the transfer of its text-based abilities.
    
    \quad$\bullet$ \emph{IntrinsicVoice.} IntrinsicVoice \cite{zhang2024intrinsicvoice} introduces a speech encoder and a streaming vocoder for the tokenization and detokenization of speech, and a GroupFormer for modeling speech and text sequences. This architecture integrates a large language model (LLM) with a GroupModel. Specifically, it uses a pre-trained HuBERT encoder \cite{hsu2021hubert} and its corresponding KMeans quantizer \cite{hassid2024textually} to process speech inputs into discrete units. These units are organized into a grouped token sequence through a group partition operation. The grouped tokens are then passed through an embedding layer and adaptor module to map these embeddings into the LLM's embedding space. The context embeddings output by the LLM are processed through a linear layer and concatenated with a specified number of learnable queries. This input is fed into a smaller non-autoregressive transformer encoder model, dubbed the "GroupModel," to predict a group of speech tokens in one step. The introduction of GroupFormer effectively improves the model's ability to handle sequences within a group, mitigates the modality gap between speech and text, accelerates inference speed, and alleviates issues associated with long-sequence modeling.

    \quad$\bullet$ \emph{Moshi.} Moshi \cite{defossez2024moshi} introduces a mini codec model with 8 codebooks at a frame rate of 12.5 Hz for speech representation, where one second corresponds to 100 speech tokens. It adopts an RQ-Transformer consisting of a Temporal Transformer and a smaller Depth Transformer as the backbone network for the LLM, hierarchically modeling multi-codebook audio tokens. Similar architectures have appeared in prior research, such as UniAudio \cite{yang2023uniaudio} and Megabyte \cite{yu2023megabyte}. The Depth Transformer models sub-sequence tokens conditioned on temporal context predicted by the Temporal Transformer. Given the smaller size of the Depth Transformer, sub-sequence generation can almost be viewed as parallel generation. This allows the model to scale to longer sequences by extending the temporal modeling capacity of the Temporal Transformer or to achieve greater depth by enhancing the hierarchical modeling capabilities of the Depth Transformer, rather than modeling the flattened sequence with a single model.
    
    \quad$\bullet$ \emph{SyncLLM.} SyncLLM \cite{veluri2024beyond} employs an auto-regressive transformer decoder for full-duplex dialogue, integrating time synchronization to align speech units with the real-world clock. It predicts interleaved speech tokens for both dialogue partners, maintaining timing with speaker tags. The model is trained on deduplicated HuBERT token sequences to enhance semantic fidelity while managing latency by anticipating user responses. Interpolation reconstructs token sequences to fit expected structures, facilitating seamless speech synthesis.


\textbf{Text-guided generation.} Some end-to-end methods like \cite{zhang2023speechgpt,zhang2024speechgpt,nachmani2023spoken,chen2024emova} use chain-of-thought reasoning, which allows guiding speech generation with the output of an underlying text LLM. However, this is fundamentally incompatible with live interactions, as the model needs to produce an entire answer as text before it starts speaking. Later methods \cite{fang2024llama,xie2024mini,defossez2024moshi} can accept user speech input and simultaneously output speech and text, ensuring high-quality responses while significantly reducing latency. Lama-Omni \cite{fang2024llama} utilizes a streaming decoder to generate text and speech tokens in parallel. Mini-Omni \cite{xie2024mini} is restructured to transfer language reasoning abilities to streaming audio output through a text-audio parallel decoding approach. Moshi \cite{defossez2024moshi} details a novel feature, the Inner Monologue, which consists of joint modeling of the textual and speech modalities on the system side to improve the quality of interactions.

\textbf{W/o text-guided generation.} Other methods achieve speech-to-speech generation without relying on text stream generation. IntrinsicVoice \cite{zhang2024intrinsicvoice} introduces a novel GroupModel that predicts a group of speech tokens in one step based on global context embeddings. SyncLLM \cite{veluri2024beyond} predicts interleaved chunks of token sequences at each time step, allowing the model to handle all conversational cues such as backchannels, overlaps, interruptions, etc.

\subsection{Discussions about Training Paradigm in Spoken Dialogue Models}
\subsubsection{Text and Speech Modality Alignment} In spoken dialogue systems, the alignment between speech and text modalities is a crucial stage. To preserve the textual intelligence of large language models (LLMs) as much as possible, nearly all current methodologies \cite{zhang2023speechgpt,mitsui2024pslm,fang2024llama,xie2024mini,xie2024miniomni2opensourcegpt4ovision,defossez2024moshi,zhang2024omniflatten} incorporate a post-training phase utilizing speech-text paired data when developing spoken dialogue models. This may involve either expanding the vocabulary to treat speech tokens as an extension of the original vocabulary or using speech adaptors to map speech embeddings to the original text latent space of the LLM, and designing multi-task training objectives to achieve alignment between text and speech modalities. For example, data from speech recognition and speech synthesis can be used to train the model's speech recognition and synthesis capabilities. Although this is an effective strategy, its implementation can still lead to a certain degree of catastrophic forgetting in LLMs due to the large volume of pre-trained text corpora and the imbalance with paired speech-text data, which can harm the model's text-based capabilities. Therefore, precise parameter design and customized optimization strategies are needed to mitigate this issue as much as possible, as demonstrated by approaches like Moshi \cite{defossez2024moshi}.

This raises a consideration: during the training phase of spoken dialogue models, is it feasible to directly utilize speech data for adaptation to text-based LLMs, thereby eliminating the necessity for speech-text paired data? This is because unlabeled speech data is abundant and easily accessible, making it convenient and beneficial for training the speech intelligence of LLMs. This approach would require us to obtain a pre-aligned speech representation with the text modality. Perhaps we can consider further exploration and experimentation in the speech tokenizer component, such as directly mapping the semantic discrete units of speech onto the text token space to achieve enforced alignment.

\subsubsection{Different Temporal Alignment Methods in Spoken Dialogue Models}
In speech and text modalities, there is often a significant mismatch in sequence lengths. Even when some speech tokenizers \cite{ji2024wavtokenizer,li2024single} employ extreme sequence compression methods, a length gap remains between the two. Temporal alignment information between speech and text has been explored in tasks like Automatic Speech Recognition (ASR) and Text-to-Speech (TTS) as demonstrated by models such as Whisper \cite{radford2023robust}, FastSpeech \cite{ren2020fastspeech}, and VITS \cite{kim2021conditional}. Recently, some spoken dialogue systems have utilized temporal alignment information to enhance model performance, yielding promising results. For instance, Spirit-LM \cite{nguyen2024spirit} uses interleaving text and speech tokens for continual pre-training on the LLaMA base model, significantly boosting the model’s performance in speech understanding and generation. Experimental visualizations demonstrate that the similarity between text and speech features is notably higher in models trained with interleaved token sequences compared to those trained without this approach. This indicates that providing the model with explicit fine-grained temporal alignment information can effectively enhance modality alignment and improve the performance of LLMs.

Mini-Omni \cite{xie2024mini} achieves parallel generation of text and speech by padding text tokens to match the length of speech tokens, allowing the LLM to implicitly learn the alignment information between speech and text tokens. This can be viewed as a form of sentence-level temporal alignment information, a method also utilized in recent speech synthesis work \cite{chen2024f5}. Moshi \cite{defossez2024moshi}, on the other hand, uses word-level speech-text temporal alignment information and special marker tokens to achieve similar parallel generation capabilities. The difference lies in that Mini-Omni fully allows the LLM to implicitly learn the alignment, whereas Moshi provides word-level alignment priors first, and then lets the model learn finer-grained alignments.

Exploring the impact of introducing different levels of temporal alignment priors on the training effectiveness of spoken dialogue models, such as sentence-level, word-level, or phoneme-level, is an intriguing area of research. Understanding how these various alignment strategies affect model performance can guide the development of more efficient and accurate systems. For instance, sentence-level alignment might offer a broader contextual understanding, while word-level or phoneme-level alignments could provide more detailed synchronization between speech and text, potentially leading to improvements in nuanced tasks like speech synthesis and understanding.

\subsubsection{Reinforcement Learning (RL) in Spoken Dialogue Models} Reinforcement Learning (RL) has proven to be an effective learning paradigm in text and image processing \cite{schulman2017proximal,sutton1999policy,wallace2024diffusion}. Recent research has shown that Direct Preference Optimization (DPO) \cite{rafailov2024direct} can be extended to music and speech generation \cite{cideron2024musicrl,zhang2024speechalign}. MusicRL \cite{cideron2024musicrl} uses Reinforcement Learning from Human Feedback (RLHF) to improve music generation by fine-tuning a pretrained model for better text adherence and audio quality. By collecting extensive human feedback, MusicRL creates a more refined and subjective music generation system. Seed-TTS \cite{anastassiou2024seed} explores RL methods, comparing external reward models like REINFORCE with simpler methods like DPO. The study highlights using REINFORCE to enhance speaker similarity and emotion controllability in the Seed-TTS system. Qwen2-Audio \cite{chu2024qwen2} uses DPO to align with human preferences by optimizing responses based on human-annotated data. This enhances its ability to follow audio instructions accurately and intelligently respond to complex audio inputs, improving its performance in audio-centric tasks. However, in the dialogue system field, reinforcement learning techniques based on human feedback \cite{huang2023survey} are rarely applied. Considering the diversity of inputs and outputs in large language models, exploring the incorporation of reinforcement learning strategies such as Proximal Policy Optimization (PPO) \cite{schulman2017proximal} can be beneficial. Additionally, considering the performance metrics for evaluating spoken dialogue systems, designing targeted reinforcement learning strategies and feedback functions to enhance different objectives is also a direction worth exploring.

\section{Streaming, Duplex, and Interaction}

 Streaming, full-duplex technology, and interactions, are crucial elements for enhancing the interactive capabilities of spoken dialogue models because they directly impact the system's responsiveness, the fluidity of natural interaction, and its ability to handle complex interactions.Unlike text language models, spoken dialogue models require real-time processing of user input. \textbf{Streaming} allows the system to instantly acquire and process speech data; \textbf{full-duplex technology} enables both the system and user to speak simultaneously, enhancing the naturalness of interaction; and \textbf{handling of interactions} provides the model with the ability to recognize and adapt to various conversational contexts, making the dialogue more intelligent and realistic.
 Building on early explorations, GPT-4o's advanced spoken dialogue capabilities have ignited a surge of research interest. With real-time voice processing and natural conversational interaction, these models offer users a seamless and efficient communication experience. However, achieving these capabilities requires deep research into model architecture, data collection, system design, and training methods. The model needs to be carefully designed and optimized in terms of real-time performance, stability, and response speed. At the same time, duplex technology is an indispensable key implementation, which ensures that the voice model has both "ears" and "mouths". Next, we will first discuss the streaming processing method in Section 5.1, then introduce the key technologies of duplex communication and explains how to handle interactation to improve user experience in Section 5.2.
\subsection{Streaming Spoken Dialogue Models}
The core of streaming speech models lies in their "real-time" and "continuous" capabilities, meaning they can process input and generate output simultaneously without waiting for complete input. This includes two main aspects:

    \quad$\bullet$ \emph{Streaming Understanding.} The model can process audio input as the user speaks, without needing to wait for the user to finish entirely, allowing it to align more naturally with the flow of conversation.

    \quad$\bullet$ \emph{Streaming Generation.} This concept refers to the model's ability to generate output without waiting for all intermediate hidden states. Instead, it can produce output progressively as processing occurs, which improves responsiveness and allows for smoother, more efficient interactions.

These streaming capabilities allow the model to perform more fluidly in real-time interactions, providing a seamless communication experience for users. We will explore streaming techniques in both end-to-end and cascaded spoken dialogue models, discussing the implementation methods of streaming in each system and highlighting their similarities and differences. 

\subsubsection{Streaming End-to-End Spoken Dialogue Models}
End-to-end streaming spoken dialogue models often leverage the knowledge of pre-trained text language models alongside an audio tokenizer, employing an tokenizer-detokenizer architecture to process and output audio signals. Based on the concepts of streaming input and output discussed above, end-to-end models also require specific design considerations to enable streaming capabilities. These designs center around the model’s input and output handling and can be distilled into three core techniques: causal convolution, causal attention mechanisms, and queue management.

    \textbf{Causal Convolution.} Causal Convolution~\cite{bai2018empirical} is a specialized form of convolution widely used in time-series processing, especially suitable for streaming speech models. The key feature of causal convolution is that the current output depends only on the current and past inputs, without being influenced by future inputs, thereby strictly respecting temporal order. Unlike regular convolution, causal convolution achieves this by "shifting" the convolution kernel to avoid accessing future information. In a one-dimensional time series, if the convolution kernel size is \(k\), a standard convolution would use data from \((t - k/2)\) to \((t + k/2)\) at the current time step \(t\). Causal convolution, however, pads the input on the left with \(k-1\) zeros so that the kernel only uses data from \(t - k + 1\) to \(t\), aligning the kernel to only consider current and past inputs. This padding ensures that each layer's output depends solely on current and prior information, maintaining causality. To further expand the model’s receptive field while preserving causality, \textbf{dilated causal convolution} can be used. This technique introduces gaps within the kernel by inserting zeros between weights, effectively expanding the convolution’s range. This allows the model to capture longer dependencies in the data without increasing latency, which is particularly useful for streaming applications. In streaming spoken dialogue models, causal convolution plays a critical role in:

\quad$\bullet$ \emph{Ensuring real-time processing.} Causal convolution allows the model to compute outputs without accessing future frames, enabling real-time processing by generating outputs as input is received, which is essential for streaming.
   
\quad$\bullet$ \emph{Reducing latency.} By not requiring future input data, causal convolution significantly lowers the latency in speech models, making it more suitable for real-time interaction applications, such as voice assistants and live translation.


    \textbf{Causal Attention.} Causal Attention is a specialized form of the attention mechanism designed to ensure that each position in a sequence can only attend to previous positions, thus preserving the temporal order crucial for streaming models. This approach ensures that the model’s current output depends only on past and present information, preventing any “leakage” of future information, which is essential for real-time processing tasks. In causal attention, the attention mask is typically used to achieve causality. By applying a mask that blocks connections to future time steps, the model restricts each token’s receptive field to only the tokens before it. Specifically, a lower triangular mask is applied to the attention matrix, setting values to negative infinity for positions corresponding to future tokens. This masking technique ensures that the model’s predictions for each time step only consider current and past inputs, thereby adhering to a strict causal structure. In streaming speech models, causal attention plays a significant role in enabling real-time interaction. Unlike standard attention, which requires access to the entire sequence, causal attention can operate incrementally. As new inputs are processed, the model can generate outputs without waiting for future context.

    \textbf{Queue Management~\cite{wu2023audiodec}.} Audio streams are typically split into frames, then processed in sequence via a queue management system that ensures real-time, orderly processing.

Some end-to-end models, such as Llama-Omni\cite{fang2024llama}, Mini-Omni\cite{xie2024mini} and Mini-Omni2\cite{xie2024miniomni2opensourcegpt4ovision}, employ non-streaming ASR model Whisper as an audio encoder components. These models have made improvements on the output side to reduce latency. 

\quad$\bullet$ \emph{Mini-Omni.} Mini-Omni use a generation strategy delayed parallel decoding is a that layer-by-layer delays during audio token generation. This allows the model to generate text and multiple audio tokens simultaneously at each step, accelerating streaming audio generation and ensuring low-latency real-time output.

\quad$\bullet$ \emph{Llama-Omni.} Llama-Omni incorporates a non-autoregressive streaming speech decoder that leverages connectionist temporal classification (CTC) to directly generate a sequence of discrete audio tokens as the response.

\quad$\bullet$ \emph{Intrinsicvoice.~\cite{zhang2024intrinsicvoice}} Intrinsicvoice introduced GroupFormer module  to group speech tokens, reducing the length of speech sequences to match that of text sequences. This approach accelerates inference, alleviates the challenges of long-sequence modeling, and effectively narrows the gap between speech and text modalities.We think they cannot be considered fully streaming because they are not designed to be streaming on the input side. 

\quad$\bullet$ \emph{Moshi.~\cite{defossez2024moshi}} In contrast, Moshi references the architecture of SpeechTokenizer to train a streaming codec from scratch, serving as the audio tokenizer-detokenizer. The entire model, including the codec, transformer, and attention mechanism, is built on a causal structure.

\quad$\bullet$ \emph{OmniFlatten.~\cite{zhang2024omniflatten}} OmniFlatten proposes chunk-based processing of text and speech along with gradual learning techniques and data handling to reduce turn-taking delays, such as response delays when users finish speaking or interrupt the system. These models have achieved true streaming capabilities and established a foundation for diverse, bidirectional interactions.
\subsubsection{Streaming Cascaded Spoken Dialogue Models}
Consistent with the above, ensuring streaming capability in a model relies on designing both input and output for streaming. Due to its cascaded nature, a cascaded model typically relies on external streaming ASR and TTS components, placing the streaming responsibility on these ASR and TTS modules.

In~\cite{wang2024full}, comparative studies were conducted on the streaming ASR model \textbf{U2++ Conformer}~\cite{wu2021u2++}, streaming TTS model \textbf{XTTS-v2}~\cite{casanova2024xtts}, non-streaming ASR \textbf{Whisper}, and non-streaming TTS \textbf{VITS}~\cite{kong2023vits2}. The combination of streaming components achieved the lowest latency and significantly contributed to interactive interruption capabilities. 

\subsection{Duplex Technology and Interaction}

\subsubsection{Duplex Technology}
The term Duplex originates from the field of communications, used to describe interaction modes between two parties in data transmission. Depending on the type of communication, duplex is divided into half-duplex and full-duplex.

With the development of audio processing and generation technology , the concept of duplex has been introduced to speech systems, especially within the context of speech language models. Here, duplex doesn’t just refer to signal transmission but emphasizes the synchronization and natural interaction in human-computer dialogue. Specifically, within model architecture, it means that the model must retain its ability to perceive external input even while generating a response---essentially, the ability to listen while speaking.

\begin{figure}[h]
    \centering

    \begin{subfigure}[b]{0.8\textwidth}
        \centering
        \includegraphics[width=\textwidth]{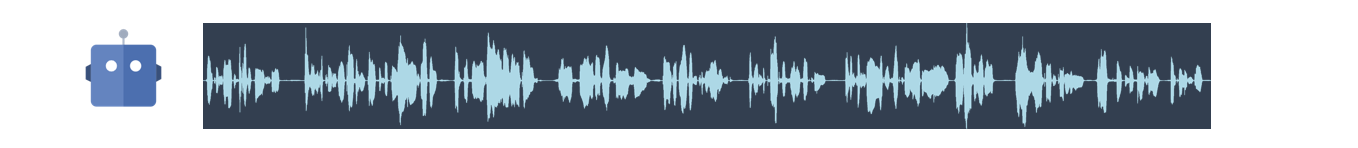}
        \caption{Simplex: One-way communication, and the direction is fixed.}
        \label{fig:simplex}
    \end{subfigure}

    \begin{subfigure}[b]{0.8\textwidth}
        \centering
        \includegraphics[width=\textwidth]{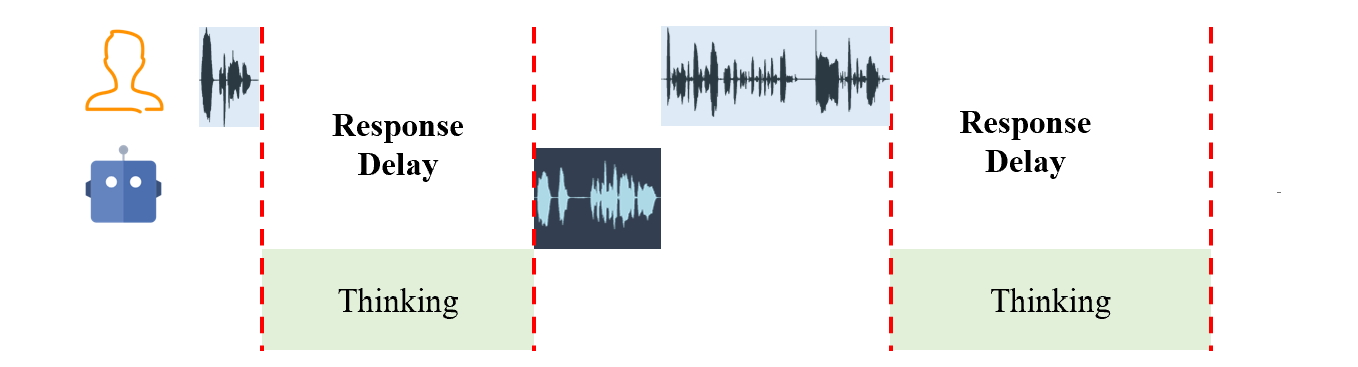}
        \caption{Half-duplex: Two-way communication, but not simultaneously.}
        \label{fig:half-duplex}
    \end{subfigure}

    \begin{subfigure}[b]{0.8\textwidth}
        \centering
        \includegraphics[width=\textwidth]{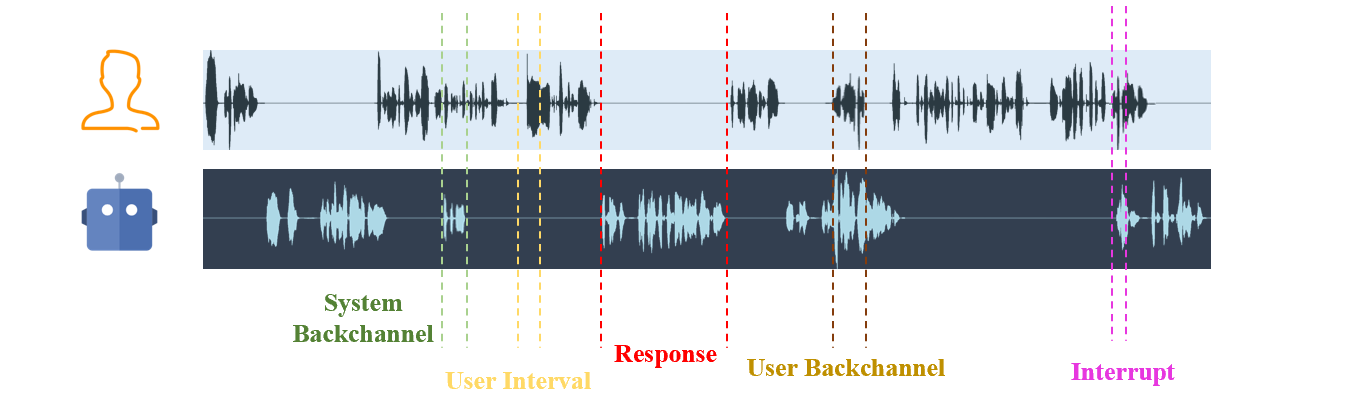}
        \caption{Full-duplex: Two-way communication, simultaneously.}
        \label{fig:full-duplex}
    \end{subfigure}

    \caption{The illustration of Simplex, Half-Duplex, and Full-Duplex.}
    \label{fig:duplex-illustration}
\end{figure}

    \textbf{Simplex.} In simplex communication, data flows in only one direction. The speaker can send data, while the listener can only receive it. As shown in Figure \ref{fig:simplex}, the robot continuously transmits audio, while the user has no ability to respond. This fixed-direction, one-way communication has the limitation of lacking interactivity.

    \textbf{Half-Duplex.} In half-duplex communication, data flows in both directions but not simultaneously. The two parties must take turns speaking and listening. As illustrated in Figure \ref{fig:half-duplex}, the user speaks first, followed by a response delay during which the robot "thinks" before replying. The robot’s response occurs only after the user has finished speaking, and vice versa. This turn-taking method is similar to using a walkie-talkie, where each party can only transmit after the other has finished, limiting efficiency.Half-duplex is a common mode in early voice interaction systems. In a typical half-duplex interaction, there are noticeable pauses in the conversation; the user and the system cannot “speak”  simultaneously, making the conversation feel less smooth, much like communication through a walkie-talkie. For example, voice assistants like Siri use wake words or button presses to trigger the dialogue and require the speaker to finish a complete sentence before responding. These systems typically adopt an ASR-LM-TTS cascaded structure and are often constrained by cascade delays and the turn-based nature of text language models. Although this interaction method is simple and easy to implement, it can feel rigid and disjointed in natural conversational settings, with notable latency. It is designed more for command execution rather than interactive communication.

    \textbf{Full-Duplex.} Full-duplex communication allows both parties to send and receive data simultaneously~\cite{ma2024language}. Figure \ref{fig:full-duplex} shows the user and robot engaging in overlapping, real-time interaction, where backchannels and interruptions are possible. This mode enables a natural, two-way conversation, where both the user and robot can speak, respond, and even interrupt each other as needed, much like a phone call.In dialogue systems, full-duplex means that the system and user can speak simultaneously and interrupt each other, making it closer to natural conversation in real life. Full-duplex large voice models allow the system not only to listen and understand the user while they speak but also to interrupt at appropriate moments or respond with backchannel cues. Moreover, the system can detect the user’s intent to interrupt and pause itself accordingly, maintaining a smooth flow in the interaction. 

The ultimate goal of a spoken dialogue moded is to make the user feel as though they are conversing with a real human friend.
Clearly, full-duplex technology is essential for achieving natural voice dialogue systems, enabling the system to send and receive audio signals simultaneously, thus facilitating real-time interaction. Unlike text-based models, it doesn’t “cover its ears” while speaking. Users and intelligent agents can interrupt each other while listening or express their attitude through non-verbal signals, such as interjections or laughter. The challenges in realizing this lie in ensuring conversational fluidity, seamless turn-taking, and precise timing of interactions. Developing a full-duplex system that can both generate and receive voice signals in complex interactive scenarios remains a key focus in academic and industrial research.
\subsubsection{Interaction}

Now that we understand duplex technology, we can further explore duplex spoken dialogue model.

We start with some concept.Turn-taking is the core concept in duplex dialogue. It refers to the process in which speakers take turns speaking in an orderly manner during a conversation, forming a pattern of turn-taking. Over the past few decades and has been extensively studied across fields such as linguistics, phonetics, and sociology. Some research ~\cite{raux2009finite,sacks1974simplest}uses a non-deterministic finite-state machine with six states to describe the turn-taking behavior between the system and the user in a spoken dialogue system (SDS). It outlines all possible states of turn-taking within an SDS, defining the objective of turn-taking as minimizing mutual silence or overlap between interlocutors, thereby improving communication efficiency. Turn-taking encompasses three fundamental concepts:

\quad$\bullet$ \emph{Turn-taking cues}~\cite{duncan1972some,duncan1974signalling}.  These include voice, rhythm, breathing, gaze, or gestures. Agents can use these cues to determine whether to take a turn from the user or to relinquish the turn.

    \quad$\bullet$ \emph{Turn-end detection or prediction.} The distinction between detection~\cite{hara2019turn,lala2017attentive} and prediction~\cite{lala2019smooth,ekstedt2020turngpt} lies in that detection determines whether the agent should take a turn at the current moment, whereas prediction decides when the turn-taking should occur in the future.
    
    \quad$\bullet$ \emph{Overlap.} This mainly involves two situations. When the user and agent’s voices overlap, if the user intends to take the turn from the agent, this behavior is defined as an \textit{interruption}~\cite{khouzaimi2016reinforcement,marge2022spoken}. If the user has no intention of taking the turn, this behavior is considered \textit{backchannel}~\cite{hara2018prediction} or a listener response, such as "uh-huh," "right."

Through these concepts, we can better understand turn-taking behavior in duplex dialogues. In summary, our interactions with voice dialogue systems can be categorized as \textit{interruptions}, \textit{backchannels}, and \textit{normal turn exchanges}.

The earliest full-duplex systems used a simple Voice Activity Detection (VAD) component to model whether the user intended to interrupt. However, this approach is inadequate for handling backchannel interaction forms, leading to frequent interruptions and introducing considerable delays.

We can briefly categorize the exploration of interactions into cascaded systems and end-to-end systems based on duplex technology. Regardless of the system type, the critical core idea is that the system must continuously track external information in real-time, analyze it, and determine the model’s operational state accordingly. An interactive voice system must meet two requirements: 1) The ability to accept external information in real-time at any moment. 2) The ability to respond to this information accurately. This includes:
 
        \quad$\bullet$ \emph{Detecting User Interactions.} When the user tries to interject or provide new information, the system can recognize this intent and immediately stop its output to allow the user to speak.
        
        \quad$\bullet$ \emph{Backchanneling During User Speech.} While the user is speaking, the system can provide brief acknowledgments like "uh-huh" or "I see" to indicate active listening, which encourages the user to continue.
        
        \quad$\bullet$ \emph{Quickly Responding After User Completion.} When the user finishes speaking, the system can promptly recognize this cue and respond without unnecessary delays, maintaining a smooth conversational flow.
        
        \quad$\bullet$ \emph{Handling Pauses in User Speech.} When the user briefly pauses, the system can interpret this as a moment of thought rather than an invitation to respond, thus avoiding premature interruptions and preserving the natural flow.
        
        \quad$\bullet$ \emph{Interrupting the User When Necessary.} In situations where the system detects critical information, it can choose to interrupt the user to provide immediate feedback. For example, if the user is speaking but the system needs to alert them to an error, it can intervene in real-time to ensure effective communication.

\textbf{Cascaded Systems.}
To enable interactive functionality, cascaded spoken dialogue models typically require explicit modeling of dialogue turns. As the core, the large language model needs effective context and turn management. Next, we introduce several representative works on interaction in cascaded systems.

\quad$\bullet$ \emph{Duplex Conversation.} In \cite{lin2022duplex}, three core modules are proposed to achieve smooth full-duplex dialogue: user state detection, response signal selection, and interruption detection. The user state detection module not only focuses on traditional turn-end detection but also identifies whether the user intends to switch turns, continue speaking, or hesitates during their speech. To achieve this, the system uses a multimodal model, taking audio and text as inputs, and incorporates features such as speech rhythm, pitch, and pauses for more accurate assessment of the user’s state, determining whether to respond immediately or wait longer.
The response signal selection module inserts small backchannel cues (such as "uh-huh" or "right") at appropriate times to simulate natural human conversation. By analyzing a large volume of real dialogues, this module extracts and trains suitable response signals for various conversation scenarios. Using multi-label classification, the system selects the optimal response for each dialogue context, significantly reducing user waiting time and enhancing conversation flow.
The interruption detection module flexibly responds to user interruptions. Unlike traditional rule-based detection methods, this system builds an end-to-end detection model with multimodal input (audio and text) that not only identifies genuine user interruptions but also avoids misinterpreting background noise or unintended voice signals as interruptions.

\quad$\bullet$ \emph{Outbound Agent System.} \cite{jin2021duplex} proposed a full-duplex dialogue scheme for outbound systems, focusing on the issues of conversational fluidity and timing of interaction in speech dialogue. This scheme uses semantic analysis to determine whether the user truly intends to interrupt the system and can handle disjointed expressions when users mention named entities. The core of this system is a full-duplex interaction finite-state machine (FSM), which retrieves text snippets from ASR results every 300 milliseconds to decide whether to interrupt. Through continuous semantic analysis of user speech, the interruption model identifies meaningful user interruptions and avoids frequent interruptions caused by brief, meaningless responses (like "uh-huh"). The model employs a pre-trained BERT-based text classifier and utilizes streaming input, ensuring that the system can process and analyze user speech in real-time as it is received. Additionally, the system includes a Discontinuous Expression module to handle user pauses when mentioning named entities. Specifically, when users hesitate over entities (such as numbers, locations, or company names), VAD may erroneously detect turn-end. 

The advent of Large Language Models  has significantly advanced generative AI development. Models like ChatGPT demonstrate strong capabilities in semantic understanding and logical reasoning, offering a simplified method to integrate various dialogue components into a unified framework, which may simplify SDS construction. GPT-4o represents a milestone for dialogue systems, showcasing a nearly human-like conversational voice model. Its flexible interaction style and interruption mechanisms make human-computer interaction more natural and fluid. However, as a commercial model, its training data and implementation details remain proprietary, making replication challenging.

 \quad$\bullet$ \emph{Full-duplex LLM.} \cite{wang2024full} proposed a full-duplex spoken dialogue models based on LLMs, enabling simultaneous reception and transmission of voice signals through a perception module, an action module, and a neural finite-state machine (FSM). The perception module uses a streaming ASR model, capturing and processing user speech in real-time with 640-millisecond intervals per time step, converting it into token inputs for the LLM. The action module, utilizing a streaming TTS model, instantly converts the LLM-generated text into audio output and can pause or resume playback as needed, ensuring the system can generate audio while receiving user input. At the core is the neural FSM, allowing the LLM to switch between "speaking" and "listening" states. Controlled by FSM signals, the system can dynamically decide to continue speaking, listen, or interrupt based on the dialogue context. Experimental results show that Wang et al.'s full-duplex streaming system reduces response latency by threefold, achieves a response time within 500 milliseconds in over 50\% of dialogues, and handles user interruptions at a rate of 96.7\%, with an interruption accuracy of 54.7\%.

\quad$\bullet$ \emph{VITA.} VITA is an open-source multimodal large language model which aimed at enhancing multimodal interaction experiences. VITA can process multiple modalities, such as video, image, text, and audio, and achieves fluid human-computer interaction through a new duplex architecture involving two simultaneously operating models: one for generating responses to user queries, and another for continuously monitoring environmental inputs. When a new user query is detected, the generation model pauses, and the monitoring model processes the new query and generates an updated response. This setup enables VITA to support audio interruption, allowing users to ask new questions during system generation, with the system immediately pausing the current response to handle new input. VITA’s perception abilities are achieved through multimodal alignment and instruction fine-tuning, enabling it to switch automatically between different inputs. Additionally, VITA employs state tokens to distinguish user input types, such as query audio, background noise, and text input, facilitating wake-free interaction. VITA's enhanced listening module prevents unnecessary user feedback from interrupting system responses, improving robustness.

\quad$\bullet$ \emph{CleanS2S.}\cite{CleanS2S}
This model employs a structured pipeline to enable responsive and flexible interactions in a spoken dialogue setting. Designed to facilitate seamless turn-taking and interruption handling, the model consists of several interconnected modules working in a coordinated sequence to optimize user experience.
Starting with user input, the system uses a Voice Activity Detection (VAD) module to continuously monitor for incoming audio signals. As soon as a user starts speaking, VAD captures the input and immediately initiates processing by sending the audio data to the Automatic Speech Recognition (ASR) module. This quick detection and response setup allows the system to react to user input without delay.
Once ASR transcribes the audio into text, the transcription is passed to the Large Language Model (LLM), which generates a relevant response based on the user’s query. Meanwhile, the model is designed to be interruption-aware. During response generation, if VAD detects a new user input (indicating an interruption or a follow-up query), the system can promptly adjust its processing flow. In this case, the LLM temporarily pauses its current task, allowing ASR to transcribe the new input, which the LLM then uses to generate an updated response. This interruption capability is achieved through the model’s layered processing design, allowing for adaptive turn-taking that feels natural and responsive.
The Text-to-Speech (TTS) module then converts the generated text response into audio, which is transmitted to the user via WebSocket. To further support interruption handling, TTS breaks down lengthy responses into smaller audio segments that are sent progressively. This segmentation allows the system to stop audio output instantly if an interruption occurs, switching to the new input without delay. Each segment is prepared and sent only after a brief VAD check, ensuring that the system is ready to pause and handle new input at any time.
This interconnected processing chain—VAD detecting input, ASR transcribing, LLM generating responses, and TTS outputting segmented audio—creates a duplex interaction framework that balances response generation and user-driven interruptions. By seamlessly coordinating these components, the model provides a fluid, real-time dialogue experience that adapts to user interactions dynamically.

\textbf{End-to-End Systems.}
In contrast, end-to-end spoken dialogue models do not require explicit modeling of dialogue turns; instead, they learn interaction modeling directly from training data. Next, we introduce several representative works on interaction in end-to-end systems.

\quad$\bullet$ \emph{dGSLM.} In end-to-end systems, the introduction of the dGSLM model marks a significant milestone in full-duplex technology development. Within the dGSLM framework, duplex technology is effectively implemented. This model demonstrates how to capture complex interactions within dialogues directly from raw audio data through generative spoken dialogue modeling, without relying on text. The core innovation of dGSLM is the dual-tower Transformer architecture, called the Dialogue Transformer Language Model (DLM), which uses a cross-attention mechanism to enable the system to process two parallel audio channels simultaneously. Through this architecture, the model not only independently generates speech for each channel but also shares information between channels using cross-attention, effectively modeling silences and interaction events. It leverages the HuBERT encoder and HiFi-GAN decoder, combined with the dual-tower DLM, and is trained on 2,000 hours of dual-channel telephone conversation audio (Fisher dataset), where each speaker in a conversation is allocated an independent audio track. The dGSLM model transforms the audio on both channels into discrete tokens using HuBERT, and the DLM model autoregressively predicts the next audio token and its duration. Finally, the HiFi-GAN\cite{kong2020hifi} decoder reconstructs the audio for both channels. This approach differs significantly from traditional text-dependent spoken dialogue models, with a particular emphasis on modeling turn-taking and backchanneling capabilities. This capability gives dGSLM a notable advantage in duplex voice interaction, better mimicking the natural dynamics of human conversation. Through its duplex model design, dGSLM represents an essential step forward in interactive capabilities and provides a foundation for further advancements.

\quad$\bullet$ \emph{Moshi.} As a novel full-duplex architecture, Moshi incorporates a rich array of design concepts. Unlike dGSLM, Moshi does not abandon the language model’s ability in text dialogue. Moshi’s architecture is based on the Helium language model and Mimi neural audio codec, both trained from scratch. Helium, as a large pre-trained text language model, provides strong reasoning capabilities, while Mimi handles audio signal encoding and decoding. To achieve real-time interaction, Moshi is designed as a multi-stream architecture, simultaneously processing "user" and "moshi" audio streams without explicitly modeling speaker turns. Moshi also introduces the "Inner Monologue" method within the "moshi" audio stream, a process that jointly models text and audio tokens during training and inference. This approach allows the model to fully utilize textual knowledge while maintaining speech-to-speech system characteristics, significantly enhancing generation quality. Mimi, a neural audio codec integrating semantic and acoustic information through residual vector quantization and knowledge distillation, captures high-quality user input audio and Moshi’s output voice efficiently. To jointly model Moshi and user audio streams alongside Moshi’s text tokens, Depth Transformer with streaming inference capabilities is employed. The Mimi encoder and decoder combine convolutional and Transformer layers, with causal convolutions, allowing for streaming operation. Moshi is pre-trained on unsupervised audio data to handle speech scenarios and then fine-tuned on the Fisher dataset to address overlapping speech and interruptions. Finally, the system is further optimized on a custom instruction-tuning dataset, ensuring robust performance across various interactive scenarios. Experimental results show that Moshi excels in speech modeling and spoken QA tasks, especially in latency, achieving a theoretical latency of 160 milliseconds and 200 milliseconds in practice, significantly lower than the typical 230 milliseconds in natural conversation, enhancing real-time interaction and conversation flow.
 
 \quad$\bullet$ \emph{Parrot.} Parrot~\cite{meng2024sd} model incorporates multiple features specifically designed to enhance interaction in spoken dialogue. It uses a dual-channel audio setup, where each channel represents a different speaker. This configuration allows Parrot to manage both sides of a conversation independently, facilitating real-time turn-taking. By distinguishing between the user’s input and the system’s response on separate channels, the model can listen and respond in parallel, creating a more natural conversational flow.
To handle simultaneous speaker inputs effectively, Parrot employs a "next-token-pair prediction" mechanism, allowing it to predict tokens for both channels in a coordinated sequence. This approach helps the model manage conversational dynamics such as overlapping speech and smooth transitions between turns, adjusting response timing based on the user’s input.
During inference, Parrot supports streaming input, enabling continuous processing of user audio on one channel while generating responses on the other. This streaming capability allows the model to respond to live spoken input in real-time, handling turn-taking, pauses, and interruptions dynamically. Unlike cascaded systems that rely on intermediate text conversions, Parrot processes audio directly, reducing latency and allowing immediate responses to spoken input.
These interaction-focused design choices make Parrot highly responsive, enabling it to manage turn-taking naturally, respond to interruptions, and handle overlapping speech,

\quad$\bullet$ \emph{Mini-Omni2.} Mini-Omni2 is an open-source multimodal large language model aimed at simulating the multimodal capabilities of GPT-4o in vision, hearing, and text, supporting real-time full-duplex interaction. Mini-Omni2 combines visual and audio encoders with a language model to enable simultaneous input and output of images, audio, and text. The model incorporates an interrupt mechanism based on instruction design for more flexible user interactions. This system uses a delayed parallel generation algorithm, allowing the model to generate text and audio responses simultaneously, greatly improving conversational real-time capabilities and response speed. To achieve full-duplex interaction, Mini-Omni2 introduces an interrupt mechanism based on a limited instruction approach, trained on a specially constructed dataset with specific irq (interrupt) and n-irq (non-interrupt) state markers for model optimization. For training Mini-Omni2’s interruption functionality, the researchers used noisy speech data synthesized with specific command phrases (such as "Stop Omni") in various voices and tones to simulate scenarios where users might issue interrupt commands. The dataset also includes background noises, such as environmental sounds, music, and other dialogues, enhancing the model’s robustness in complex environments. During training, Mini-Omni2 controls output flow through irq and n-irq state markers, generating these markers in real-time to determine whether to continue output. In this way, the model can immediately halt generation based on user instructions and switch to "listening" mode in real-time dialogue. The training data consists of long audio streams from which the model extracts and encodes user commands like "Stop Omni." Researchers inserted interrupt commands at various time points, marking data after the insertion point as irq (interrupt) and data before as n-irq (non-interrupt). This labeling method ensures that the model learns to accurately identify interrupt commands in complex audio inputs and respond appropriately.

\quad$\bullet$ \emph{SyncLLM.} SyncLLM achieves full-duplex dialogue and interruption capabilities through multi-stream interleaving and chunk processing. SyncLLM divides the conversation's audio stream into fixed-sized chunks, each corresponding to a specific time interval. The model alternates between generating user and system speech segments within each time step (chunk), ensuring real-time system responses while processing user speech input. To maintain temporal synchronization with the user, SyncLLM predicts the user’s speech at each time step before generating each system chunk, using it as context to infer the system’s next response. This mechanism enables the system to keep pace with the conversation even with network latency. The chunk method allows SyncLLM to handle both user and system audio streams simultaneously, supporting complex dialogue features like speech overlap, interruption, and real-time feedback. Additionally, by using de-duplicated speech token sequences and periodic synchronization markers, the model efficiently performs chunk-level real-time inference, making conversation more fluid and natural.

\quad$\bullet$ \emph{OmniFlatten.} Similar to SyncLLM, the OmniFlatten model achieves full-duplex and interruption functionality primarily through multi-stream data processing and progressive training. To enable full-duplex dialogue, the model adopts a multi-stream architecture that interleaves the user’s speech stream with the assistant’s speech and text streams into a single sequence for training, simplifying multimodal modeling and enhancing real-time capability. The model first aligns the text language model with modality through multitask supervised fine-tuning, enabling it to understand and generate both speech and text, ensuring basic capability for handling speech and text simultaneously. Through a progressive training process, OmniFlatten attains full-duplex capability in three stages: initial training for half-duplex dialogue, then removing the user’s text stream to support real-time prediction with multi-stream data, and finally removing the assistant’s text stream to enable pure speech stream generation. These steps reduce reliance on text and decrease latency, allowing the system to generate voice responses while receiving user speech input. By using a block-by-block generation strategy, OmniFlatten divides the input and output speech sequences into fixed-size blocks, processing each segment in turn. This effectively implements streaming processing, ensuring low latency and high responsiveness in full-duplex dialogue, thereby providing a more natural response to user interruptions.

\quad$\bullet$ \emph{Freeze-Omni.} To support duplex dialogue, Freeze-Omni~\cite{xiong2024freeze} uses a chunk-level state prediction mechanism for natural turn-taking. When the user begins speaking, a voice activity detection module identifies the audio input, prompting the model to process the audio chunk by chunk. After processing each chunk, the model's classification layer predicts the conversation state to determine the next action. There are three possible states: State 0, where the model continues listening for more input, assuming the user hasn’t completed their turn; State 1, where the model interrupts to provide an immediate response if a quick acknowledgment or feedback is needed; and State 2, where the model has completed processing the current user input and is ready to generate and output a response, thus transitioning smoothly into the response phase without further listening. This chunk-wise state prediction enables the model to decide effectively when to respond and when to continue listening, enhancing its ability to handle natural conversational cues and support interactive dialogue.

\subsubsection{Discussions about streaming and interaction}

Significant progress has been made in dialogues models, particularly in real-time interaction and semantic understanding, with notable achievements in streaming processing and full-duplex interaction. Current systems exhibit strong technical capabilities in reducing response latency, enhancing interruption handling, and improving the naturalness of conversation. However, existing spoken dialogues models still lack a unified system that can handle all forms of interaction seamlessly.
Future research could explore new frameworks to better manage both user interruptions and the system’s ability to interrupt users, making interactions more natural. Additionally, standardized benchmarks for evaluating interaction capabilities remain underdeveloped. A unified evaluation benchmark would provide a consistent method for assessing and comparing the performance of different models, thereby advancing the development of more intelligent and responsive interaction systems.

\section{Training Resources and Evaluation}

\subsection{Training resources}

Training a spoken dialogue system is a complex, multi-stage process, with each stage relying on specific datasets to achieve distinct training objectives and enhance system performance. This section provides an in-depth analysis of the training resources about the spoken dialogue models, showcasing the data collection and processing methods at each stage and illustrating how these elements contribute to the system's intelligence. It further reveals how key steps, from foundational architecture to fine-tuning, shape the intelligent development of dialogue systems.

To address the limitations of existing training spoken dialogue data and leverage the knowledge and reasoning capabilities of mature text-based models, many approaches involve \textit{Continue Training} on pre-trained text language models. This training paradigm encompasses nearly all data types required to build a spoken dialogue system. The following sections focus on analyzing data acquisition and processing methods under this training flow, covering the following core stages: \textit{Text Language Model Pre-training}, \textit{Post-Train for Audio Modal Adaption}, \textit{Post-Train for Dual-Stream Audio Processing}, \textit{Enhancing Conversational Abilities and Instruction Tuning}. We have listed commonly used datasets for training in Table \ref{traindataset}. However, current spoken dialogue models lack exploration in music and sound. To support future development in spoken dialogue systems, we provide a list of common music and sound datasets in the appendix \ref{music datasets} as a reference.

\subsubsection{Training resources about Text LLM Pre-training}

Text Language Model pre-training serves as the foundational stage for spoken dialogue models. Through unsupervised learning on large-scale text data, the model acquires knowledge of vocabulary, grammar, and contextual relationships, gaining essential knowledge and reasoning capabilities. Most spoken dialogue systems are built upon pre-existing open-source text language models (such as Llama~\cite{touvron2023llama}, Palm~\cite{anil2023palm}, etc). Although we does not delve into this stage in detail, it provides a solid foundation for the model’s natural language understanding and generation capabilities.

\subsubsection{Training resources about Post-Train for Audio Modal Alignment}

After establishing a text-based foundational model, the system possesses essential knowledge and reasoning abilities. In this stage, we introduce the audio modality, enabling the text language model to understand and generate speech while minimizing any potential loss of textual knowledge. This process is known as \textit{modal adaption} or \textit{modal alignment}. This multimodal structure incorporates an audio encoder with a codebook, helping the model recognize linguistic, emotional, and tonal information in speech. The audio decoder supports the generation of natural and fluent speech output, while audio signal embeddings and special token types (e.g., speaker-distinguishing tokens for Synchronous LLM, task-distinguishing tokens for OmniFlatten, and state tokens for VITA) are added to the vocabulary of the text language model.

The primary goal at this stage is to align information from different modalities into a unified space or representation, allowing the model to correlate and comprehend such information. Consequently, the model is often trained on cross-modal tasks such as TTS , ASR , and audio captioning. The datasets used include numerous paired audio and text samples to ensure effective conversion between modalities. Commonly used TTS and ASR datasets include Aishell-3~\cite{shi2020aishell}, LibriTTS~\cite{zen2019libritts}, TED-LIUM~\cite{rousseau2012ted}, VoxPopuli~\cite{wang2021voxpopuli}, Librispeech ~\cite{panayotov2015librispeech}, MLS~\cite{pratap2020mls}, Wenetspeech~\cite{zhang2022wenetspeech}, Gigaspeech~\cite{chen2021gigaspeech}, VCTK~\cite{veaux2013voice}, LJSpeech~\cite{ljspeech17}, Common Voice~\cite{ardila2019common}, and others. For audio captioning, Wavcaps~\cite{mei2024wavcaps} are frequently used. Some speech datasets require ASR model transcription to generate corresponding text.

In this phase, the emphasis is placed on capturing and generating audio features and aligning them with text in vector space, rather than focusing on dialogue functionality.Therefore, the data typically consists of single-channel audio, which can be used after resampling. Notably, in some works, it is essential to ensure word-level alignment between text tokens and audio tokens (e.g., Spirit-LM, Moshi, and OmniFlatten), achievable through tools like the Whisper-timestamped package or other alignment tool. In Moshi, to prevent catastrophic forgetting, half of the training time is allocated to text data, highlighting the importance of balancing text and audio data during training.
\begin{table}[h]
\centering
\caption{Datasets used in the various training stages}
\label{traindataset}

\resizebox{1\linewidth}{!}{%
\begin{tabular}{llcccl}
\hline
\textbf{Stage} & \textbf{Task} & \textbf{Dataset} & \textbf{Size} & \textbf{URL} & \textbf{Modality} \\ \hline
\multirow{14}{*}{Modal Alignment} 
    & Mandarin ASR    & AISHELL-1\cite{bu2017aishell}      & 170 hrs   & \url{https://www.openslr.org/33/}       & Text, Speech \\
    & Mandarin ASR    & AISHELL-2\cite{du2018aishell}      & 1k hrs   & \url{https://github.com/kaldi-asr/kaldi/tree/master/egs/aishell2}       & Text, Speech \\
    & Mandarin TTS    & AISHELL-3\cite{shi2020aishell}      & 85 hrs, 88,035 utt., 218 spk.   & \url{https://www.aishelltech.com/aishell_3}       & Text, Speech \\
    & TTS             & LibriTTS\cite{zen2019libritts}       & 585 hrs                         & \url{https://www.openslr.org/60/}                 & Text, Speech \\
    & ASR             & TED-LIUM\cite{rousseau2012ted}       & 452 hrs                         & \url{https://lium.univ-lemans.fr/ted-lium3/}      & Text, Speech \\
    & ASR             & VoxPopuli\cite{wang2021voxpopuli}      & 1.8k hrs                        & \url{https://github.com/facebookresearch/voxpopuli} & Text, Speech \\
    & ASR             & Librispeech\cite{panayotov2015librispeech}    & 1,000 hrs                       & \url{https://www.openslr.org/12}                  & Text, Speech \\
    & ASR             & MLS\cite{pratap2020mls}            & 44.5k hrs                       & \url{https://www.openslr.org/}                    & Text, Speech \\
    & TTS             & Wenetspeech\cite{zhang2022wenetspeech}    & 22.4k hrs                       & \url{https://wenet.org.cn/WenetSpeech/}           & Text, Speech \\
    & ASR             & Gigaspeech\cite{chen2021gigaspeech}     & 40k hrs                         & \url{https://github.com/SpeechColab/GigaSpeech}   & Text, Speech \\
    & ASR             & VCTK\cite{veaux2013voice}           & 300 hrs                         & \url{https://paperswithcode.com/dataset/voice-bank-demand} & Text, Speech \\
    & TTS             & LJSpeech\cite{ljspeech17}       & 24 hrs                          & \url{https://keithito.com/LJ-Speech-Dataset/}     & Text, Speech \\
    & ASR             & Common Voice\cite{ardila2019common}   & 2,500 hrs                       & \url{https://commonvoice.mozilla.org/zh-CN}       & Text, Speech \\
    & Audio Caption   & Wavcaps\cite{mei2024wavcaps}        & 400k clips                      & \url{https://github.com/XinhaoMei/WavCaps}        & Text, Speech \\
    & ASR             & LibriLight\cite{kahn2020libri}     & 60k hrs                         & \url{https://github.com/facebookresearch/libri-light} & Text, Speech \\ 
    & ASR             & PeopleSpeech\cite{galvez2021people}     & 30k hrs                         & \url{https://huggingface.co/datasets/MLCommons/peoples_speech} & Text, Speech \\ 
    & Mandarin ASR             & KeSpeech\cite{tang2021kespeech}     & 1,542 hrs                         & \url{https://github.com/KeSpeech/KeSpeech} & Text, Speech \\
    & TTS            & Emilia\cite{he2024emilia}     & 	101k hrs                        & \url{https://huggingface.co/datasets/amphion/Emilia-Dataset} & Text, Speech \\ 
    \hline

\multirow{7}{*}{Dual-Stream Processing} 
    & Instruction     & Alpaca\cite{maeng2017alpaca}         & 52,000 items                    & \url{https://huggingface.co/datasets/tatsu-lab/alpaca} & Text + TTS \\
    & Instruction     & Moss           & -                               & \url{https://huggingface.co/fnlp/moss-moon-003-sft} & Text + TTS \\
    & Instruction     & BelleCN        & -                               & \url{https://github.com/LianjiaTech/BELLE/tree/main} & Text + TTS \\
    & Dialogue        & UltraChat\cite{ding2023enhancing}      & 1.5 million                     & \url{https://github.com/thunlp/UltraChat}          & Text + TTS \\
    & Instruction     & Open-Orca\cite{OpenOrca}      & -                               & \url{https://huggingface.co/datasets/Open-Orca/OpenOrca} & Text + TTS \\
    & Noise          & DNS~\cite{reddy2001interspeech} & 2425 hrs                       & \url{https://github.com/microsoft/DNS-Challenge} & Noise  data \\
    & Noise           & MUSAN \cite{snyder2015musan}         & -                               & \url{https://www.openslr.org/17/}                  & Noise data\\ \hline

\multirow{4}{*}{Conversation Fine-Tune} 
    & Dialogue        & Fisher         & 964 hrs                         & \url{https://catalog.ldc.upenn.edu/LDC2004T19}     & Text, Speech \\
    & Dialogue        & GPT-Talker\cite{liu2024generative}     & -                               & \url{https://github.com/AI-S2-Lab/GPT-Talker}      & Text, Speech \\
    & Instruction     & INSTRUCTS2S-200K & 200k items                    & \url{https://github.com/ictnlp/LLaMA-Omni}         & Text + TTS \\
    & Instruction     & Open Hermes    & 900k items                      & \url{https://ollama.com/library/openhermes}        & Text + TTS \\ \hline

\end{tabular}%
}
\end{table}

\subsubsection{Training resources about Post-Train for Dual-Stream Dialogue Processing}

To ensure that the model possesses the ability to “listen while speaking”. Most research such as Moshi~\cite{defossez2024moshi} and OmniFlatten~\cite{zhang2024omniflatten} has implemented a dual audio-stream model: one audio stream generates model output, while the other captures user audio. The objective of this training phase is to enable the model’s dual-stream processing without requiring complex human-computer interaction modeling. Consequently, text dialogue data can be converted to speech and processed into dual-track audio format. However, text dialogue data typically contains content unsuitable for TTS conversion to speech (such as code, formulas, URLs) or long, formal dialogue passages that do not align with spoken language, as real dialogue is often more concise. Therefore, when synthesizing from text dialogue data, it is necessary to preprocess the text data. High-quality, open-source text dialogue data is first collected, including datasets like Alpaca~\cite{maeng2017alpaca}, Moss, BelleCN, ultraChat~\cite{ding2023enhancing}, and Open-Orca~\cite{OpenOrca}. To ensure suitability for speech synthesis (TTS), heuristic rules are applied to filter out samples with high proportions of non-text elements (such as code, mathematical expressions), samples exceeding 200 words, and samples containing rare symbols.

After filtering the text, TTS models~\cite{du2024cosyvoice} are used to synthesize speech for each turn in the dialogues. For consistent voice effects, the model audio stream maintains a uniform voice, while the user audio stream is sampled with varied voices to enhance the model's robustness. The synthesized dialogue audio is arranged using simulation strategies to achieve natural timing, such as turn-taking, well-timed interruptions, and pauses to maintain fluency and naturalness. The final dialogue audio is organized in dual-channel format: the conversation begins with a user utterance, followed by alternating user and assistant turns. After each user turn, the assistant responds immediately; upon completion of the assistant’s turn, a sampled pause length is introduced to simulate the natural rhythm of alternating dialogue. To better simulate real scenarios, further data augmentation can be applied. For example, random gain adjustments can be applied to the user audio stream, and background noise randomly selected from datasets like MUSAN~\cite{snyder2015musan} and DNS~\cite{reddy2001interspeech} can be added to the user audio channel (OmniFlatten). To simulate echo effects from a user’s microphone, portions of the audio stream can be scaled down and added to the user’s audio stream with random delays between 100 to 500 milliseconds, along with reverberation-like enhancements, helping the model adapt to real-world environments.

\subsubsection{Training resources about Enhancing Conversational Abilities and Instruction Tuning}

While the foundational model has been established, there remains a gap between this and a complete dialogue system. The above model utilizes non-overlapping dialogue audio, where one party remains silent while the other speaks, failing to fully simulate real conversational dynamics. Some speech datasets, such as \textit{Generative Expressive Conversational Speech Synthesis}~\cite{liu2024generative} and \textit{Fisher}, contain dialogues from real-world settings, providing a basis for modeling interruptions and backchannels scenarios in voice dialogue systems.

Currently, there is no suitable dataset for real-world speech instructions. Most approaches use synthetic methods based on text instruction data to perform \textit{instruction tuning} in this stage. Common text instruction datasets include \textit{Open Hermes} and \textit{moss-002-sft-data}, though they face similar challenges as text dialogue data, such as unsuitability for TTS conversion and inconsistency with spoken language conventions. Following the synthetic processes provided by Moshi and Llama-Omni, this aims to generate instruction data in the format of (SpeechInstruction, TextInstruction, TextResponse, SpeechResponse).

The first method is synthetic generation from scratch. Contexts and summaries are first generated by sourcing high-quality text data from sources like Wikipedia and StackExchange, producing thematic paragraphs as the dialogue foundation, referred to as “context.” Based on these contexts, dialogue summaries are generated. Next, a specific prompt template guides the generation of complete dialogues, including context and requesting dialogues around the theme with roles as user and system. The model is prompted to exhibit knowledge on the topic and include interruptions (backchannels) and brief turn-taking, simulating the natural flow of conversation. To enhance dialogue diversity, additional instructions involving speech emotion and role-playing can be generated, requesting dialogues in specific tones or styles. Furthermore, dialogues containing spelling errors or misinformation are synthesized to train the system in handling scenarios where user clarification or repetition is required. Single-turn interactions on basic mathematics, grammar, and factual questions are also generated to ensure the system can handle simple factual tasks. Finally, scenarios involving ethical or NSFW requests are created to train the system in declining to answer under such conditions.

The second method involves filtering and refining existing text instruction datasets. Initially, open-source text language models paraphrase text instructions to match spoken language traits, adding fillers like “uh” and “um” to mimic natural speech tone, while converting numbers and symbols into spoken language to ensure the instructions are concise and conversational. Generated text responses are also optimized to meet TTS output requirements, removing lengthy expressions and complex grammatical structures to make content clear and concise for TTS output. After adjusting the instruction and response text, a TTS system converts the text to audio.

\subsection{Evaluation}
\label{eval}
Fair and comprehensive evaluation of spoken dialogue models presents a multifaceted challenge. On the one hand, the field of spoken dialogue still lacks publicly available test sets, comprehensive evaluation metrics, and established benchmarks. On the other hand, assessing the performance of spoken dialogue systems requires consideration from multiple perspectives. Basic aspects include the quality of generated speech, robustness, dialogue naturalness and accuracy, as well as response speed and generation time. Beyond these, more advanced evaluations are needed to assess multi-turn dialogue capabilities (such as long-form speech editing), interaction abilities, and the system's proficiency in audio and music understanding and generation. Given these requirements, and in line with the comprehensive expectations for spoken dialogue systems outlined in Section \ref{section21}, we will evaluate these systems from two angles: common evaluations and advanced evaluations. Specifically, we will assess eleven key factors: speech generation quality, text intelligence, speech intelligence, audio and music generation, audio and music understanding, multilingual capability, context learning, interaction capability, streaming latency, multimodal capability, and the safety of dialogue systems. Finally, we will list the current benchmarks and summarize the common conclusions derived from them.

\begin{table}[]
\renewcommand{\arraystretch}{1.4}
\caption{This table  provides a comprehensive overview of the different components used to evaluate dialogue systems, including various abilities, common tasks, representative benchmarks, and corresponding metrics. The abilities include Text Intelligence, Speech Quality, Audio Understanding and Generation, Music Understanding and Generation, Multilingual Capability, Context Learning, Interaction Capability, Multimodal Capability, Security, and Speech Intelligence. The table aligns these tasks with widely used benchmarks such as VoiceBench, SUPERB, AudioBench, AirBench, SpokenWOZ, SD-EVAL, SuperCLUE, and MMAU, highlighting the dimensions they assess. To ensure comprehensive evaluation some metrics are defined: \textbf{MT-Metrics}, which evaluate the quality of generated outputs using semantic and syntactic similarity; \textbf{Acc-Metrics}, which measure recognition performance using precision, recall, and F-score; \textbf{Subjective Metrics}, which assess creative and generative tasks like speech quality and audio generation. This structured framework provides a holistic view of benchmarks, tasks, and evaluation criteria for assessing diverse model capabilities.}

\resizebox{1\linewidth}{!}{%
\begin{tabular}{cccccccccccc}
\Xhline{1.5pt} 
\multirow{2}{*}{\textbf{Level}} &
  \multirow{2}{*}{\textbf{Ability}} &
  \multirow{2}{*}{\textbf{Task}} &
  \multicolumn{8}{c}{\textbf{Benchmark}} &
  \multirow{2}{*}{\textbf{Metric}} \\
 & 
   & 
   & 
  \multicolumn{1}{c|}{VoiceBench} & 
  \multicolumn{1}{c|}{SUPERB} & 
  \multicolumn{1}{c|}{AudioBench} & 
  \multicolumn{1}{c|}{AirBench} & 
  \multicolumn{1}{c|}{SpokenWOZ} & 
  \multicolumn{1}{c|}{SD-EVAL} & 
  \multicolumn{1}{c|}{SuperCLUE} &
  \multicolumn{1}{c|}{MMAU} &
  \\ \Xhline{1.5pt} 
\multirow{5}{*}{Basic}     & \multirow{3}{*}{Text Intelligence} & Reasoning             & \redcross & \redcross & \redcross & \redcross & \greencheck & \redcross & \greencheck &  \greencheck & Acc-Metrics      \\
                           &                                    & Instruction Following & \greencheck & \redcross & \greencheck &  \redcross & \greencheck & \redcross & \greencheck & \redcross & MT-Metrics       \\
                           &                                    & Conversational QA     & \greencheck & \redcross & \greencheck & \redcross & \greencheck & \redcross & \greencheck & \greencheck & MT-Metrics       \\ \cline{2-12}
                           & Speech Quality                     & MOS, WER Evaluation   & \redcross & \greencheck & \redcross &  \redcross & \redcross & \redcross & \redcross  & \redcross  & MOS, WER         \\ \cline{2-12}
                           & Streaming Latency                  & Real-Time Dialogue    & \redcross & \redcross & \redcross & \redcross & \redcross & \redcross & \greencheck  & \redcross & Real-Time Factor \\ \Xhline{1.2pt} 
\multirow{16}{*}{Advanced} & \multirow{5}{*}{Audio U\&G}        & Audio Classification  & \redcross & \greencheck & \greencheck& \greencheck & \redcross & \greencheck & \redcross & \greencheck & Acc-Metrics      \\
                           &                                    & Sound Event Detection & \redcross & \greencheck & \redcross & \greencheck & \redcross & \greencheck & \redcross & \greencheck & Acc-Metrics      \\
                           &                                    & Audio Captioning      & \redcross & \redcross & \greencheck & \greencheck & \redcross & \greencheck & \redcross &  \redcross & MT-Metrics       \\ 
                           &                                    & Audio-Motivated Creative Writing & \redcross & \redcross & \redcross  & \redcross & \redcross & \redcross & \redcross &  \redcross & Subjective Metrics \\ 
                           &                                    & Audio Generation   & \redcross & \redcross & \redcross & \redcross & \redcross & \redcross & \redcross  & \redcross & MOS, FD, IS, KL, FAD, CLAP Score \\ \cline{2-12}                            
                           & \multirow{3}{*}{Music U\&G}        & Music Captioning      & \redcross & \redcross & \redcross & \greencheck & \redcross & \redcross & \redcross & \redcross & MT-Metrics       \\
                           &                                    & Music Classification  & \redcross & \redcross & \redcross & \redcross  & \redcross & \redcross & \redcross & \greencheck & Acc-Metrics      \\
                           &                                    & Music Synthesis       & \redcross & \redcross & \redcross & \greencheck & \redcross & \redcross & \redcross & \redcross & MOS              \\ \cline{2-12}
                           & Multilingual Capability            & Speech Translation    & \redcross & \greencheck & \redcross  & \redcross & \redcross & \redcross & \redcross  & \redcross & MT-Metrics       \\ \cline{2-12}
                           & Context Learning                   & Context-Aware QA      & \redcross & \redcross & \redcross & \redcross & \redcross & \redcross & \greencheck & \redcross & MT-Metrics       \\ \cline{2-12} 
                           & Interaction Capability             & Interaction Events    & \redcross & \redcross & \redcross  & \redcross & \redcross & \redcross & \greencheck &  \redcross & Statistic-Method \\ \cline{2-12} 
                           & Multimodal Capability              & Multimodal QA         & \redcross & \redcross & \redcross & \redcross & \redcross & \redcross & \redcross & \redcross & MT-Metrics \\ \cline{2-12} 
                           & Security                           & Attack Events         & \greencheck & \redcross & \redcross & \redcross & \redcross & \redcross & \greencheck & \redcross & Attack Success Rate \\

                           \cline{2-12}
                            & \multirow{4}{*}{Speech Intelligence}          & Speaker Info & \greencheck & \greencheck & \greencheck & \greencheck & \redcross & \redcross & \redcross  & \redcross & Acc-Metrics \\
                            &                                    & Paralinguistic info Classification & \greencheck & \greencheck & \greencheck & \greencheck & \redcross & \greencheck & \redcross &  \greencheck & Acc-Metrics \\ 
                            &                                    & Conditioned response & \redcross & \redcross & \redcross & \greencheck & \redcross & \greencheck & \redcross &  \redcross & MT-Metrics \\
                            &                                    & Controllable Style Generation & \redcross & \redcross & \redcross & \redcross & \redcross & \redcross & \greencheck &  \redcross & MT-Metrics \\ 
                            \Xhline{1.5pt}
\end{tabular}}
\label{table:benchmark}
\end{table}

\subsubsection{Common Evaluation}
\label{eval_text}


\textbf{Text Intelligence.} As shown in Figure \ref{fig:main} (a), text intelligence refers to the fundamental understanding and generation capabilities of the spoken dialogue model. When evaluating text intelligence, the focus is solely on the semantic content generated by the model, without considering other aspects such as timbre, emotion, or style. In practical evaluations of this kind, some spoken dialogue models output only text~\cite{shu2023llasm, tang2023salmonn, chu2023qwen, chu2024qwen2, xue2023chat}, while others generate both text and speech~\cite{defossez2024moshi,xie2024mini,xie2024miniomni2opensourcegpt4ovision}, or only speech~\cite{zhang2024omniflatten}. Regardless of the output format, we are concerned only with the generated text or the transcribed text from the speech when evaluation the text intelligence in the spoken dialogue models. There are typically two categories of metrics and benchmarks used to assess text intelligence, MT-Metrics and Acc-Metrics. The details are outlined as follows:

\quad$\bullet$ \emph{ACC-Metrics.} A common approach to evaluating text intelligence is to use benchmarks typically~\cite{talmor2018commonsenseqa, liang2022holistic, zellers2019hellaswag, clark2018think, sakaguchi2021winogrande, chen2021evaluating, zhong2023agieval, mishra2021cross, wang2022super, feng2022mmdialog} employed for large language models, such as the classic MMLU~\cite{hendrycks2020measuring} and GSM-8K \cite{cobbe2021training}. These benchmarks often include complex multiple-choice questions, which assess the model's reasoning abilities through Acc-Metrics. Acc-Metrics refers to metrics that measure recognition accuracy, such as accuracy, F-score, and Mean Average Precision (mAP). It is noteworthy that these benchmarks often evaluate the text-based intelligence of spoken dialogue models from various perspectives. For example, MMLU \cite{hendrycks2020measuring} and GSM-8K \cite{cobbe2021training} are more focused on LLM's core knowledge, Flan \cite{longpre2023flan, wei2021finetuned} and Self-instruct \cite{wang2022self} are more focused on LLM's instruction following capability, CoQA \cite{reddy2019coqa} and OpenAssistant \cite{kopf2024openassistant} are more focused on LLM's conversational capability. These benchmarks often contain questions and corresponding answers. Most of these questions are close-ended questions with short answers, so that they can have good generalization ability, any model that can generate text answers can be evaluated with these benchmarks and accuracy and F-Score can be easily adopted as the evaluation metrics. 

\quad$\bullet$ \emph{MT-Metrics.}  With the development of the LLMs, LLMs can follow instructions to accomplish many complex problems, so the scope of the evaluation was further expanded to include open-ended questions. These open-ended questions often lack standard answers, therefore it's difficult to measure them by common ACC-Metrics. A common approach is to measure the grammatical similarity between generated and reference utterances using the metrics used to measure grammatical similarity in mechanical translation (e.g. BLEU \cite{papineni2002bleu}, METEOR \cite{banerjee2005meteor}, ROUGE \cite{lin2004rouge}). We collectively refer to these evaluation metrics as \textbf{MT-Metrics}. However, these metrics have certain limitations since one meaning has many different ways to convey. So there are some metrics like BertScore \cite{zhang2019bertscore} focus on evaluating the semantic similarity between two sentences. And there are also been some methods utilizing LLM to judge the effectiveness of the responses which focusing on human preference \cite{zheng2023judging, liu2023g}. The results of these large model-based especially GPT4o-based ratings of evaluation metrics demonstrated a high degree of correlation with human.


\textbf{Speech Quality.} Speech quality is one of the fundamental aspects for evaluating the performance of spoken dialogue systems, as it is closely tied to the experience of users. There are two common dimensions for assessing speech quality: the clarity and naturalness (expressiveness and prosody) of the generated audio, and the robustness of the generated speech, such as the presence of missing or extra words. The former is typically evaluated by using subjective MOS (Mean Opinion Score) ratings, while the latter is commonly assessed by using WER (Word Error Rate) or CER (Character Error Rate) metrics.

\textbf{Streaming Latency.} 
In addition to evaluating the quality of text understanding and generated speech, the speed at which a spoken dialogue system generates speech responses is also crucial. This necessitates the ability to stream both the comprehension and generation of speech in real time, achieving an effect of generating speech while speaking~\cite{zhang2024intrinsicvoice,defossez2024moshi,fang2024llama}. To assess the streaming performance of a model, one typically measures the time taken to generate the first token of speech (i.e., the waiting time after the user finishes speaking) and calculates the overall Real-Time Factor (RTF) of the spoken dialogue model's response. The RTF value is obtained by dividing the total duration of the speech segment generated by the model by the time taken by the model to generate that response.

\subsubsection{Advanced Evaluation}
\label{eval_SI}

\textbf{Speech Intelligence.} 
Evaluating the speech intelligence of spoken dialogue systems is one of the key aspects. The definition of speech intelligence in spoken dialogue systems is discussed in detail in Section \ref{section212}. Given that speech intelligence encompasses a wide range of application scenarios, we address the evaluation separately for the understanding and generation components during the assessment.

\quad$\bullet$ \emph{Understanding.} Ordinary cascaded spoken dialog models based on ASR getting text input will loss many paralinguistic information like speaking style, accent, emotion, etc. Thus many spoken dialogue models \cite{xue2023chat, lin2024paralinguistics, lin2024advancing} devoted into helping dialog models understand the paralinguistic information. Evaluating this capability can start from two aspects: a) the accuracy of the paralinguistic information's understanding, b) the ability of \textbf{automatically} generating appropriate and coherent content responses and acoustic information based on the varying acoustic input. \textbf{For the former}, since the classes of the paralinguistic information are always limited, for example, sentiments are generally categorized as neural, negative, positive. So the researchers always use Accuracy or F-Score to evaluate the models' paralinguistic information understanding capability. Recently, there are many studies \cite{goel2024audio, busso2008iemocap, poria2018meld, xue2023chat, lin2024advancing, firdaus2020meisd, busso2016msp} available for researchers to use in identifying speech emotions in the dialogue scenes. In addition to recognizing speech emotions, recent benchmarks \cite{ao2024sd, yang2024air} has also begun to investigate the influence of speaker age, accent, and other factors on the evaluation of spoken dialogue models. \textbf{For the latter}, recent work~\cite{xue2023chat} has increasingly focused on the possibility of generating appropriate content responses based on acoustic information from the input. The current evaluation methods usually transcript the output audio into text through Automatic Speech Recognition and then evaluate the relevance between generated content and the reference content in the internal dataset. Evaluations are usually conducted in text, so commonly used evaluation metrics are as the same as in the section \ref{eval_text}, like BLEU and METEOR, which are used to measure the similarity between two sentences. Currently, there is limited research exploring whether spoken dialogue models can autonomously generate appropriate acoustic responses based on varying acoustic information, making it a promising area for future investigation.


\quad$\bullet$ \emph{Generation.} In the generation component, evaluating the speech intelligence of spoken dialogue systems primarily focuses on controllability, i.e., the ability of the dialogue model to respond in a user-specified style and timbre in the zero-shot scenarios. There are various dimensions to assess style, such as pitch, speech rate, energy, emotion, and accent, among others. ACC-metrics can be used to evaluate whether the spoken dialogue model can generate speech in the desired style. Additionally, the evaluation of voice cloning capabilities within the model can borrow metrics from the zero-shot TTS domain~\cite{wang2023neural,shen2023naturalspeech,ji2024mobilespeech,wang2024ham}, using speaker similarity indices~\cite{chen2022wavlm}. Currently, there are few models that explore the generation of speech intelligence in spoken dialogue systems, and this area warrants further refinement and exploration in future work.

\textbf{Audio Understanding and Generation.}
In real-world scenarios, the broader definition of speech modality encompasses not only clear human speech but also a wide range of natural sounds such as dog barking and bird chirping, all of which can be considered forms of audio. Evaluating the ability of spoken dialogue models to understand and generate such audio is a critical aspect of assessing the model’s performance.

\quad$\bullet$ \emph{Audio Understanding.} On the audio comprehension side, various sub-tasks are commonly employed to measure a system's capacity to understand audio, including tasks such as Audio Captioning (AudioCap) \cite{kim2019audiocaps}, Sound Event Detection (SED) \cite{mesaros2021sound}, audio classification, and audio-motivated creative writing~\cite{chu2023qwen}, among others. The core of these tasks lies in evaluating the model’s ability to process and interpret the complex acoustic information embedded within the audio. For tasks like audio classification and SED, which involve fixed outputs, evaluation is relatively straightforward, typically using objective metrics such as accuracy or Mean Average Precision (mAP). However, for the AudioCap task, the problem is generally open-ended, meaning there are no fixed answers. As a result, existing evaluation methods are primarily based on measuring the similarity between the generated text and the reference text, using traditional metrics such as BLEU \cite{papineni2002bleu} and METEOR \cite{banerjee2005meteor}, or newer evaluation approaches involving large language models such as GPT-4o \cite{zheng2023judging}. In the case of audio-motivated creative writing, where the objective is to generate inventive descriptions from a given audio input, evaluation typically relies on subjective measures, given the divergent nature of the creative process involved.

\quad$\bullet$ \emph{Audio Generation.} Additionally, on the audio generation side, producing high-quality audio should be considered an advanced capability for a conversational spoken dialogue model. However, as most current spoken dialogue systems lack the ability to generate audio, this remains an area for further exploration in the future end-to-end spoken dialogue systems. The evaluation of generated audio can draw from methods used in the text-to-audio domain~\cite{huang2023make2,huang2023make}. Typically, such evaluations focus on the quality of the generated audio itself, using metrics such as Mean Opinion Score (MOS) and the similarity between generated and target audio. Objective evaluation metrics for audio similarity often include Fréchet Distance (FD), Inception Score (IS), Kullback-Leibler (KL) divergence, Fréchet Audio Distance (FAD), and CLAP score. Specifically, Fréchet Audio Distance (FAD) \cite{kilgour2018fr} is adapted from the Fréchet Inception Distance (FID) to the audio domain and serves as a reference-free perceptual metric that quantifies the distance between the generated and ground-truth audio distributions. The Inception Score (IS) is an effective metric that evaluates both the quality and diversity of generated audio. KL divergence is computed at the paired sample level between generated and ground-truth audio, based on the label distribution and averaged to produce a final result. Fréchet Distance (FD) evaluates the similarity between the generated and ground-truth audio distributions. FD, KL, and IS are built upon the PANNs model \cite{kong2020panns}, which takes mel-spectrograms as input. In contrast, FAD uses VGGish \cite{hershey2017cnn} as an audio classifier, processing raw audio waveforms as input. The CLAP score, adapted from the CLIP score \cite{hessel2021clipscore}, is a reference-free metric used to assess audio-text alignment and strongly correlates with human perception.

\textbf{Music Understanding and Generation.} In advanced spoken dialogue models, the evaluation of music modality understanding and generation follows a methodology similar to that used for audio modality. Unlike Audio Understanding, which only requires a general description of the events that occur in the audio, Music Understanding requires appreciating the style and genre of music, understanding its keys, themes, and other rich information. For classification, emotion recognition tasks in music, common metrics such as accuracy can be used. For music captioning task, MusicCaps \cite{agostinelli2023musiclm} offers a general dataset for evaluating a model's music understanding capability. For music analysis, Nsynth \cite{engel2017neural} provides rich note data information. In terms of evaluation for music generation, subjective Mean Opinion Score (MOS) assessments or measures of similarity between generated music and target music are commonly used.

\textbf{Multilingual Capability.} The ability to speak multiple languages is also required for a spoken dialogue model, but most current models \cite{gong2023listen,lin2024advancing, lin2024paralinguistics, shu2023llasm, wang2023blsp,   xue2023chat} only focus on English and Chinese. A naive idea is to directly evaluate spoken dialogue models' capability in speech-to-speech or speech-to-text translation tasks \cite{ jia2022cvss,wang2020covost1}. These evaluations can be done with common machine learning metrics like BLEU \cite{papineni2002bleu} or BertScore \cite{zhang2019bertscore}. However, evaluating the capability of translation is insufficient to measure the model's multilingual conversational ability, and further exploration is still needed in this area of evaluation. Explicitly requiring a spoken dialogue model to perform speech translation is not a typical use case in conversational scenarios. In most cases, when a user asks a question in a different language or with a distinct accent, the model is expected to automatically respond in the same language that the user is using. In this context, it seems more reasonable to evaluate the accuracy of the model’s generated speech in terms of language identification, combined with subjective human assessments, as a more intuitive and appropriate evaluation method.

\textbf{Context Learning.} The context learning capability is crucial for maintaining the coherence of an entire conversation. Similar to a memory function, the challenge lies in how to preserve this capability when relying solely on speech. Typically, the evaluation of a spoken dialogue model's context learning ability depends on specific long-duration dialogue test sets, after which standard MT-Metrics or Acc-Metrics used in text intelligence evaluations can be applied. For instance, a model's context learning capability can be assessed by evaluating its QA performance based on the given context \cite{lipping2022clotho}. However, it is important to note the relevance of editing scenarios in long-duration spoken dialogues. In real spoken dialogue scenarios, the users will modify some certain key information, the model needs to promptly understand and respond accordingly, e.g., the users offer wrong information for solving a problem and modify the condition in the next dialog. So how to evaluate the model's online understanding ability is still needed further study.

\textbf{Interaction Capability.} 
Interactive ability is also an essential metric for assessing the advanced capabilities of spoken dialogue systems. As illustrated in Figure \ref{fig:main} (b), basic interactive ability refers to the system's capacity to allow users to interrupt the conversation at any time. In this context, it is crucial to evaluate whether the spoken dialogue model can promptly comprehend the user's new input and halt its current response. This is commonly measured using accuracy. Furthermore, it is important to assess whether the model can generate a coherent and appropriate response based on the new input, which ties back to previous evaluation standards related to text and speech intelligence.

In addition, in real-world scenarios, beyond basic interruptions, various discourse markers such as "okay", "haha" are often used to indicate interaction. Current spoken dialogue systems~\cite{nguyen2023generative} typically track the frequency of these markers as a standard evaluation metric. Looking ahead, it may be valuable to assess whether future spoken dialogue models can effectively and appropriately interrupt human speakers, which could also represent a key dimension for evaluation the interaction capability.



\textbf{Multimodal Capability.} Spoken dialogue models primarily focus on the audio modality for both input and output. However, considering the close coupling between video and audio modalities in practical applications of dialogue systems, recent advancements in spoken dialogue models have incorporated the understanding of video and images in the input stage~\cite{fu2024vita,li2024baichuan,park2024let} , indicate that future spoken dialogue models need to simultaneously understand visual information and audio information to achieve real-time Audio-Visual Understandings. The evaluation of such models generally still focuses on the evaluation of dialogue quality, that is, whether the generated dialogue and the reference dialogue are similar. Therefore, this aspect can still be evaluated using metrics such as BLEU \cite{papineni2002bleu} and METEOR \cite{banerjee2005meteor} to assess sentence semantic similarity. However, research in this area also focuses on the understanding of visual information, and how to evaluate the model's correct understanding of real-time visual information in dialogue is also a difficulty, still can be a future benchmark direction.

\textbf{Security.} Security is also an integral part of the evaluation, how to ensure that the output of the model complies with ethical and social norms is a critical aspect. Spoken dialogue models may encounter security issues such as harmful content generation, privacy pitfalls, bias, and adversarial attacks. There has been considerable research progress in evaluating text modalities \cite{dong2024attacks}. The commonly used metric is to evaluate the attack success rate of injection attacks and so on. However, there are relatively few evaluation methods in the field of speech modality. How to construct a dataset for attacking spoken dialogue models, avoid poisoning of speech data, and evaluate the model's speech defense capabilities as benchmarks are required further research in the field of spoken dialogue model evaluation in the future.

\subsection{Benchmark}

We list the common benchmarks for evaluating voice dialogue systems in the table\ref{table:benchmark}, and briefly introduce each benchmark in this section.

\quad$\bullet$ \emph{VoiceBench.}
VoiceBench's~\cite{chen2024voicebench} Key evaluation dimensions include general knowledge, instruction-following ability, and safety compliance. The benchmark incorporates both synthetic and real spoken instructions to simulate diverse speaker styles, environmental conditions, and content variations.
It challenges systems with tasks involving accent adaptability, handling noisy environments, and robustness against content irregularities such as grammatical errors, disfluencies, and mispronunciations. Additionally, it explores the systems' resilience under varying speaker characteristics (age, pitch, and speaking speed) and environmental challenges like reverberation, background noise, and far-field effects.

\quad$\bullet$ \emph{SUPERB.\cite{yang2021superb}}
The benchmark evaluates speech processing models across multiple dimensions, including content recognition, speaker modeling, semantic understanding, and paralinguistic analysis. Tasks in content recognition cover phoneme recognition, automatic speech recognition, keyword spotting, and query-by-example spoken term detection, focusing on transcription and content detection accuracy. Speaker modeling involves tasks like speaker identification, automatic speaker verification, and speaker diarization to assess speaker-related features. Semantic understanding includes intent classification and slot filling, testing models' ability to infer high-level meaning directly from raw audio. Paralinguistic analysis focuses on emotion recognition, capturing models' ability to interpret affective cues from speech. The evaluation framework uses publicly available datasets and conventional metrics to provide a standardized testbed for assessing generalizability and task-specific performance.

\quad$\bullet$ \emph{AudioBench.}
AudioBench~\cite{wang2024audiobench} evaluates spoken dialogue models across three primary dimensions: speech understanding, audio scene understanding, and voice (paralinguistic) understanding. It encompasses eight distinct tasks and leverages 26 datasets, including seven newly developed datasets. The evaluation emphasizes models' ability to handle instruction-following tasks conditioned on audio signals, addressing aspects such as speech recognition accuracy, environmental sound interpretation, and paralinguistic feature extraction (e.g., emotion, gender, accent).


\quad$\bullet$ \emph{AirBench.}
AIR-Bench~\cite{yang2024air} assesses the capabilities of Spoken dialogue models to understand and interact based on various audio types, including human speech, natural sounds, and music. It consists of two primary components: a foundation benchmark with 19 specific audio tasks and over 19,000 single-choice questions, and a chat benchmark featuring more than 2,000 open-ended audio-prompted questions. The foundation benchmark evaluates fundamental skills such as speech recognition, acoustic scene classification, and music genre identification, focusing on specific subtasks to diagnose model weaknesses. The chat benchmark tests the models' ability to handle complex, real-world audio-based queries, including mixed audio with varying loudness and temporal offsets.
AIR-Bench introduces a novel audio mixing strategy to simulate complex real-world scenarios and employs GPT-4-based evaluation to judge model-generated hypotheses against reference answers. 

\quad$\bullet$ \emph{SpokenWOZ.}
SpokenWOZ~\cite{si2024spokenwoz} evaluates task-oriented dialogue (TOD) systems in spoken scenarios, addressing challenges unique to spoken conversations, such as incremental processing, disfluencies, incomplete utterances, and Automatic Speech Recognition (ASR) noise. It introduces novel metrics to assess performance in tasks like cross-turn slot detection and reasoning slot detection, which require integrating information across multiple turns and reasoning from implicit cues. The benchmark encompasses multi-domain, human-to-human dialogues with diverse speech characteristics, testing systems on both textual and auditory inputs through large-scale annotated datasets with over 200,000 utterances and 249 hours of audio

\quad$\bullet$ \emph{SD-EVAL.}
SD-Eval~\cite{ao2024sd} evaluates spoken dialogue models across multiple dimensions, focusing on both spoken understanding and response generation beyond textual content. It assesses models' abilities to process three key types of information embedded in speech: content (e.g., linguistic meaning), paralinguistic cues (e.g., emotion, accent, age), and environmental context (e.g., background sounds).
The benchmark consists of four sub-tasks—emotion, accent, age, and environment—constructed from diverse datasets and totaling 7,303 utterances spanning 8.76 hours.

\quad$\bullet$ \emph{SuperCLUE.}
SuperCLUE evaluates spoken dialogue systems across four main dimensions: voice interaction, general capabilities, scenario applications, and response speed. Key metrics include interruption recognition, speech tone adjustment, semantic understanding, naturalness of speech, and memory accuracy. Additionally, it measures real-time data retrieval, reasoning ability, compliance with commands, and multilingual translation accuracy. Scenario-specific applications like emotional counseling, health consultations, and customer service are assessed for precision and effectiveness. The final aspect is response timeliness, focusing on latency and delay management.However, this benchmark is not open source and focuse on Mandarine ability


\quad$\bullet$ \emph{MMAU.}
 MMAU~\cite{sakshi2024mmau} evaluates spoken dialogue models across multiple dimensions, encompassing 27 distinct tasks divided into reasoning and information extraction categories. It assesses models on their ability to comprehend and reason about speech, sound, and music by leveraging advanced cognitive skills and domain-specific knowledge. Key evaluated areas include temporal event reasoning, speaker role mapping, emotional tone interpretation, eco-acoustic knowledge, phonemic stress pattern analysis, and melodic structure interpretation. It examines not just basic recognition or transcription capabilities but also models' proficiency in complex reasoning, contextual understanding, and the ability to extract and apply world knowledge. Additionally, MMAU scrutinizes performance consistency across varying difficulty levels, testing systems' depth of reasoning and robustness in real-world audio scenarios.



\section{Conclusion}
In this work, we systematically review the research related to spoken dialogue models, categorizing it according to two paradigms: cascaded spoken dialogue models and end-to-end spoken dialogue models. Additionally, we provide a detailed overview of the core technologies behind spoken dialogue models, including speech representation, training paradigms, streaming duplex systems, and interaction mechanisms. In the speech representation module, we classify and explain the representations from both the input and output perspectives, focusing on different types of semantic and acoustic representations. In the training paradigm module, we thoroughly discuss five modalities of alignment for spoken dialogue models, multi-stage training strategies, model architectures, and generation paradigms. Following this, we provide an in-depth analysis of streaming input and output for spoken dialogue models, as well as the related duplex interaction technologies. Finally, we compile key training resources, evaluation metrics, and benchmarks relevant to spoken dialogue models. We specifically address the evaluation of different levels of intelligence in spoken dialogue models across various scenarios. It is important to note that, given that spoken dialogue models are a relatively new and emerging technology, many aspects such as semantic and acoustic representations, still lack well-established paradigms. Therefore, at the end of each section, we include a dedicated discussion module to explore these open issues. We hope that this survey will contribute to the further development of the field of spoken dialogue systems.

\bibliographystyle{plain}
\bibliography{neurips_2024}

\appendix

\section{Resources about Music and Sound Datasets}
\label{music datasets}

This section lists commonly used music and sound datasets. These datasets cover different modalities, including environmental sounds, music, and emotional sounds, and provide some help for the development of future voice dialogue systems. The table \ref{table:soundmusicdataset} shows the basic information of each dataset, including the dataset name, number of samples, dataset link, and modality type.

\begin{table}[h]
\centering
\caption{Music and Non-Speech Sound Datasets}
\label{table:soundmusicdataset}

\resizebox{1\linewidth}{!}{%
\begin{tabular}{lcccl}
\hline
\textbf{Dataset} & \textbf{Size} & \textbf{URL} & \textbf{Modality} \\ \hline

ESC-50 \cite{piczak2015esc}             & 2,000 clips (5s each)           & \url{https://github.com/karoldvl/ESC-50}             & Sound \\
UrbanSound8K \cite{salamon2014dataset} & 8,732 clips (<=4s each)         & \url{https://urbansounddataset.weebly.com/urbansound8k.html} & Sound \\
AudioSet \cite{gemmeke2017audio}  & 2000k+ clips (10s each)                   & \url{https://research.google.com/audioset/}           & Sound \\
TUT Acoustic Scenes 2017 \cite{mesaros2016tut} & 52,630 segments            & \url{https://zenodo.org/record/400515}                & Sound \\
Warblr      & 10,000 clips                    & \url{https://warblr.net/}                              & Sound \\
FSD50K \cite{fonseca2021fsd50k}          & 51,197 clips (total 108.3 hours) & \url{https://zenodo.org/record/4060432}              & Sound \\
DCASE Challenge \cite{mesaros2017dcase}  & varies annually                 & \url{http://dcase.community/}                          & Sound \\
IRMAS \cite{bosch2012comparison}         & 6,705 audio files (3s each)     & \url{https://www.upf.edu/web/mtg/irmas}               & Music \\
FMA  \cite{fma_dataset} & 106,574 tracks            & \url{https://github.com/mdeff/fma}                    & Music \\
NSynth \cite{engel2017neural}            & 305,979 notes                   & \url{https://magenta.tensorflow.org/datasets/nsynth}  & Music \\
EMOMusic               & 744 songs                       & \url{https://cvml.unige.ch/databases/emoMusic/}        & Music \\
MedleyDB \cite{bittner2014medleydb} & 122 multitrack recordings     & \url{https://medleydb.weebly.com/}                     & Music \\
MagnaTagATune    & 25,863 clips (30s each)         & \url{https://mirg.city.ac.uk/codeapps/the-magnatagatune-dataset} & Music \\
MUSDB~\cite{rafii2017musdb18} & 150 songs & \url{https://paperswithcode.com/dataset/musdb18} & Music \\
M4Singer~\cite{zhang2022m4singer} & 700 songs & \url{https://github.com/M4Singer/M4Singer} & Music \\
Jamendo &  600k songs & \url{https://www.jamendo.com/?language=en} & Music \\ 

\hline

\end{tabular}%
}

\end{table}

\section{Open-source Spoken Dialogue Models}

In this section, we provide a comprehensive list of publicly available and open-source spoken dialogue models in Table \ref{table:opensource_model}.

\begin{table}[]
\renewcommand{\arraystretch}{1.4}
\centering
\caption{A comprehensive list of publicly available  spoken dialogue models and their URL}
\resizebox{1\linewidth}{!}{%
\begin{tabular}{cc}
\hline
\textbf{Model} & \textbf{URL}                                      \\ \hline
AudioGPT       & \url{https://github.com/AIGC-Audio/AudioGPT}       \\
SpeechGPT      & \url{https://github.com/0nutation/SpeechGPT}              \\
Freeze-Omni & \url{https://github.com/VITA-MLLM/Freeze-Omni}              \\
Baichuan-Omni  & \url{https://github.com/westlake-baichuan-mllm/bc-omni} \\
GLM-4-Voice       & \url{https://github.com/THUDM/GLM-4-Voice }      \\
Mini-Omni     & \url{https://github.com/gpt-omni/mini-omni  }      \\
Mini-Omni2     & \url{https://github.com/gpt-omni/mini-omni2  }      \\
FunAudioLLM    & \url{https://github.com/FunAudioLLM}             \\
Qwen-Audio    & \url{https://github.com/QwenLM/Qwen-Audio }             \\
Qwen2-Audio    & \url{https://github.com/QwenLM/Qwen2-Audio }             \\
LLaMA3.1       & \url{https://www.llama.com}              \\
Audio Flamingo & \url{https://github.com/NVIDIA/audio-flamingo}             \\
Spirit LM      & \url{https://github.com/facebookresearch/spiritlm }             \\
dGSLM          & \url{https://github.com/facebookresearch/fairseq/tree/main/examples/textless_nlp/dgslm}              \\
Spoken-LLM     & \url{https://arxiv.org/abs/2305.11000}              \\
LLaMA-Omni     & \url{https://github.com/ictnlp/LLaMA-Omni }             \\
Moshi          & \url{https://github.com/kyutai-labs/moshi}          \\
SALMONN        & \url{https://github.com/bytedance/SALMONN}              \\
LTU-AS         & \url{https://github.com/YuanGongND/ltu}             \\
VITA           & \url{https://github.com/VITA-MLLM/VITA}             \\
SpeechGPT-Gen  & \url{https://github.com/0nutation/SpeechGPT}              \\ 
WavLLM &
\url{https://github.com/microsoft/SpeechT5/tree/main/WavLLM}\\
Westlake-Omni  & \url{https://github.com/xinchen-ai/Westlake-Omni}              \\ MooER-Omni  & \url{https://github.com/MooreThreads/MooER}              \\ Hertz-dev  & \url{https://github.com/Standard-Intelligence/hertz-dev}              \\ Fish-Agent  & \url{https://github.com/fishaudio/fish-speech}              \\ SpeechGPT2   & \url{https://0nutation.github.io/SpeechGPT2.github.io/}              \\ \hline
\end{tabular}}
\label{table:opensource_model}
\end{table}

\section{Open-source Codec Models}

In this section, we provide a comprehensive list of publicly available and open-source codec models, as shown in Table ~\ref{tab:all_codec}.

\begin{table}[htbp]
\centering
\caption{A comprehensive list of publicly available codec models and their URL}
\label{tab:all_codec}
\resizebox{1\linewidth}{!}{%
\begin{tabular}{cc}
\toprule
\textbf{Model}                    & \textbf{URL}  \\
\midrule

Encodec \cite{defossez2022high}                  & \url{https://github.com/facebookresearch/encodec}          \\
SoundStream \cite{zeghidour2021soundstream}              & \url{https://github.com/wesbz/SoundStream}                 \\
DAC \cite{kumar2024high}                     & \url{https://github.com/descriptinc/descript-audio-codec}  \\
WavTokenizer  \cite{ji2024wavtokenizer}           & \url{https://github.com/jishengpeng/WavTokenizer}  \\
SpeechTokenizer \cite{zhang2023speechtokenizer} & \url{https://github.com/ZhangXInFD/SpeechTokenizer}        \\
SNAC  \cite{siuzdak2024snac}           & \url{https://github.com/hubertsiuzdak/snac}  \\
SemantiCodec \cite{liu2024semanticodec}             & \url{https://github.com/haoheliu/SemantiCodec-inference}   \\
Mimi \cite{defossez2024moshi} &
\url{https://github.com/kyutai-labs/moshi} \\
HiFi-Codec \cite{yang2023hifi}              & \url{https://github.com/yangdongchao/AcademiCodec}        \\
FunCodec  \cite{du2024funcodec}               & \url{https://github.com/modelscope/FunCodec }              \\
APCodec  \cite{ai2024apcodec}                & \url{https://github.com/YangAi520/APCodec/tree/main}       \\
AudioDec \cite{wu2023audiodec}                 & \url{https://github.com/facebookresearch/AudioDec}         \\
FACodec  \cite{ju2024naturalspeech}                & \url{https://github.com/lifeiteng/naturalspeech3\_facodec} \\
Language-Codec  \cite{ji2024language}         & \url{https://github.com/jishengpeng/Languagecodec}         \\
XCodec  \cite{ye2024codecdoesmatterexploring}           & \url{https://github.com/zhenye234/xcodec}  \\
TiCodec  \cite{ren2024fewer}           & \url{https://github.com/y-ren16/TiCodec}  \\
SoCodec  \cite{guo2024socodec}           & \url{https://github.com/hhguo/SoCodec}  \\
FUVC  \cite{zheng2024fuvc}           & \url{https://github.com/z21110008/FUVC}  \\
HILCodec \cite{ahn2024hilcodec}           & \url{https://github.com/aask1357/hilcodec}  \\
LaDiffCodec \cite{yang2024generative}           & \url{https://github.com/haiciyang/LaDiffCodec}  \\
LLM-Codec \cite{yang2024uniaudio}           & \url{https://github.com/yangdongchao/LLM-Codec}  \\
SpatialCodec \cite{xu2024spatialcodec}           & \url{https://github.com/XZWY/SpatialCodec}  \\
BigCodec \cite{xin2024bigcodec} &
\url{https://github.com/Aria-K-Alethia/BigCodec} \\
SuperCodec \cite{zheng2024supercodec} &
\url{https://github.com/exercise-book-yq/Supercodec} \\
RepCodec \cite{huang2023repcodec} & 
\url{https://github.com/mct10/RepCodec} \\
EnCodecMAE \cite{pepino2023encodecmae} & 
\url{https://github.com/habla-liaa/encodecmae} \\ 
MuCodec \cite{xu2024mucodec} & 
\url{https://github.com/xuyaoxun/MuCodec} \\ 
SPARC \cite{cho2024articulatory} & 
\url{https://github.com/Berkeley-Speech-Group/Speech-Articulatory-Coding} \\
BANC \cite{ratnarajah2023m3} &
\url{https://github.com/anton-jeran/MULTI-AUDIODEC} \\
SpeechRVQ \cite{shechtman24_interspeech} & 
\url{https://huggingface.co/ibm/DAC.speech.v1.0} \\
QINCo \cite{huijben2024residual} & 
\url{https://github.com/facebookresearch/Qinco} \\
SimVQ \cite{zhu2024addressing} &
\url{https://github.com/youngsheen/SimVQ} \\

\bottomrule
\end{tabular}}
\end{table}

\end{document}